 \documentclass[final,1p,times]{elsarticle}

\usepackage[utf8]{inputenc}
\usepackage[english]{babel}
\usepackage{caption}
\usepackage{subcaption}
\usepackage{booktabs}
\usepackage{siunitx}
\usepackage{float}
\usepackage{hyperref}
\usepackage{geometry}
\usepackage{array}
\usepackage{tabularx}
\usepackage{adjustbox}
\usepackage{rotating}
\usepackage{threeparttable}
\usepackage{makecell}
\usepackage[ruled,vlined]{algorithm2e}
\usepackage{amsmath}
\usepackage{longtable}
\usepackage{graphicx}%
\usepackage{multirow}%
\usepackage{amsmath,amssymb,amsfonts}%
\usepackage{amsthm}%
\usepackage{mathrsfs}%
\usepackage[title]{appendix}%
\usepackage{xcolor}%
\usepackage{textcomp}%
\usepackage{manyfoot}%
\usepackage{listings}%
\usepackage{natbib}
\usepackage{tabularx}
\usepackage{ragged2e}
\usepackage{xcolor}
\usepackage{siunitx}
\usepackage{lineno}

\newcommand{\wmedia}[1]{\includegraphics[width=1\textwidth]{#1}} 
\newcommand{\LD}{\ensuremath{\text{LD}}}
\newcommand{\MOID}{\ensuremath{\text{MOID}}}

\newcommand{\tildH}{\ensuremath{\widetilde{H}}}
\newcommand{\Dv}{\ensuremath{\Delta v}}
\newcommand{\Dvzero}{\ensuremath{\Delta v_0}}

\newcommand{\dV}{\Delta v}
\newcommand{\vect}{\mathbf}
\journal{Acta Astronautica}

\begin{document}

\begin{frontmatter}
	
	
	\title{The Feasibility of Potentially Hazardous Asteroids Flybys Using Multiple Venus Gravity Assists}
	
	
	\author{Vladislav Zubko} 
	
	\affiliation{organization={Space Research Institute (IKI) of the Russian Academy of Sciences},
		addressline={Profsoyuznaya st. 84/32}, 
		city={Moscow},
		postcode={117997}, 
		state={},
		country={Russian Federation}}
	
	\begin{abstract}
		This work develops low-energy spacecraft (SC) trajectories using Venus gravity assists to study asteroids during heliocentric transfer segments between planetary encounters. The study focuses on potentially hazardous asteroids (PHAs) as primary exploration targets. This paper proposes a method for calculating SC trajectories that enable asteroid flybys after a Venus gravity assist. The method involves formulating and solving an optimization problem to design trajectories incorporating flybys of selected asteroids and Venus. Trajectories are calculated using two-body dynamics by solving the Lambert problem. A preliminary search for candidate asteroids uses an algorithm to narrow the search space of the optimization problem. This algorithm uses the V-infinity globe technique to connect planetary gravity assists with resonant orbits. The resonant orbit in this case serves as an initial approximation for the SC's trajectory between two successive planetary flybys. Four flight schemes were analyzed, including multiple flybys of Venus and asteroids, with the possibility of an SC returning to Earth. The proposed solutions reduce flight time between asteroid approaches, increase gravity assist frequency, and enhance mission design flexibility. The use of Venus gravity assists and resonant orbits ensures a close encounter with at least one asteroid during the SC's trajectory between two consecutive flybys of Venus, and demonstrates the feasibility of periodic Venus gravity assists and encounters with PHAs. The developed method was applied to construct trajectories that allow an SC to approach both co-orbital asteroids with Venus and PHAs via multiple Venus gravity assists. An additional study was carried out to identify asteroids accessible during the Earth-Venus segment in launch windows between 2029 and 2050.
	\end{abstract}

	
\end{frontmatter}




\section*{Introduction}
\label{sec:intro}
The study of asteroids located in the inner region of the Solar System, with orbits interior to the Earth one, is crucial for understanding planetary evolution and assessing potential threats to Earth. Currently, astronomical catalogs contain over 1 million minor bodies, but only approximately 3,000 of them have been spectrally classified. This highlights the need for new methods to research and classify asteroids, utilizing both ground-based facilities and spacecraft (SC).

Spacecraft missions for asteroid exploration are gaining increasing relevance. These include flyby studies and sample collection operations with subsequent sample return from the 81P/Wild 2 comet and the (25143) Itokawa, (162173) Ryugu, and (101955) Bennu asteroids to Earth. To date, 20 missions targeting 39 small bodies have been completed\footnote{Based on an analysis of data from Jet Propulsion Laboratory (JPL) database \url{https://ssd.jpl.nasa.gov/sb/targets.html} (access date: 01.03.2025)}.  However, the active development of new concepts and projects in this area continues, accompanied by a growing number of publications and projects from various space agencies.

The exploration of Near-Earth Asteroids (NEAs) represents a particularly significant focus within asteroid research. In recent decades, landmark sample return missions including Hayabusa \cite{kawaguchi2008hayabusa}, Hayabusa-2 \cite{tsuda2013system}, and OSIRIS-ReX (changed to OSIRIX-APEx) \cite{lauretta2017osiris} have been successfully completed. More recently, the DART equiped with CubeSat LICIACube \cite{rivkin2021double} asteroid redirection mission to the (65803) Didymos -- Dymorphos system has demonstrated new capabilities for planetary defense, with follow-up missions such as Hera \cite{michel2025hera} coupled with two CubeSat Milani \cite{pugliatti2022milani} and Juventas \cite{goldberg2019juventas} currently in development. This growing interest in asteroid exploration has also enabled smaller, more focused missions like M-ARGO \cite{walker2017miniaturised} and NEAScout \cite{mcnutt2014near, johnson2017near}, highlighting the expanding diversity of approaches to studying these celestial bodies. 

When designing trajectories for SC missions to asteroids, various strategies can be employed depending on a variety of factors. These factors include the number of target bodies, the use of gravity assists, and the type of propulsion system (chemical or electric). Typically, these trajectories are obtained by solving an optimization problem to find trajectories that meet criteria such as minimizing fuel consumption, minimizing flight time, or finding a balance when strict limits are imposed on any of these parameters.

The approaches can be broadly categorized as follows:

\begin{enumerate}
	\item Approaches involving direct or multi-revolution transfer trajectories from Earth to asteroids without gravity assists \cite{ball2009small, michel2010orbital, michel2014marcopolo, mcnutt2014near, johnson2017near, cataldi2022design}.
	
	\item Approaches utilizing flight schemes that employ a combination of Earth and Venus gravity assists with subsequent entry into a heliocentric orbit with a significantly elongated aphelion. Such schemes often include a series of maneuvers near Venus and/or Earth. Examples of such trajectories are presented in works \cite{bender1975multi, sukhanov1996trajectory, rivkin2020multi, sukhanov2024s, Bellome2024, Stikkle2024}; a similar strategy was implemented in the Lucy mission \cite{olkin2024mission}.
	
	\item Approaches utilizing the unstable dynamics of the few-body problem on various transfer segments, particularly --- in the vicinity of the Sun-Earth system libration points. Aspects of the practical application of such dynamics are described the works of \cite{walker2017miniaturised, de2020asteroid, mascolo2022optimal, Chen2023, pupkov2025construction, ivanyukhin2023designing, canalias2004assessment}.
	
	\item Approaches using invariant manifolds in the vicinity of libration points in the Sun-Earth system for guidance and control of a station to be sent to study a comet/asteroid that passes near Earth \cite{yamaguchi2025trajectory, sanchez2021esa, rebelo2024optimizing,jones2024comet, pupkov2025construction}.
	
\end{enumerate}

It is important to note that the approaches described above can be applied to the design of either spacecraft trajectories or asteroid trajectories. The latter case is relevant for planetary defense or asteroid exploration concepts where the asteroid itself is the controlled body, as exemplified in \cite{eismont2013possibility, ceriotti2016control, wang2025hazardous}.s

The approaches in the first group, subject to stringent $\Delta v$ constraints\footnote{$\Delta v$ denotes a change in the characteristic velocity after the execution of a rocket-powered maneuver.}, primarily focus on trajectories that are suitable for exploring the near-Earth asteroids (NEAs) \cite{michel2010orbital, michel2014marcopolo, ball2009small}. The second group focuses on Main Belt asteroids, which are the largest population of minor bodies. The approaches of this group focus on finding SC trajectories for encountering with Main Belt asteroids \cite{Urrutxua2015, Sanchez2018, Ionescu2022, Bolin2023, Bellome2024, Stikkle2024}. The third group also targets near-Earth asteroids (NEAs), as it relies on gravitational effects from the $n$-body problem to achieve exhaust velocity enough to escape from Earth's sphere of influence (SOI) \cite{de2020asteroid, mascolo2022optimal, Chen2023, pupkov2025construction, ivanyukhin2023designing, canalias2004assessment}.

Research from \cite{bender1975multi, sukhanov1996trajectory} has demonstrated that high-thrust propulsion systems, combined with gravity assists from Venus, Earth, and in some cases Mars, enable targeting multiple objects during Earth-to-Earth transfers at sufficiently low cost. This approach requires only a single $\Delta v$ maneuver\footnote{$\Delta v$ denotes a change in the characteristic velocity resulting from a rocket-powered maneuver.} for the initial flight to Venus, with negligible impulses required during subsequent gravity assist maneuvers. These studies showed that the patched-conic approximation significantly simplifies the initial search for candidate asteroids. A similar analysis was performed in \cite{Bellome2024}, where the authors applied Bellman's optimality principle to identify trajectories for multiple encounters with Main Belt asteroids using an SC equipped with low-thrust engines.

Recent studies \cite{simonov2025, Zubko2023a, Zubko2024a} have demonstrated the possibility of an asteroid encounter during the SC flight from Earth to Venus. In the article \cite{simonov2025}, ballistic schemes for Venus with possible launch periods between 2030 and 2040 were investigated. The analysis revealed that feasible trajectories for an asteroid encounter during the Earth-Venus transfer segment exist. In \cite{Zubko2023a, Zubko2024a}, the authors analyzed a flight scheme in which, after a Venus gravity assist, the SC enters a resonant orbit\footnote{Based on an analysis of data from the JPL database \url{https://ssd.jpl.nasa.gov/tools/sbdb_query.html} (access date: 01.03.2025)}, allowing an asteroid flyby before the next encounter with the planet. The analysis used the patched-conic approximation and solved the Lambert problem for each trajectory, with flybys of one of 120 target objects. It showed that at least one flyby was possible within the proposed scheme. The launch dates for missions in previous studies were chosen to coincide with the planned Venusian missions between 2029-2035: \textit{Venera-D} \cite{Eismont2019, Zasova2019, Zubko2023b}, \textit{DAVINCI+} \cite{Garvin2022}, \textit{VERITAS} \cite{Smrekar2016}, and \textit{EnVision} \cite{Widemann2020}.

It should be noted that, despite similar ideas being presented in previous research \cite{Zubko2023a, Zubko2024a}, where it was shown that an asteroid can be encountered by a spacecraft flying along a resonant 1:1 orbit, their scope was inherently limited to a single opportunistic flyby during a transfer trajectory designed for a specific planetary objective - landing on the surface of Venus. Therefore, those studies were more focused on reaching the desired region on Venus than on asteroid flybys.

This research focuses on the trajectories of the second group mentioned above. The main focus is on the trajectory for the SC using gravity assist maneuvers near Venus, which enables the SC to encounter asteroids while remaining near Venus's orbit, facilitating the simultaneous study of both the small bodies and the planet. This approach has the potential to significantly increase the effectiveness of multiple SC missions to asteroids and other planetary bodies for scientific purposes, thereby enhancing the scientific return of missions to the inner Solar System.

Four flight schemes were considered:

\begin{enumerate}
	\item asteroid flyby without gravity assists with a return to Earth: Earth $\to$ asteroid $\to$ Earth;
	
	\item asteroid flyby with gravity assists at Venus and  without a return to Earth: Earth $\to$ Venus $\to$ asteroid $\to$ Venus;
	
	\item multiple asteroid flybys without return to Earth: Venus $\to$ 1st asteroid $\to$ Venus $\to$ 2nd asteroid $\to$ Venus $\to$ $p$-th asteroid \ldots $\to$ Venus;
	
	\item multiple asteroid flybys with a return to Earth: Earth $\to$ Venus $\to$ asteroid $\to$ Venus \ldots $\to$ $p$-th asteroid $\to$ Venus  $\to$ Earth.
\end{enumerate}

Similarly to previously mentioned studies \cite{sukhanov2024s, Bellome2024, Urrutxua2015}, this study uses the Lambert problem to determine the flight trajectories of SC between each pair of bodies. However, this research focuses primarily on asteroid exploration within a region close to the orbit of Venus, where the asteroid population is distinctive. For this reason, the research uses a methodology that includes a primary search for asteroid targets using the V-infinity globe approach, which connects gravity assist maneuvers with resonant orbits. This provides initial data for a subsequent search, using patched-conic techniques, for optimal flight trajectories.

Within this study, PHAs\footnote{Potentially Hazardous Asteroids or PHAs are a special group of near-Earth asteroids for which the Minimum Orbit Intersection Distance (MOID) is less than 0.05 astronomical units, and the absolute magnitude is less than 22.} are considered as target objects for exploration. These objects are significant because of their potential threat to the Earth, making it particularly important to observe them from close distances.

An initial selection of asteroids from the Jet Propulsion Laboratory (JPL) database was performed. Subsequently, trajectories ensuring flybys with minimal $\Delta v$ costs, including launch and interplanetary maneuver costs, were determined. Then, the Earth $\to$ asteroid $\to$ Earth schemes were compared with the scheme proposed in this work, which includes multiple Venus flybys.

The comparison showed that the multi-stage scheme does not significantly increase flight time and, in some cases, allows all flybys to be completed in a shorter time. The feasibility of returning the SC to Earth after a series of maneuvers was also demonstrated. For example, the SC trajectory using Earth $\to$ Venus $\to$ 1st asteroid $\to$ Venus $\to$ 2nd asteroid $\to$ Venus $\to$ Earth scheme was calculated. It was shown that such a return is possible in most cases of a transfer to an asteroid involving a Venus gravity assist.

The analysis revealed that using a 1:1 resonant orbit with the Venus orbit (having a period of $\approx$224.699 days) significantly reduces the transfer duration to asteroids. Calculations show that in some cases, a Venus gravity assist reduces the total impulse $\Delta v$ compared to direct transfers when launch windows coincide. Moreover, transfers between Earth and Venus limit opportunities for flybys of asteroids, because the orientation of the SC orbits in this case depends on the position of both planets at the initial and terminal points, respectively. The use of resonant orbits allows aligning the SC's motion plane with the orbits of the target objects, thereby increasing the mission efficiency.

Another motivation for studying a particular region of space, namely the vicinity of the orbit of Venus, is the potential to discover unknown asteroids in an orbital resonance with Venus. A further significant objective is the investigation of the dust envelope located near the orbital distance of Venus \cite{dobritsa2025}. Currently, only five such asteroids are known \cite{simonov2025} and about 20 are suspected to be captured in resonance with the Venus orbital motion \cite{pan2025attempt}. Therefore, the exploration of the SC orbits identified in the current paper could yield interesting results that can greatly contribute to the state of the art in this field. Therefore, the exploration of the SC orbits identified in the current paper could yield interesting results that can greatly contribute to the state of the art in this field.

Therefore, this work's novel contribution is twofold. First, it proposes, explores and validates the feasibility of Venus-asteroid-Venus flight schemes. Second, it presents techniques for calculating the initial parameters needed to optimize these trajectories in terms of total $\Delta v$. Additionally, the catalog of the optimal SC trajectories with asteroid flybys using multiple Venus gravity assists is provided for 2029 and 2031 launch windows.

The paper is structured as follows: the Introduction establishes the study's relevance and reviews different trajectory design approaches. The first section, Calculation Techniques (Section \ref{sec:sc1}), is divided into subsections detailing the calculation of SC trajectories using the patched-conic approximation (Section \ref{sec:sc11}), the construction of trajectories for multiple flybys (Section \ref{sec:sc12}), and the specific technique for searching candidate asteroids (Section \ref{sec:sc13}). The second section, Results (Section \ref{sec:sc2}), presents the findings in subsections covering the selection of the target asteroid group (Section \ref{sec:sc21}), a comparative analysis of different flyby schemes (Section \ref{sec:sc22}), an example of a multi-asteroid flyby trajectory and an analysis of trajectories that return to Earth (Section \ref{sec:sc23}), refining results in high-fidelity model (Section \ref{sec:sc232}), a study of combined flight with Venus resonant asteroids and PHAs (Section \ref{sec:sc25}), and a study of flybys on the Earth-Venus segment (Section \ref{sec:sc24}). The summary of the advantages of the proposed methods as well as used flight schemes is provided in Conclusion (Section \ref{sec:con}), followed by a section for Acknowledgments.

\section{Methods}
\label{sec:sc1}
\subsection{Interplanetary Flight Trajectory Design Using the Patched-Conic Approximation}
\label{sec:sc11}
For all considered flight schemes, the patched-conic method was used \cite{de2007comparison}. According to this method, the SC transfer trajectory between each pair of celestial bodies was considered as a conic section. The SOIs of these celestial bodies are contracted to a point in heliocentric motion. The SC transfer trajectory between two bodies was determined by solving the Lambert problem. This problem determines the heliocentric velocity vectors at the initial and terminal points of the motion based on known heliocentric position vectors for the initial and terminal positions, as well as the time of flight. In this work, Izzo's method was used to solve the Lambert problem \cite{izzo2015revisiting}.

Since the Izzo method allows calculating the transfer trajectory between two celestial bodies, even when the angular distance exceeds $2\pi$, to avoid encountering with solutions involving multiple revolutions around the Sun, the maximum number was limited to four in this work.

The SC trajectory was calculated using the following algorithm:

\begin{enumerate}
	\item The required number $p$ of asteroid and $k$ (with $k > p$) planetary encounters was defined.
	\item $s+1$ transfer segments between each pair of celestial bodies were defined. The number of transfer segments is determined by $s = p + k$, and an additional segment is added corresponding to the Earth-Venus segment.
	\item The SC position vectors at the initial time, planet flyby, and asteroid flyby at times $t_j$ were defined as follows: $\vect{r}_0(t_0), \vect{r}_1(t_1), \dots, \vect{r}_j(t_j), \dots, \vect{r}_s(t_s), \vect{r}_{s+1}(t_{s+1})$.
	\item The Lambert problems for each segment to obtain the velocities was solved and velocities at launch and each body encounter were obtained: 
	\[\vect{v}_0(t_0), \vect{v}_1(t_1), \dots, \vect{v}_j(t_j), \dots, \vect{v}_s(t_s), \vect{v}_{s+1}(t_{s+1}).\]
	\item The asymptotic velocities at each planet flyby were calculated:
	\[
	\vect{v}_{\infty,0} = \vect{v}_0 - \vect{v}_{pl,0},\quad
	\vect{v}_{\infty,1} = \vect{v}_1 - \vect{v}_{pl,1},\quad
	\dots,\quad
	\vect{v}_{\infty,j} = \vect{v}_j - \vect{v}_{pl,j}
	\dots,\quad
	\vect{v}_{\infty,s} = \vect{v}_{s} - \vect{v}_{pl,s}
	\]
	\item The following cost function was defined:
	\begin{align}
		G &= \dV_{tot}(t_0, t_1, \dots, t_{s+1}) = \dV_0(t_0, t_1) \nonumber \\
		&+ \sum_{j=1}^{s-1} \left[ \dV_{\pi}(t_j, t_{j+1}) + \dV_t(t_j, t_{j+1}) + \dV_{ast}(t_{j+1}, t_{j+2}) \right]
		\label{eq:functional}
	\end{align}
	where the components are defined as follows:
	\begin{itemize}
		\item $\dV_0 =  \sqrt{2\frac{\mu_{\text{E}}}{r_{\text{LEO}}} + (v_{\infty, 0}^{+})^2} - \sqrt{\frac{\mu_{\text{E}}}{r_{\text{LEO}}}}$ is the impulse required to launch from a low Earth orbit (LEO) with a radius of 6571 km; $\mu_{\text{E}}$ -- gravitational parameter of the Earth.
		\item $\dV_{\pi, j} = \sqrt{2\frac{\mu_{\text{pl}}}{r_{\pi, j}} + (v_{\infty, j}^{-})^2} - \sqrt{2\frac{\mu_{\text{pl}}}{r_{\pi, j}} + (v_{\infty, j}^{+})^2}$ is the impulse at the planet's periapsis during its flyby, where $r_{\pi, j}$ is the periapsis radius; $\mu_{\text{pl}}$ -- gravitational parameter of the planet.
		\item $\dV_{t, j}$ is the impulse required to adjust the velocity asymptote (in cases where the turn angle exceeds the natural one);
		\item $\dV_{ast, j} = \| \vect{v}^{-}_{\text{ast, j}} - \vect{v}^{+}_{\text{ast, j}} \|$ is the velocity change during asteroid flyby.
		The impulses described above were applied if the following condition for an unpowered (free) flyby was violated:
		\begin{align}
			A_{\text{min}} \leq \arccos{\frac{\vect{v}^{-}_{\infty,j} \cdot \vect{v}^{+}_{\infty,j}}{{v}_{\infty,j}^2}} \leq A_{\text{nat}}
			\label{eq:constr}
		\end{align}
		
		where $A_{\text{min}} = \arcsin \left(1 / \left(1 + r_{SOI} v_{\infty,j}^2 / \mu_{\text{pl}}\right)\right)$ and  $A_{\text{nat}} = \arcsin \left(1 / \left(1 + r_{\pi,\min} v_{\infty,j}^2 / \mu_{\text{pl}}\right)\right)$ are the minimum and natural turn angles of the asymptotic velocity, achieved at the distance of the planet's SOI ($r_{SOI}$) and minimum planet flyby distance ($r_{\pi, \min}$) correspondingly.

	\end{itemize}
	\item The goal was to find the trajectory that minimizes:
	\[
	G^* = \min_{ t_0, t_1, t_2, t_3, t_4, \dots, t_{s+1}} \Delta v_{tot}.
	\]
	A two-stage optimization algorithm was applied:
	\begin{itemize}
		\item A global search was performed to locate the region containing the global minimum of $G^*$. This search was performed using a meta-heuristic genetic algorithm from the evolutionary class of optimization techniques -- differential evolution \cite{price2013differential}.
		
		\item A search for the exact minimum of the cost function $G^*$ in the previously found region using a gradient-based Broyden-Fletcher-Goldfarb-Shanno algorithm \cite{morales2002numerical}.
	\end{itemize}
	\item After the optimization procedure, the following constraints are checked:
	\begin{align}
		\begin{cases}
			\dV_0 \leq \dV_0^{\lim} \\
			\sum_{j=1}^{s-1} (\dV_{\pi, j} + \dV_{t, j}) \leq l_1 \, \\
			\sum_{j=1}^{s-1} (\dV_{ast, j+1}) \leq l_2
			\label{eq:restr}
		\end{cases}
	\end{align}
	where $\dV^{lim}_0$ is the limit on the launch impulse value, adopted according to technical constraints on existing upper stages of launchers; $l_1$ and $l_2$ are constraints on the total impulse for planetary flybys and asteroids, respectively.
\end{enumerate}
In this work, $\dV_0$ value was set to 4.1 km/s, reflecting the current development of the rocket industry, which allows the use of liquid oxygen/hydrogen upper stages. The theoretical limit for this impulse is about 4.53 km/s, while the more realistic impulse is close to 3.49--3.62 km/s (for existing rockets)\cite{mazzetti2016paraffin}. A value of $\dV^{lim}_0 = 4.1$ km/s was chosen, consistent with current upper stage capabilities, although a detailed launch performance analysis was beyond the scope of this study.

Equation (\ref{eq:restr}) imposes constraints on the total impulse for planetary flybys (the second constraint) and for asteroid flybys (the third constraint). The constraints on the total impulse for planetary and asteroid flybys depend on the number of asteroids in the flight scheme. For a scheme including one asteroid flyby, both constraints are set to $10^{-3}$ km/s, while for five or more asteroids, their values are taken as $l_1=0.1 \dV^{lim}_0$ and $l_2=0.025 \dV^{lim}_0$. The value $l_2 = l_1/5$, selected empirically, ensures that an asteroid flyby requires a significantly lower impulse than a planetary flyby. 

It can be noticed that for trajectories satisfying the condition $\sum_{j=1}^{2} (\Delta v_{\pi, j} + \dV_{t, j} + \Delta v_{ast, j+1}) \leq 10^{-3}$ km/s, the duration between successive planetary gravity assists fall into one of two categories:

\begin{equation}
	\begin{aligned}
		& \Delta t_{j, j+2} = \sum_{j=1}^{2} (t_{j+1} - t_j) \neq nP_{pl}, \\
		& \Delta t_{j, j+2} = \sum_{j=1}^{2} (t_{j+1} - t_j) \approx nP_{pl}. \\
	\end{aligned}
	\label{eq:resonance}
\end{equation}

where $P_{pl}$ is the planet's orbital period.

The above relations imply that for a transfer trajectory obtained by connecting transfer segments initiating and terminating on the same celestial body, the heliocentric trajectory of the SC can be approximated by corresponding $m:n$ resonant orbits connecting at the moment of the asteroid approach. Based on this, one can formulate the simplification to the cost function.

\textbf{Simplification 1.} The cost function (\ref{eq:functional}) is modified assuming that the SC transfer duration between two successive Venus flybys is known and exactly equals $nP_{pl}$ days, then $t_{j+2} = nP_{\text{pl}} + t_j$. Thus, the original cost function (\ref{eq:functional}) can be rewritten as

\begin{align}
	\widetilde{G} = \dV_0(t_0, t_1) + \sum_{j=1}^{s-1} \left[ \dV_{\pi}(t_j, t_{j+1}) +\dV_{t}(t_j, t_{j+1})+ \dV_{ast}(t_{j+1}, t_j + nP_{\text{pl}}) \right]
	\label{eq:simplified}
\end{align}

The obtained cost function is then minimized $ \widetilde{G} \to \min$ over the set of time moments $t_0, t_1, \dots, t_{s}$.

These rules for transfer segment sequence and duration (indexed by $j$) apply to SC trajectories that include an Earth-to-Venus segment. Alternatively, when such a segment is absent, for example, in an Earth-asteroid-Earth flight scheme, the number of segments are reduced by one and the algorithm described in this section can be applied similarly.

\subsection{Construction of Trajectories for Planetary and Several Asteroid Flybys}
\label{sec:sc12}

The method established in Section \ref{sec:sc11} is suitable for determining an optimal trajectory for a predefined sequence of bodies. It allows us to easily solve the problem of determining the optimal trajectory when the SC passes only one asteroid. However, when the number of encountered asteroids exceeds one, it becomes necessary to conduct a preliminary search for suitable candidate asteroids and determine the range of possible flyby dates.

This section provides an algorithm for the preliminary selection of candidate asteroids at each flyby of the planet. The geometric relationship is used to approximate the SC trajectory with a resonant orbit. A V-infinity globe approach, which connects gravity assists and resonant orbit design in a two-body problem, is used to determine the initial turn angle. This initial angle is determined by the requirement for close encounters between the SC and an asteroid on a resonant orbit.

Assuming, to a first approximation, that the eccentricity of Venus's orbit is negligible, a set of unit vectors describing a coordinate system suitable for asymptotic velocity transformations are written as:

\begin{equation}
	\mathbf{\xi}^0 = \frac{\mathbf{v}_{pl}}{\left\| \mathbf{v}_{pl} \right\|}, \quad
	\mathbf{\zeta}^0 = \frac{\mathbf{r}_{pl} \times \mathbf{v}_{pl}}{\left\| \mathbf{r}_{pl} \times \mathbf{v}_{pl} \right\|}, \quad
	\mathbf{\eta}^0 = \xi^0 \times \zeta^0
	\label{eq:syst}
\end{equation}

where $\mathbf{\zeta}^0$ points along the planet velocity vector, $\mathbf{\zeta}^0$ is directed along the angular momentum vector, and $\mathbf{\eta}^0$ completes the dextral orthonormal triad.

\begin{figure}[H]
	\centering
	\includegraphics[scale=0.5]{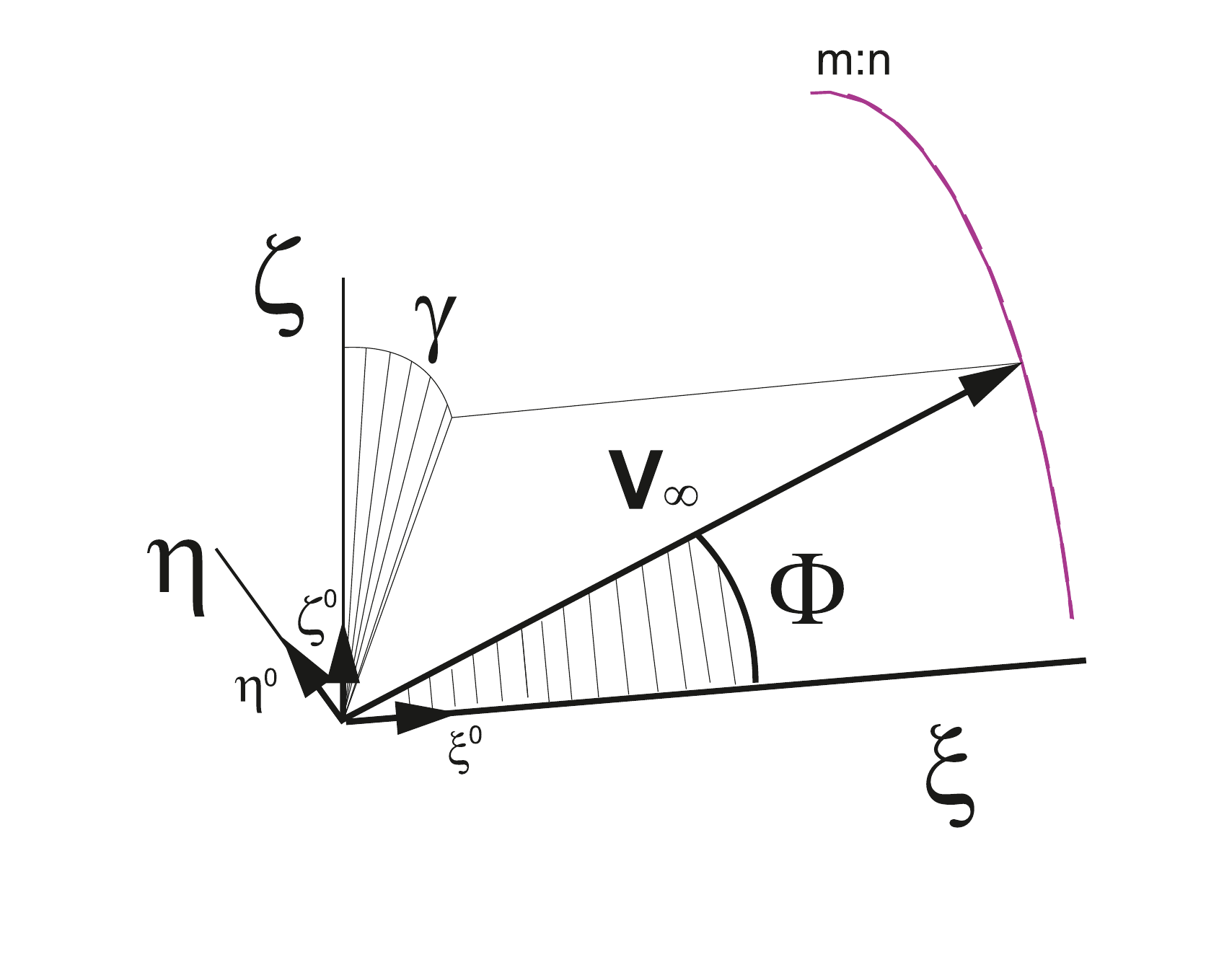} 
	\caption{SC asymptotic velocity at Venus flyby in the  $\xi \zeta \eta$ coordinate system}
	\label{fig:frame1}
\end{figure}

The asymptotic velocity vector of the SC at the Venus flyby is shown in Figure \ref{fig:frame1}. In the coordinate system (\ref{eq:syst}), the vector $\mathbf{v}_{\infty}$ can be defined by introducing angles  $\Phi, \gamma$ \footnote{
	The angles $\Phi, \gamma$ introduced in this work follow \cite{Zubko2023a, Zubko2022}), while in other papers these angles may be denoted differently, usually they are called \"{O}pic variables \cite{masat2022different} and denoted $\theta, \psi$; in \cite{strange2007mapping2} the notations $\alpha, \kappa$ are used; in \cite{sukhanov2025} the notations $\xi_1, \xi_2$ are used.
} as follows:

\begin{equation}
	\mathbf{v}_{\infty} = v_{\infty} \left[\cos \Phi,\  \sin \Phi \sin \gamma,\  \sin \Phi \cos \gamma \right]^T.
	\label{eq:vect}
\end{equation}

Therefore, the SC state vector (heliocentric position vector of SC $\mathbf{r}_{pl},$ and heliocentric velocity vector $\mathbf{v}_{pl}$) after Venus gravity assist can be written as:

\begin{equation}
	[\mathbf{r}_{sc}, \mathbf{v}_{sc}] = [\mathbf{v}_{pl}, \mathbf{v}_{pl} + \Xi \mathbf{v}_{\infty}].
	\label{eq:vect2}
\end{equation}

where $\Xi$ is the state transition matrix from $\xi \zeta \eta$ to Heliocentric ecliptic J2000 coordinate system.

Assuming that $\mathbf{v}_{\infty} = const$ during the Venus flyby, the m:n resonant orbit is completely determined by $\Phi, \gamma$, where $\Phi$ completely determines the $m:n$ ratio, and $\gamma$ determines the orientation of the resonant orbit relative to the planet's orbital plane at $\Phi = const$. Thus, further in the work, by fixing the value of $\Phi$, the resonant orbit is defined by the parameter $\gamma$.

Note that the theoretical investigation of resonant orbit design was conducted in \cite{strange2007mapping2, masat2022different} while the general case of resonant and non-resonant double planetary flybys was widely described and analyzed in the paper \cite{sukhanov2025}.

\subsection{Algorithm for Rapid Search for the Candidate Asteroids}
\label{sec:sc13}

\begin{figure}[H]
	\centering
	\includegraphics[scale=0.8]{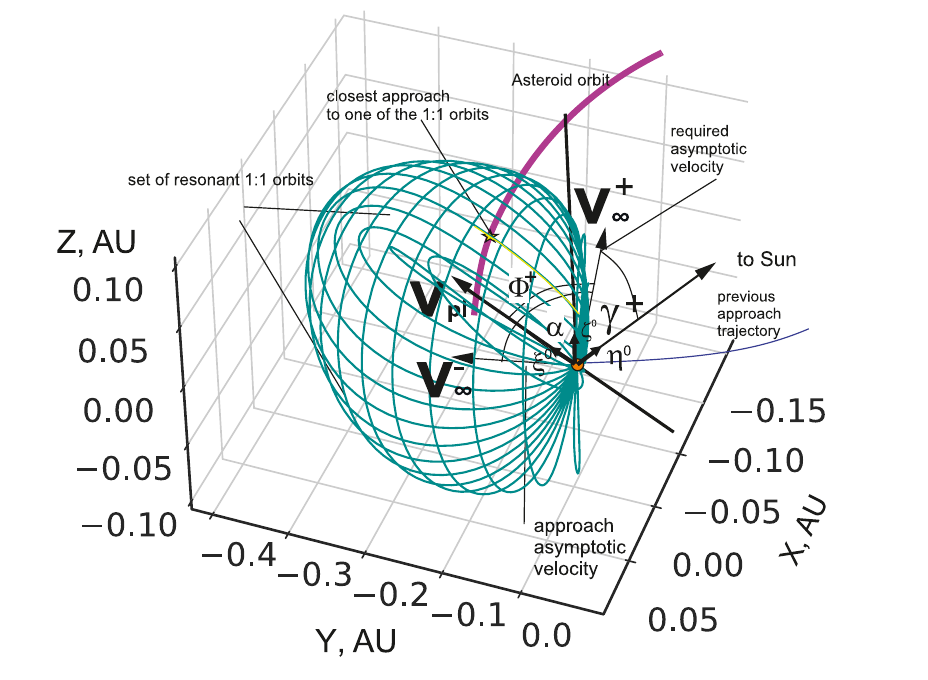} 
	\caption{The geometric interpretation of the proposed technique}
	\label{fig:frame1a}
\end{figure}

Resonant orbits serve as an initial guess for constructing the SC trajectory, allowing an asteroid flyby (defined as $\Delta v \leq 1$ m/s). Therefore, to select candidates from the entire set of asteroids, the following algorithm is used:

The core idea of the proposed approach is illustrated in Fig.~\ref{fig:frame1a}. The figure depicts the set of resonant orbits formed by possible resonances available at the current $v_\infty$ value during the Venus approach. An asteroid whose orbit intersects this set of orbits can have a close encounter with one of the resonant trajectories. Therefore, if such an encounter occurs at a sufficiently small distance—small enough to be compensated during the full optimization process described in Section~\ref{sec:sc11}—then the asteroid is included in the set of candidate asteroids.

The algorithm for finding suitable candidate asteroids, based on the core idea presented and illustrated above, is provided in \ref{alg:asteroid_selection}.

\scriptsize
\begin{algorithm}[H]
	\caption{Selection of candidate asteroids and construction of trajectories for multiple asteroids flybys}
	\label{alg:asteroid_selection}
	\KwIn{Set of PHAs $\mathcal{A}$, time of SC and Venus encounter $t_j$; position
		and velocity of SC in heliocentric coordinates $\textbf{r}(t_j)_{},\textbf{v}(t_j)_{}$; maximum allowed \MOID value:   $\rho^{\max}$; maximum allowed value of minimum distance between SC and an asteroid on interval $(t_j, t_{j+2})$, where $t_{j+2}=t_j+nP_{pl}$; $\Delta r_{\text{min}}^{\text{lim}}$ maximum allowable minimum approach distance between SC and an asteroid; $A_{\text{nat}}$; $\Phi (m:n, v_{\infty, j}) = const$}
	
	\KwOut{Set of candidate asteroids $\mathcal{A}_c$}
	
	\ForEach{asteroid $a \in \mathcal{A}$}{
		Compute the minimum MOID between the SC's orbit and the asteroid's orbit over $\gamma$: $(\rho_{\text{moid}}^{\text{sc-ast}})_{\min} = \min\limits_{\gamma} \rho_{\text{moid}}^{\text{sc-ast}}(\gamma)$\; \ $\tilde{\theta}_{j+1} = f(\theta_j, (\rho_{\text{moid}}^{\text{sc-ast}})_{\min})$ is the true anomaly of the SC at the minimal intersection point between the SC and asteroid orbits \;
		\ $\tilde{t}_{j+1} = f(\tilde{\theta}_{j+1})$ is the time in UTC at which the SC reaches the specified position $\theta_j$ \;
		$\tilde{\gamma}^{+} = \arg\min\limits_{\gamma} \rho_{\text{moid}}^{\text{sc-ast}}(\gamma)$ \ is parameter at which the minimum value of the $\MOID$ parameter is achieved among all resonant orbits of a given $m:n$\;
		\If{$\rho_{\text{moid}}^{\text{sc-ast}} \leq \rho_{MOID}^{\max}$}{
			Determine $ \Delta r_{\min}^{\text{sc-ast}} = \min\limits_{\begin{array}{cc}
					\tilde{\gamma}^{+}-\gamma^{lim} \leq \gamma \leq \tilde{\gamma}^{+}+\gamma^{lim}  \\
					\tilde{t}_{j+1}-\delta t^{lim} \leq t \leq t_{enc}+\delta t^{lim}
			\end{array}}  \|\mathbf{r}(t, \gamma) - \mathbf{r}_{\text{ast}}(t, \gamma)\|$\;
			\If{$ \Delta r_{\min}^{\text{sc-ast}} < \Delta
				r_{\text{min}}^{\text{lim}}$}
			
			{
				Check the condition:
				\[
				A_{\text{min}} \leq \arccos(\tilde{\mathbf{v}}_{\infty}^{-} \cdot \tilde{\mathbf{v}}_{\infty}^{+}) \le A_{\text{nat}};
				\]
				where $\tilde{\mathbf{v}}_{\infty}=\frac{\vect{v}_{\infty}}{v_{\infty}}$;\
				
				\If{constraint is satisfied}{
					Add $a$ to $\mathcal{A}_c$\;
					Determine $[t_{j+1}^{*}, \gamma^{+*}] = \arg \min\limits_{\begin{array}{cc}
							\tilde{\gamma}^{+}-\gamma^{lim} \leq \gamma \leq \tilde{\gamma}^{+}+\gamma^{lim}  \\
							\tilde{t}_{j+1}-\delta t^{lim} \leq t \leq t_{enc}+\delta t^{lim}
					\end{array}} \|\mathbf{r}(t, \gamma) - \mathbf{r}_{\text{ast}}(t, \gamma)\||$\;
				}
			}
		}
	}
	
	\Return{$\mathcal{A}_c$}\;
\end{algorithm}

\normalsize

It should be noted that the threshold values for the algorithm when used in this work framework were selected empirically through multiple simulations.The motivation for choosing empirical values stems from the highly nonlinear dependence of between $\Delta r$, $\rho_{moid}$ and $v_\infty$, $\gamma$, even at a fixed $\Phi$ value. In the considered case, the true anomaly at which the closest encounter occurs is a critical parameter for selecting appropriate threshold values for $\Delta r$ and $\rho_{moid}$. However, this specific problem was beyond the scope of the current research, and the final threshold values were chosen to ensure a comprehensive initial search. The objective of this paper was to first identify suitable asteroids, then subject this set to the optimization problem, and finally filter them to obtain optimal trajectories. Therefore, the purpose of this algorithm is to narrow the search space for optimal trajectories that would ensure nearly impulse-free flybys of asteroids and, preferably, planets.

The following empirical values were used for the limiting parameters: $\Delta r^{\text{lim}}_{\min} = 5$ million km, $\rho_{moid}^{\max} = 5 $ million km, $\gamma^{lim}=15^{\circ}$, and $\delta t^{lim}$ = 15 days. The parameter $\rho_{moid}^{\text{sc-ast}}$ (also designated as $\rho_{moid}$) represents the minimum distance between the orbits of the SC and the asteroid and was calculated using an analytical algorithm from \cite{Bonnano2000}. After compiling the candidate set, the steps from Section \ref{sec:sc11} were repeated for each asteroid in $\mathcal A_c$, with the optimization problem size increased by two additional segments.

\textit{Remark.} This work uses an analytical approximation of the MOID parameter (designated as AMOID in \cite{Bonnano2000}), which allows for approximating the characteristic change in the minimum distance between the orbits of the SC and the asteroid near local minima. However, since this approximated parameter (AMOID) can differ significantly from the true MOID, the search algorithm accounts for this potential error by using an initially overestimated value for $\rho_{\text{moid}}^{\text{sc-ast}}~\leq~\rho^{\max}$. Furthermore, the value of $\rho_{\text{moid}}^{\text{sc-ast}} \leq \rho^{\max}$ is chosen considering that the minimum MOID and  $\Delta r_{\min}^{\text{sc-ast}}$ may occur at different values of $ \gamma $.

\begin{figure}[H]
	\centering
	\wmedia{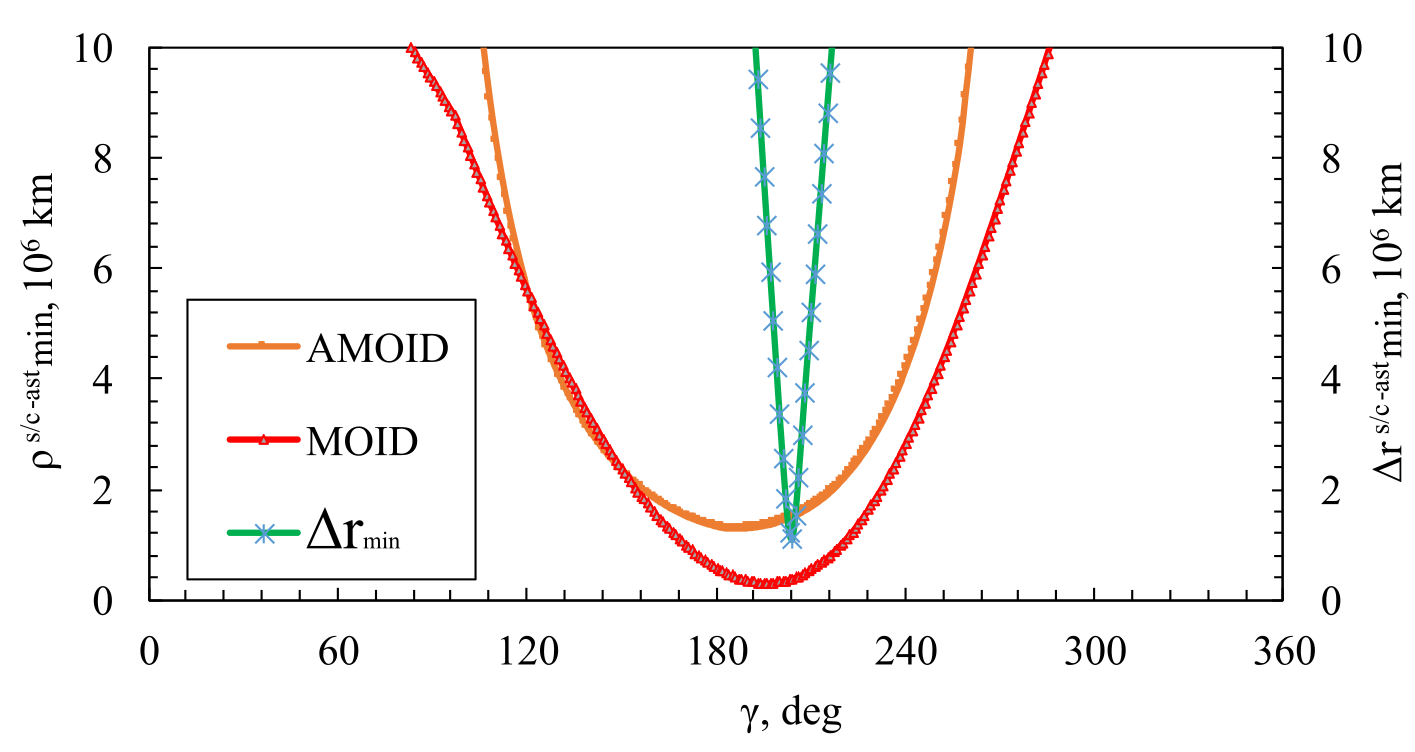} 
	\caption{Comparison of the AMOID, MOID curves and the minimum distance between the SC and the asteroid as a function of the 1:1 resonant orbit, as a function of the parameter $\gamma$.}
	\label{fig:amoid_comparison}
\end{figure}

For example, Fig.~\ref{fig:amoid_comparison} shows three different curves: AMOID, MOID, and $\Delta r_{\min}^{\text{sc-ast}}$ plotted against $\gamma$  for the 2000 DO1 asteroid. Figure \ref{fig:amoid_comparison} shows that the AMOID curve generally approximates the MOID values near the minimum, while the extremum of $\Delta r_{\text{sc-ast}}^{\text{min}}$ is different (shifted several degrees to the right in $\gamma$)

\section{Results}
\label{sec:sc2}
\subsection{Selection of Target Group of Asteroids From International Catalogs}
\label{sec:sc21}

\begin{figure}[H]
	\centering
	\wmedia{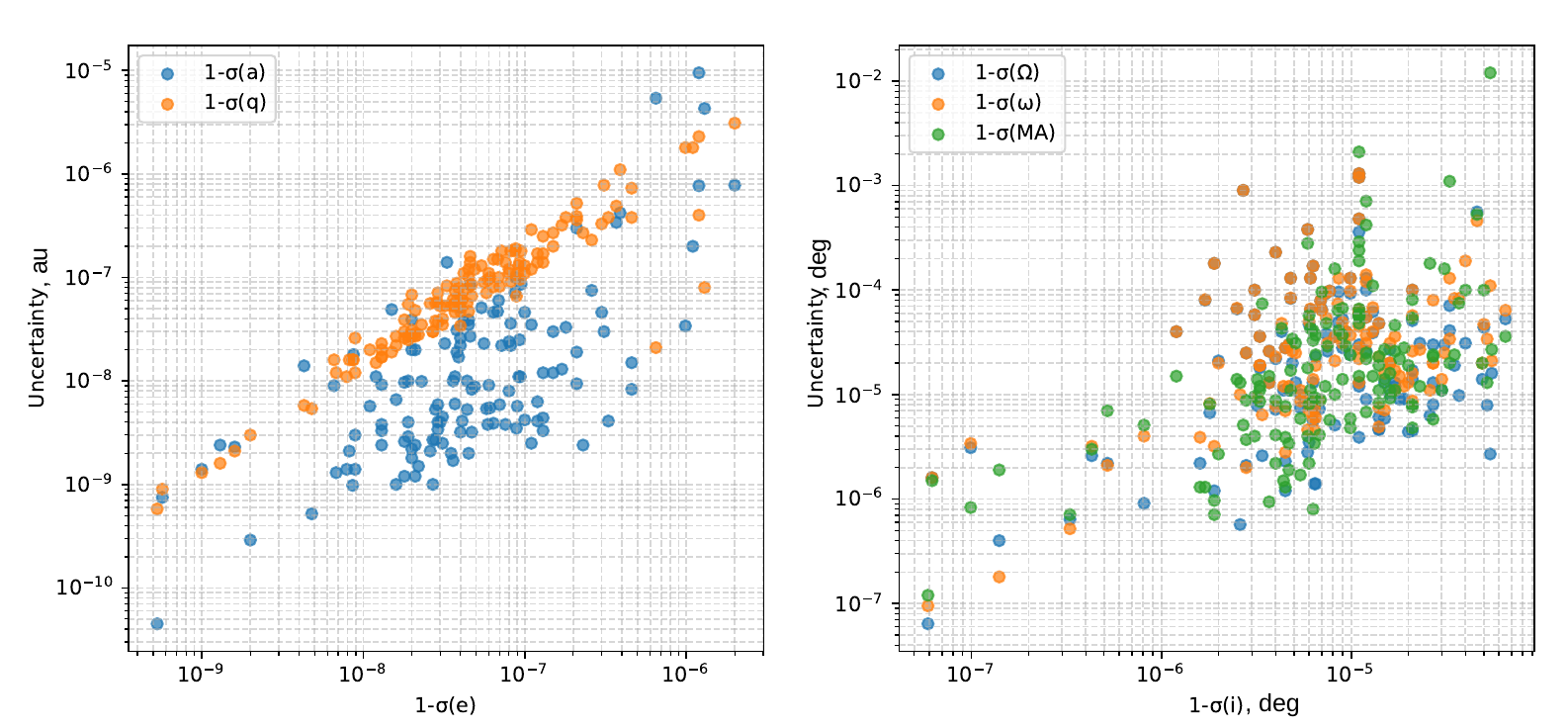}
	\caption{Distribution of uncertainties in orbital parameters for the selected asteroid group.}
	\label{fig:uncertainties}
\end{figure}

To identify suitable asteroid targets for the proposed Venus flyby missions, a comprehensive search using the NASA Jet Propulsion Laboratory (JPL) Navigation and Ancillary Information Facility (NAIF) database was conducted, which contains orbital data for over 1.5 million small Solar System bodies. The selection process employed the following sequential filtering criteria:

\begin{itemize}
	\item \textbf{Initial PHA Selection.} From the total database of 1,467,232 asteroids, at first  about 2,500 objects is classified as Potentially Hazardous Asteroids (PHAs) according to JPL and Minor Planet Center databases.
	
	\item \textbf{Earth MOID Restriction.} To focus on objects with the highest Earth-impact risk, asteroids with Minimum Orbit Intersection Distance $\rho_{moid} \leq 1.05 \LD$ (where $1 \LD = 384,000$ km) were selected. This criterion identified 139 candidate asteroids that formed the primary target set for initial mission design.
	
	\item \textbf{Orbit Determination Precision.} To ensure reliable trajectory design, the strict uncertainty thresholds to orbital elements were applied:
	\begin{itemize}
		\item Semi-major axis: $1$-$\sigma(a) \leq 10^{-5}$ au
		\item Perihelion distance: $1$-$\sigma(q) \leq 10^{-5}$ au  
		\item Eccentricity: $1$-$\sigma(e) \leq 10^{-5}$
		\item Angular elements (inclination, longitude of ascending node, argument of perihelion, mean anomaly): Total angular uncertainty $\leq 10^{-4}$ rad
	\end{itemize}
	Application of these uncertainty filters to the complete PHA dataset resulted in 2,030 well-characterized objects. The intersection of this refined set with the MOID-based selection yielded to the final target group of 139 asteroids. Figure \ref{fig:uncertainties} illustrates the distribution of orbital parameter uncertainties across the selected asteroids.
\end{itemize}

\begin{figure}[H]
	\centering
	\wmedia{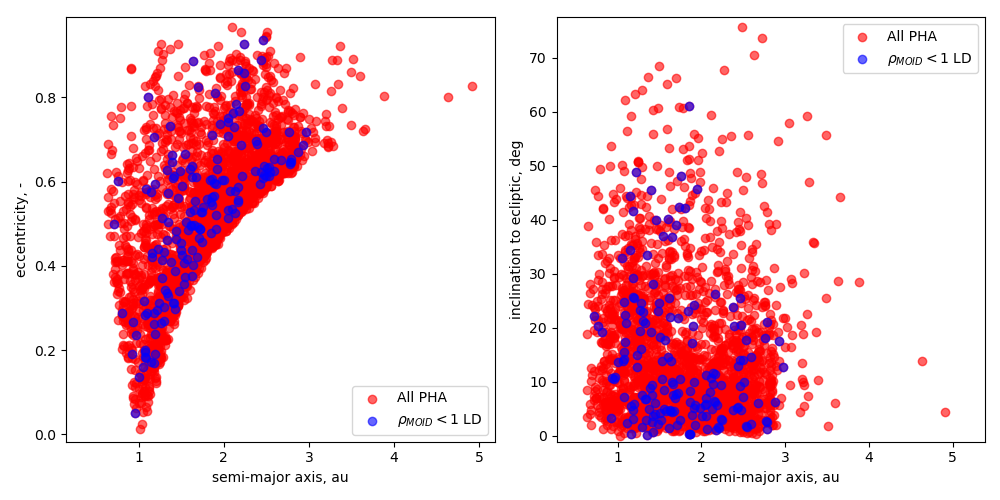}
	\caption{Distribution of orbital parameters for selected asteroids, showing semi-major axis versus eccentricity of their heliocentric orbits.}
	\label{fig:orbital_params}
\end{figure}

The target group includes both numbered and unnumbered asteroids to ensure comprehensive mission planning. While unnumbered asteroids currently present greater orbital uncertainty, they represent valuable future targets as orbit determinations improve. Figure \ref{fig:orbital_params} shows the distribution of orbital parameters for the final selection of 139 asteroids.

Although asteroids satisfying $\rho_{\MOID} \leq 1.05 \LD$ have orbits that bring them near Earth's orbital path, the actual collision risk remains low. Only a small subset of these 139 objects are projected to approach within one lunar distance in the foreseeable future. Table \ref{tab:moid1} provides representative examples of close-approach parameters for selected asteroids, including minimum approach distances to both Earth and the Moon, where available\footnote{https://naif.jpl.nasa.gov/naif/ (Access date 25.07.2025)}.

\begin{table}[h]
	\centering
	\caption{Asteroids with $\rho_{MOID} \leq 1.05 \LD$ and $d_{min} < 10$ LD}\label{tab:moid1}
	\footnotesize
	\begin{tabular}{@{}ccccc@{}}
		\toprule
		Name & Celestial Body & Date and time of approach & $d_{min}$ & $\rho_{MOID}$ \\
		-&   - & (UTC) & (LD) & (LD) \\
		\midrule
		(35396) 1997 XF11 & Earth & 26.10.2028 06:44 & 2.417 & 0.158 \\
		1997 XF11 & the Moon & 26.10.2028 07:39 & 3.451 & - \\
		(279744) 1998 KM3 & Earth & 27.10.2095 01:03 & 5.099 & 0.983 \\
		(308635) 2005 YU55 & the Moon & 07.11.2075 23:50 & 0.601 & - \\
		2005 YU55 & Earth & 08.11.2075 07:24 & 0.53 & 0.182 \\
		2009 XT6 & Earth & 17.12.2112 01:15 & 7.229 & 0.866 \\
		(835579) 2011 SM68 & Earth & 17.10.2072 15:25 & 1.877 & 0.267 \\
		2011 SM68 & the Moon & 17.10.2072 17:36 & 1.713 & - \\
		2016 FG60 & Earth & 22.07.2112 17:59 & 8.253 & 0.124 \\
		\bottomrule
	\end{tabular}
\end{table}

In Table \ref{tab:moid1}, $d_{min}$ is the minimum approach distance of the asteroid to Earth on the given date.

As shown in Table \ref{tab:moid1}, only a small fraction of the selected 139 objects approach Earth closer than 1 LD and thus pose little danger in the coming century. Nevertheless, constant refinement of the physical characteristics and orbits of these objects is a critical parameter in assessing their danger to the planet in the future. 

In cases where flybys of several asteroids are studied, an extended catalog of objects is used. This includes all known PHAs as of July 2025 that satisfy the constraints above - 2030 objects.

\subsection{Trajectories of Potentially Hazardous Asteroids Flybys}
\label{sec:sc22}

This section compares two schemes for asteroid flybys:

\begin{enumerate}
	\item E-V-A-V = Earth-Venus-Asteroid-Venus,
	\item E-A-E = Earth-Asteroid-Earth.
\end{enumerate}

To simplify the scheme descriptions, the following designations are used: 'P' for a planetary flyby (Venus), and 'A' for an asteroid flyby.

For the study, 139 asteroids were selected (see Section \ref{sec:sc21}). Trajectories were constructed for each of these asteroids according to the schemes described in Introduction Section, using the methods outlined in Sections \ref{sec:sc12} and \ref{sec:sc13}. Trajectories that did not satisfy the constraints in Eq. (\ref{eq:restr}) were excluded from further analysis. Two possible launch periods were considered: 2029 and 2031. The rationale for concentrating on these years is that many missions were planned to be launched to Venus, as described in the introduction.

In this section, comparisons are made between the times of flight from a low Earth orbit to encounter with Venus after asteroid flybys. Also, comparisons of $\Delta v_0$ values and asymptotic velocities at planetary and asteroid flybys are presented.

\subsubsection{General Analysis of Results}
\label{sec:sc221}

\begin{figure}[H]
	\centering
	\begin{minipage}[b]{0.45\textwidth}
		\centering
		\includegraphics[width=\textwidth]{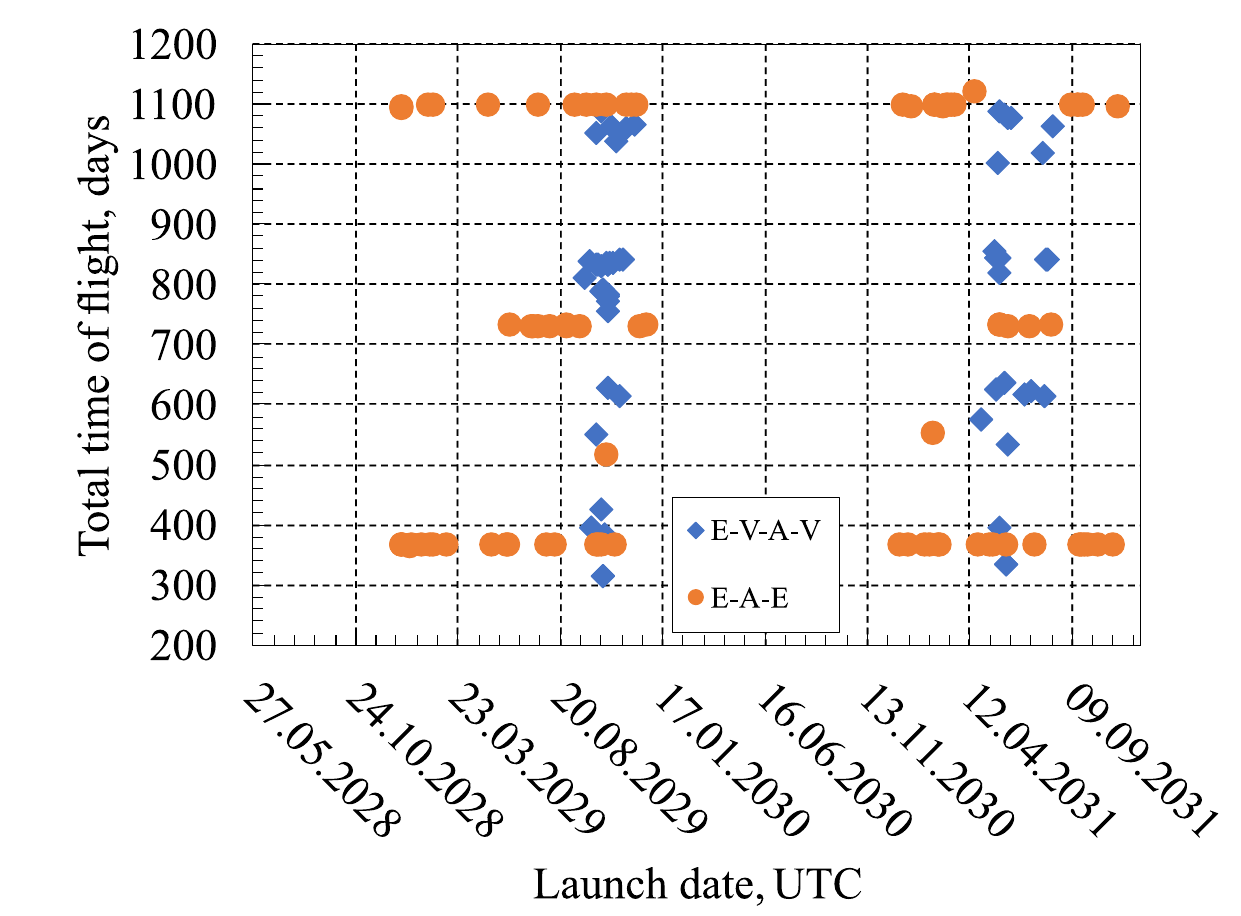}
		\captionof{figure}{Total transfer duration for the two studied schemes vs the launch date                                                                                                                                            }
		\label{fig:transfer_time}
	\end{minipage}
	\hfill
	\begin{minipage}[b]{0.5\textwidth}
		\centering
		\includegraphics[width=\textwidth]{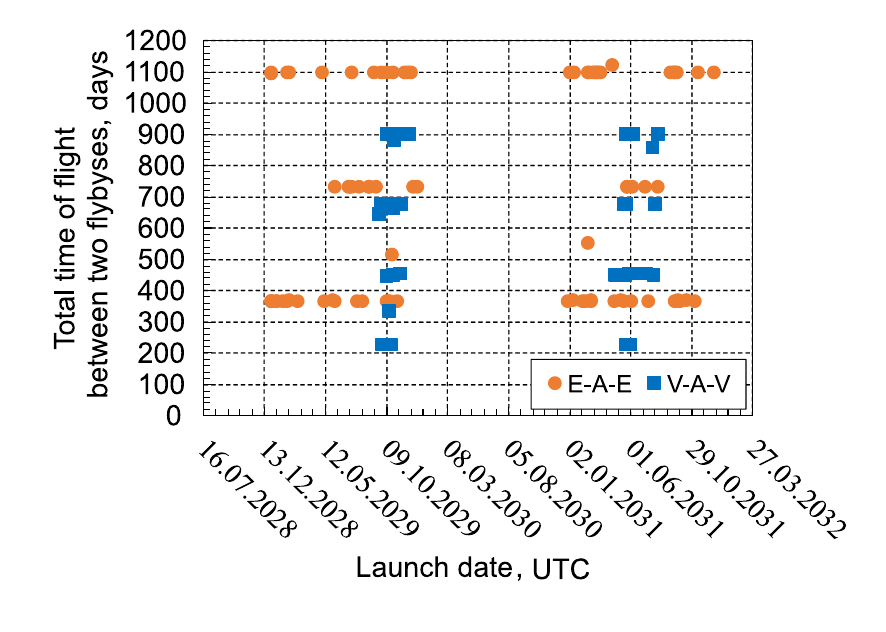}
		\captionof{figure}{Transfer duration between two successive planetary flybys for the two studied schemes versus the launch date}
		\label{fig:second_figure}
	\end{minipage}
\end{figure}

Fig.~\ref{fig:transfer_time} shows the total transfer duration for each of the considered schemes as a function of the launch date. The diagram shows results only for those asteroids (out of 139) for which a flyby was possible under the restrictions of Eq. (\ref{eq:restr}). Table~\ref{tab:resc} shows the number of optimal flight trajectories that satisfy the constraint in Eq. (\ref{eq:restr}). These were found for launch periods where asteroid flybys are possible under the studied schemes. Most complete data characterizing flight trajectories found during the study period for launch dates are provided in Tables \ref{tab:evav} and \ref{tab:eae} in Appendix \ref{sec:app}.


Figure \ref{fig:transfer_time} shows the distribution of flight times from the initial point to the terminal point, including flybys of specific asteroids from the selected set. Analysis of this figure shows that the duration of the first trajectory (E-V-A-V) is generally shorter than the second (E-A-E). This is due to the shorter orbital period of Venus, which reduces the time between successive planetary encounters, thereby increasing the frequency of gravity assists. This provides a basis for finding subsequent candidate asteroids that can be encountered after the next Venus flyby.

It is possible to classify all calculated trajectories according to $m:n$ resonance on the P-A-P segment, based on the calculated flight time between two successive planetary encounters. The total flight duration presented in Fig. \ref{fig:transfer_time}, allows such analysis only for the E-A-E scheme. For the E-V-A-V scheme, the duration of the V-A-V segment can be calculated by subtracting the E-V duration from the total time of flight (see Fig. \ref{fig:second_figure}). Therefore, for each optimal trajectory using the method from Section \ref{sec:sc11}, the classification has been outlined in Table \ref{tab:resc}.

\begin{table}[h]
	\centering
	\caption{Distribution of trajectories on the P-A-P transfer segment by resonant ratios $m:n$}
	\begin{tabular}{lccccc}
		\toprule
		& \multicolumn{2}{c}{V-A-V} & \multicolumn{2}{c}{E-A-E} \\
		\cmidrule(lr){2-3} \cmidrule(lr){4-5}
		$m:n$ \text{/} launch year & 2029 & 2031 & 2029 & 2031 \\
		\midrule
		$1:1$ & 4 & 2 & 28 & 21\\
		$3:2$ & 21 & 5 & - & - \\
		$2:1$ & 8 & 14  & 13 & 15\\
		no resonance & 2 & -  & 1 & 2\\
		total & \multicolumn{2}{c}{56} & \multicolumn{2}{c}{80} \\
		\bottomrule
	\end{tabular}
	\label{tab:resc}
\end{table}

Most of the transfer trajectories found involved asteroid flybys in $m:n$ resonant orbits, as seen from Fig. \ref{fig:transfer_time} and Table \ref{tab:resc}. The V-A-V flight trajectory corresponding to a $3:2$ resonant orbit was the most frequently occurring for launches in 2029. The most frequently encountered resonant ratio allowing a flyby of an asteroid when launched directly from Earth was $1:1$ for launches in both 2029 and 2031.

In the case of the E-A-E scheme, the frequent occurrence of $1:1$ resonant orbits as a result of patching together the E-A and A-E trajectory parts can be easily explained by the initial small MOID between Earth's orbit and the asteroid's. Therefore, orbits with semi-major axes close to that of Earth are chosen. For the E-V-A-V scenario, the semi-major axes should be chosen such that they intersect with the Venus orbit. Since the initial MOIDs for orbits of Venus and the asteroids vary, the $m:n$ ratio varies more in this case.

\subsubsection{Comparison of E-A-E and E-V-A-V Schemes}
\label{sec:sc222}

This section compares the E-A-E and E-V-A-V flight schemes, using parameters $\Delta v_0$ and $v_\infty$, for asteroid flybys, vs. the launch date. A separate analysis of $v_\infty$, at Venus, for the corresponding scheme, was also performed for the E-V-A-V scheme. 

To perform a detailed analysis of the E-V-A-V scheme, a comparison was made between this scheme and the Earth-Venus (E-V) transfer. Typically, solutions to the Lambert problem for the E-V scheme are divided into two trajectory families. The first family includes trajectories for which the angular distance of flight is less than $\pi$; the second family includes those where this distance is greater than $\pi$. These trajectory families facilitate the analysis of the initially studied E-V-A-V scheme.

In total, using the method described in Section \ref{sec:sc11}, optimal trajectories were found for flybys of 56 asteroids using the E-V-A-V scheme and for 80 asteroids using the E-A-E scheme, for two launch windows in 2029 and 2031.

\begin{figure}[H]
	\centering
	\wmedia{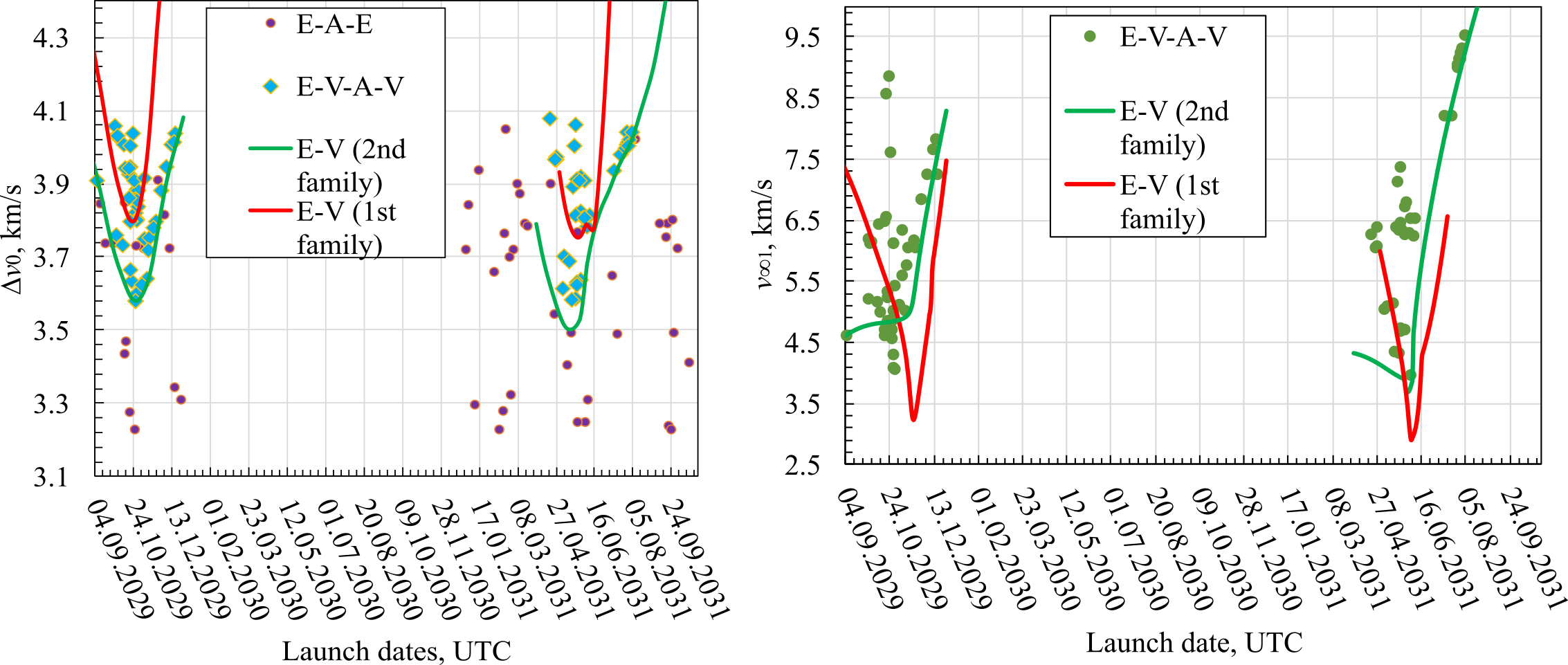} 
	\caption{(a) Dependence of $\Dvzero$ on launch date for the E-A-E and E-V-A-V schemes; (b) Dependence of $v_{\infty, 1}$ for Venus flyby in the E-V-A-V scenario}
	\label{fig:deltaV0_comparison}
\end{figure}

Fig.~\ref{fig:deltaV0_comparison}(a) compares the required $\Delta V_0$ for transfer to asteroids using the E-A-E and E-V-A-V schemes. The $\Delta v_0$ requirements for single-revolution E-V transfers (for both solution families) are also shown in Fig.~\ref{fig:deltaV0_comparison}(a).

Figure \ref{fig:deltaV0_comparison}(a) shows that the E-A-E scheme typically requires lower $\Dvzero$ than the E-V-A-V scheme. This is expected since the $\Dvzero$ for Venus-assisted transfers cannot be lower than that of optimal single-revolution Earth-Venus transfers, which provide the theoretical minimum. The observed minimum $\Dvzero$ for E-V-A-V trajectories is approximately 3.6 km/s, closely matching the optimal direct Earth-Venus transfer requirement.

Analysis of the $\Dvzero$ distribution reveals that E-V-A-V values cluster near the minima of single Earth-Venus transfers. Most E-V-A-V $\Dvzero$ values fall between the first and second solution family curves of direct transfers, with approximately half exceeding the second family's optimal values. This indicates that in these cases, faster transfers to Venus are required to enable subsequent asteroid encounters compared to optimal direct Earth-Venus trajectories.

The asymptotic velocities $v_{\infty}$ at Venus approach (Fig. \ref{fig:deltaV0_comparison}(b) for E-V-A-V trajectories align closely with optimal direct Earth-Venus transfers. However, the requirement for impulse-free asteroid flybys generally results in $v_{\infty}$ values exceeding both transfer families' optima. The few cases with lower $v_{\infty}$ correspond to trajectories requiring higher $\Dvzero$ for faster Venus transfers, as noted in the $\Dvzero$ analysis.

\begin{figure}[H]
	\centering
	\wmedia{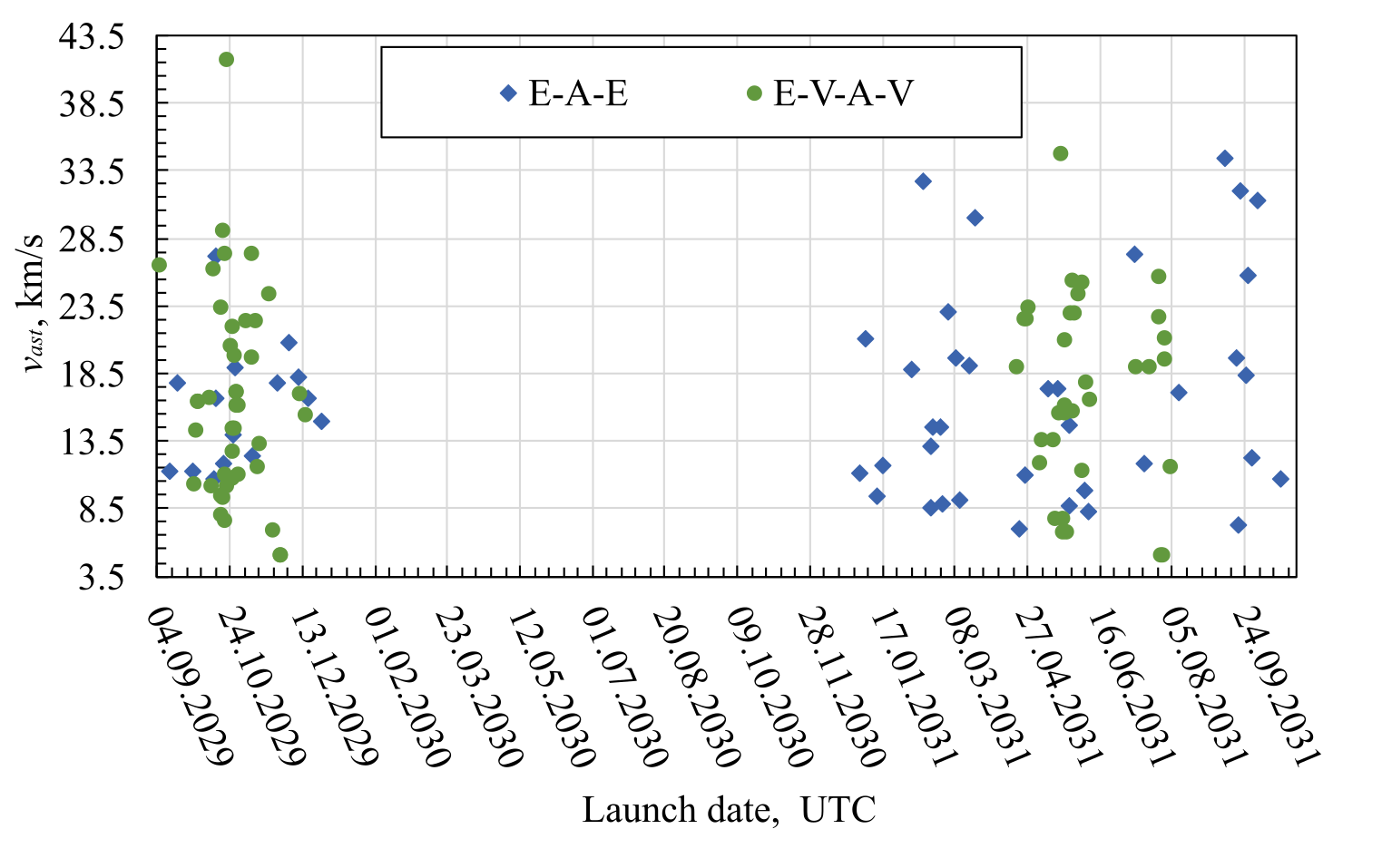} 
	\caption{Dependence of $v_{ast}$ on launch date within the launch windows in 2029 and 2031.}
	\label{fig:flyby_speed}
\end{figure}

Asteroid flyby velocities $v_{ast}$ (equivalent to $v_{\infty, 2}$ in E-V-A-V schemes) range from 2.3 to 44~km/s for both mission architectures (Fig. \ref{fig:flyby_speed}). The minimum $v_{ast}$ values for E-V-A-V trajectories occurred during flybys of asteroids 2005 YU55 (2.3 km/s) and 2011 DV (6.975 km/s), with spacecraft motion on the Venus-Venus segments following $2:1$ and $1:1$ resonant orbits, respectively.

To summarize this section:
\begin{enumerate}
	\item It was demonstrated that flybys of PHAs are feasible while the spacecraft moves along resonant orbits with Venus. This approach does not increase the total flight time and actually reduces the duration between successive planetary encounters compared to direct Earth-asteroid transfers.
	
	\item Analysis of $\Delta v_0$ for both schemes revealed that while E-A-E transfers generally require lower launch impulses for this specific asteroid subgroup, the E-V-A-V scheme provides critical advantages in mission flexibility. Specifically, it enables shifted launch windows and creates additional launch opportunities to reach target asteroids while maintaining the capability for spacecraft return and subsequent maneuvers.
	
	\item The E-V-A-V scheme demonstrates close alignment with optimal direct Earth-Venus transfers in terms of total $\Delta v$ consumption (approximately equal to $\Delta v_0$ in the considered cases) and asymptotic velocity $v_{\infty}$ at Venus flybys, validating its efficiency within the Venus gravity assist based mission architecture.
	
	\item It was demonstrated that Simplification 1 (introduced in Section~\ref{sec:sc11}) can effectively reduce the optimization parameter space for missions involving asteroid flybys with return to the same planetary body. This simplification is justified by the observation that optimal asteroid encounters typically occur outside the planet's orbital plane, with the complete trajectory between planetary encounters connected through resonant orbits at the asteroid flyby points.
\end{enumerate}

\subsubsection{Construction of Multiple-Asteroid Flyby Trajectories Using Venus Gravity Assists}
\label{sec:sc223}

This section extends the Earth-Venus-Asteroid-Venus (E-V-A-V) scheme to include multiple asteroid flybys, where each additional asteroid introduces two additional V-A-V segments. The methodology employs geometric relationships within a two-body dynamics framework (see Sec.~\ref{sec:sc13}), focusing on the 2029-2031 launch windows. 

The initial conditions for constructing multi-asteroid trajectories are derived from the single-asteroid flyby cases analyzed in the previous section, which are summarized in Table \ref{tab:num3}. 

Among all considered trajectories, those enabling nearly impulse-free asteroid flybys while operating on 1:1 resonant orbits are particularly advantageous (Table \ref{tab:num3}). These solutions combine minimal energy requirements with the capability for repeated encounters over relatively short time intervals, making them ideal candidates for extended mission sequences.

\begin{table}[h]
	\centering
	\caption{Schemes for flybys of asteroids on a 1:1 resonant orbit}
	\scriptsize
	
	\begin{tabular}{@{}cccccccccc@{}}
		\toprule
		Name & \multicolumn{1}{c}{Launch Date} & \multicolumn{3}{c}{Duration, days} & \multicolumn{3}{c}{$v_{\infty}$, km/s} & $\tildH_\pi,10^4 \text{km}$ \\
		& {(DD.MM.YYYY)} & $T_1, \text{days}$ & $T_2, \text{days}$ & $T_3, \text{days}$ & $v_{\infty, 0}, \text{km/s}$ & $v_{\infty, 1}, \text{km/s}$ & $v_{\infty, 2}, \text{km/s}$ & \\
		\midrule
		1997 XF11 & 03.10.2029 & 170.808 & 76.067 & 148.633 & 3.185 & 4.799 & 10.196 & 25.216 \\
		2005 YU55 & 30.08.2029 & 184.778 & 96.003 & 128.697 & 4.256 & 4.585 & 11.062 & 16.036 \\
		2007 JY2 & 10.09.2029 & 180.439 & 80.708 & 143.992 & 3.933 & 4.707 & 10.858 & 9.134 \\
		2013 ED28 & 21.10.2029 & 89.751 & 99.955 & 124.745 & 4.212 & 8.554 & 29.02 & 16.59 \\
		2007 RU9 & 06.05.2031 & 161.333 & 32.701 & 191.999 & 2.952 & 5.024 & 11.839 & 24.635 \\
		2009 XT6 & 24.05.2031 & 171.533 & 193.218 & 31.482 & 2.993 & 4.659 & 20.9 & 16.42 \\
		\bottomrule
	\end{tabular}
	
	\label{tab:num3}
\end{table}

Table \ref{tab:num3} presents the parameter $\tilde{H}_\pi$, the flyby altitude of the SC above a sphere with a radius equal to the planet's mean radius: $\tilde H_\pi = (r_\pi - R_\text{pl})$ in thousands of km, where $r_\pi$ is the periapsis radius of the SC's orbit, and $R_\text{pl}$ is the mean radius of the planet. The value is expressed in thousands of kilometers; $v_\infty, j = 0, 1, 2$ are asymptotic velocities at launch from Earth, Venus flyby, and asteroid flybys.

From Table \ref{tab:num3}, one can see that the lowest $\Delta v_0$ values in the transfer scheme in which the V-A-V segment is traversed on a 1:1 resonant orbit are achieved when the E-V transfer takes place along the second family trajectory, except for the 2013 ED28 asteroid. For this asteroid, the best transfer is with an E-V segment corresponding to a first-family trajectory.


\subsubsection{Initial Conditions and Candidate Selection}
\label{sec:sc2231}

Initial conditions for designing multi-asteroid trajectories are derived from the single-asteroid flyby cases presented in Table~\ref{tab:num3}. Table~\ref{tab:num4} summarizes the post-flyby parameters following the second Venus gravity assist, which serve as input conditions for Algorithm~\ref{alg:asteroid_selection}.

\begin{table}[h]
	\centering
	\caption{Candidate asteroids and initial data for calculating trajectories after the second Venus gravity assist}
	\begin{tabular}{ccccc}
		\hline
		Name & $v_{\infty, 3}$, km/s & $\gamma (t_3)^{-}$, deg & $\Phi$, deg & $t_3$, UTC \\
		\hline
		1997 XF11 & 4.801 & 155.420 & 93.919 & 03.11.2030 6:20 \\
		2005 YU55 & 4.586 & -171.821 & 93.743 & 14.10.2030 0:55 \\
		2007 JY2 & 4.702 & 125.677 & 93.842 & 20.10.2030 10:15 \\
		2013 ED28 & 8.556 & -63.562 & 96.969 & 31.08.2030 13:52 \\
		2007 RU9 & 5.023 & 8.127 & 94.1 & 26.05.2032 16:41 \\
		2009 XT6 & 4.655 & 45.157 & 93.8 & 24.06.2032 4:30 \\
		\hline
	\end{tabular}
	\label{tab:num4}
\end{table}

These data serve as initial conditions for the selection procedure \ref{alg:asteroid_selection}. These data includes asteroids international designation, as well as the flyby date and the initial asymptotic velocity at Venus. The candidate asteroids for subsequent arcs after the last approach to Venus are listed in Table \ref{tab:num5}


\begin{table}[h]
	\centering
	\caption{Pairs of candidate asteroids for subsequent flybys}
	\scriptsize
	\begin{tabular}{@{}cccccc@{}}
		\toprule
		N & \makecell{Asteroid-1\\} & \makecell{Asteroid-2\\} & Date $t_4$ UTC & $\gamma (t_3)^{+}$, deg & $\Delta r_{\min}$, km \\
		\midrule
		1 & 1997 XF11 & (454101) 2013 BP73 & 26.02.2031 19:38 & 179.536 & 2494128.4 \\
		2 & 1997 XF11 & (405212) 2003 QC10 & 29.01.2031 1:10 & 217.045 & 3954815.1 \\
		3 & 2005 YU55 & 1998 KM3 & 19.12.2030 8:23 & 129.219 & 1945531.6 \\
		4 & 2005 YU55 & 2013 BP73 & 25.02.2031 13:56 & 179.304 & 4572853.2 \\
		5 & 2007 JY2 & (279744) 1998 KM3 & 17.12.2030 8:05 & 115.199 & 588853 \\
		6 & 2007 JY2 & 2013 BP73 & 25.02.2031 22:40 & 179.518 & 2646238 \\
		7 & 2013 ED28 & (508967) 2004 VC17 & 13.10.2030 15:04 & 68.757 & 2454350.3 \\
		8 & 2007 RU9 & (152685) 1998 MZ & 21.09.2032 13:00 & 48.974 & 362259.98 \\
		9 & 2007 RU9 & 2013 JL22 & 07.09.2032 0:14 & 77.265 & 1586178.3 \\
		10 & 2009 XT6 & 1998 MZ & 26.09.2032 22:40 & 70.168 & 2300541.3 \\
		\bottomrule
	\end{tabular}
	\label{tab:num5}
\end{table}

\subsubsection{Trajectory Optimization and Refinement}
\label{sec:sc2232}

Trajectories combining flybys of several asteroids from Table \ref{tab:num5} were refined using the method from Section \ref{sec:sc11}. Table \ref{tab:num6} presents the refined parameters for the trajectories of flybys of candidate asteroids in Table \ref{tab:num5}.

The candidate pairs from Table~\ref{tab:num5} were refined using the optimization method from Section~\ref{sec:sc11}. Table~\ref{tab:num6} presents the optimized parameters, where the asterisk (*) denotes values obtained after optimization.

\begin{table}[h]
	\centering 
	\caption{Refined parameters of asteroid flybys obtained using the method from section~\ref{sec:sc11} based on data from Table~\ref{tab:num5}}
	\begin{tabular}{cccccc}
		\hline
		\makecell{Asteroid-1\\} & \makecell{Asteroid-2\\} & $t_4$, UTC & $t_4^{*}$, UTC & $\gamma (t_3)^{+}$, deg & $\gamma (t_3)^{*+}$, deg \\
		\hline
		1997 XF11 & 2013 BP73 & 26.02.2031 19:38 & 26.02.2031 09:28 & 179.536 & 179.719 \\
		2005 YU55 & 1998 KM3 & 19.12.2030 8:23 & 18.12.2030 23:51 & 129.219 & 123.066 \\
		2007 JY2 & 1998 KM3 & 17.12.2030 8:05 & 17.12.2030 11:43 & 115.199 & 117.352 \\
		2013 ED28 & 2004 VC17 & 13.10.2030 15:04 & 12.10.2030 02:47 & 68.757 & 66.296 \\
		2007 RU9 & 1998 MZ & 21.09.2032 13:00 & 21.09.2032 17:28 & 48.974 & 47.994 \\
		2007 RU9 & 2013 JL22 & 07.09.2032 0:14 & 07.09.2032 18:16 & 77.265 & 73.320 \\
		2009 XT6 & 1998 MZ & 26.09.2032 22:40 & 26.09.2032 02:57 & 70.168 & 79.185 \\
		\hline
		\footnotetext{\small Note: the symbol $*$ indicates refined parameters determined within the method from Section~\ref{sec:sc21} .}
	\end{tabular}

	\label{tab:num6}
\end{table}
\small Note that the symbol $*$ in Table \ref{tab:num6} indicates refined parameters determined within the method from section~\ref{sec:sc11} .

As shown in the table above, the total $\Dv$, calculated as the sum over all transfer segments except $\Dvzero$, in the best cases does not exceed 1 m/s, and only in two cases amounts to 95 m/s and 136 m/s for the flyby of asteroids 2005 YU55 and 2009 XT6, respectively. Moreover, in both of the latter cases, the largest $\Dv$ increase occurs during the final Venus flyby. The algorithm identified 10 candidate asteroids; feasible trajectories were found for 7 of them—5 of which were nearly impulse-free.

\subsubsection{Extended Multi-Asteroid Mission Sequences}
\label{sec:sc2233}

\begin{figure}[H]
	\centering
	\wmedia{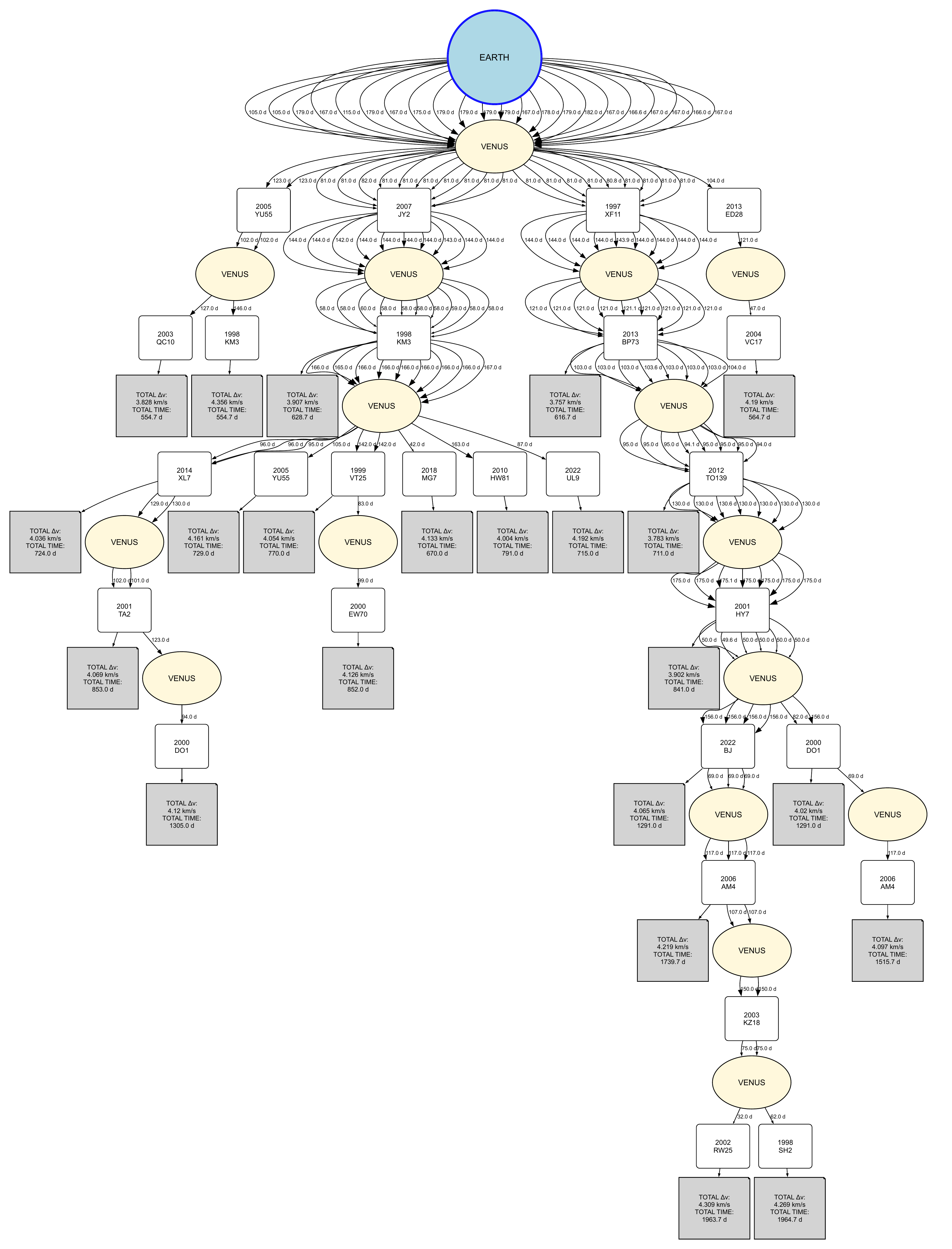} 
	\caption{The tree of opportunities for asteroid flybys on a 1:1 resonant orbit includes up to eight asteroids per flight. Note that all flights end with a Venus encounter. For simplicity, the last Venus node has not been shown in this scheme.}
	\label{fig:sfm}
\end{figure}

Let us provide an example of a search for flight trajectories involving flybys of up to 8 asteroids. For this, the baseline for designing asteroid flyby sequences was only asteroids from the selected set when the number of flybys was less than 3. After the asteroid flybys exceeded 3, the entire set of PHAs, consisting of 2,030 asteroids, was used. The search results are presented in Fig. \ref{fig:sfm}.

Building upon the four single-asteroid flyby scenarios identified for the 2029 launch window, paired asteroid combinations to construct multi-target sequences were systematically enumerated. This approach revealed numerous viable scenarios with two or more asteroid encounters within the original 139-object dataset. By further extending our search to the complete PHA catalog of 2,030 objects (as of June 2025), trajectories accommodating up to eight asteroid flybys were found. The expanded catalog inclusion was necessary to access asteroids with MOID values exceeding our initial 1.05 $\LD$ threshold, thereby increasing the pool of potential targets for extended mission sequences.

\begin{figure}[H]
	\centering
	\wmedia{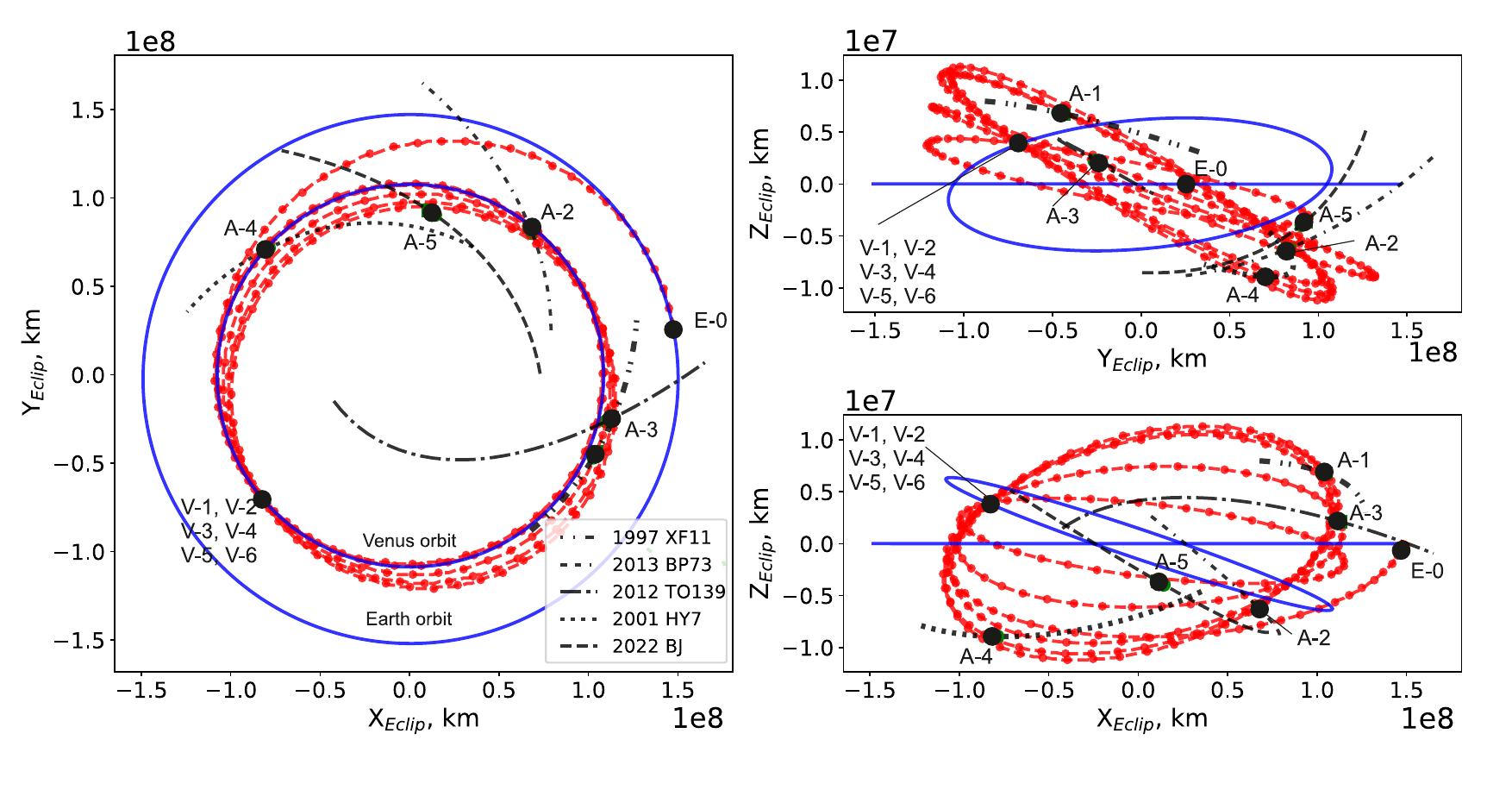} 
	\caption{The trajectory of the SC of five Venus flybys and asteroid flybys on a 1:1 resonant orbit.}
	\label{fig:trajectory_example}
\end{figure}

As a representative example, the construction of a spacecraft trajectory with five asteroid flybys (see Figure \ref{fig:sfm}) was provided, which represents the maximum number of consecutive encounters achievable for the 2029 launch while maintaining minimal expansion beyond the initial 139-asteroid set. The five-asteroid transfer trajectory is illustrated in Fig. \ref{fig:trajectory_example}, with its principal characteristics summarized in Table \ref{tab:p5}.

\begin{table}[h]
	\centering
	\caption{Characteristics of transfer trajectories with flyby of five asteroids}
	\label{tab:p5}
	\scriptsize
	\begin{tabular}{@{} l l l c S[table-format=3.3] S[table-format=1.3] r S[table-format=2.2] S[table-format=2.3] @{}}
		\toprule
		\textbf{Segment} & \textbf{Celestial Body} & \textbf{Path} &
		{\makecell[c]{\textbf{Launch/} \\ \textbf{Maneuver Date}}} & {\makecell[c]{\textbf{Duration} \\ \textbf{(days)}}} & {\makecell[c]{\textbf{$\Delta v$} \\ \textbf{(km/s)}}} & {\makecell[c]{\textbf{Flyby Alt. } \\ \textbf{ $\tildH_{\pi}$ (km)}}} & {\makecell[c]{\textbf{Comm. Angle} \\ \textbf{(E-SC-S, deg)}}} & {\makecell[c]{\textbf{Flyby $v_{\infty}$} \\ \textbf{(km/s)}}} \\
		\midrule

		\textbf{Launch} & Earth & Earth (E-0) &\text{01.10.2029}& {-} & \textbf{3.757} & 0.2 & {$\leq$ 1} & 3.462 \\
		1 & Venus & E-0 $\to$ V-1 &\text{17.03.2030}& 166.710 & 0.005 & 26607 & 42.68 & 5.070 \\
		2 & 1997 XF11 & V-1 $\to$ A-1 &\text{06.06.2030}& 80.829 & 0.000 & {-} & 82.71 & 10.081 \\
		3 & Venus & A-1 $\to$ V-2 &\text{28.10.2030}& 143.871 & 0.003 & 46095 & 174.79 & 5.070 \\
		4 & 2013 BP73  & V-2 $\to$ A-2&\text{26.02.2031} & 121.098 & 0.000 & {-} & 105.26 & 21.987 \\
		5 & Venus & A-2 $\to$ V-3 &\text{09.06.2031}& 103.602 & 0.026 & 13136 & 39.30 & 5.030 \\
		6 & 2012 TO139 & V-3 $\to$ A-3 &\text{12.09.2031}& 94.060 & 0.001 & {-} & 1.38 & 37.189 \\
		7 &Venus & A-3 $\to$ V-4 &\text{20.01.2032}& 130.640 & 0.111 & 10672 & 99.52 & 4.853 \\
		8 & (216523) 2001 HY7 & V-4 $\to$ A-4 &\text{13.07.2032}& 175.147 & 0.000 & {-} & 150.45 & 14.121 \\
		9 & Venus & A-4 $\to$ V-5 &\text{01.09.2032}& 49.553 & 0.125 & 3324 & 119.77 & 5.097 \\
		10 & 2022 BJ & V-5 $\to$ A-5&\text{04.02.2033} & 156.517 & 0.000 & {-} & 50.49 & 26.658 \\
		11 & Venus & A-5 $\to$ V-6&\text{13.04.2033} & 68.183 & {-} & {-} & 15.39 & 5.096 \\
		\midrule
		
		\textbf{Total} & & & &  \textbf{1290.210} & \textbf{4.028} & \multicolumn{3}{l}{} \\
		& & &  & {\small(\textasciitilde 3.5 yrs)} & & \multicolumn{3}{l}{} \\
		
		\bottomrule
	\end{tabular}
\end{table}
Note that in Table \ref{tab:p5}, the angle E-SC-S denotes the Earth-SC-Sun communication angle at the moments of encounter with celestial bodies.

The data in Table \ref{tab:p5} indicate that maneuvers V-3 and V-4 reduce the SC's orbital energy relative to Venus, while maneuver V-5 increases it, providing the necessary correction to the heliocentric orbit radius. The total $\Delta v$ for these maneuvers is 270 m/s, or 0.08 $\Delta V_{0}^{\lim}$. Furthermore, the magnitudes of the asteroid flyby maneuvers (A-1 through A-5) are relatively small, with a total magnitude of less than 5 m/s. Analysis of the time of flight between successive Venus flybys confirms that the SC's motion occurs along 1:1 resonant orbits, validating the initial choice of the 1:1 resonance as the primary design assumption. Therefore, for this example, the optimal SC trajectory from launch (E-0) to the final Venus encounter (V-6) consists of an Earth-Venus transfer segment followed by motion along 1:1 resonant orbits between successive Venus flybys.

Table~\ref{tab:p5} indicates that the Earth-Spacecraft-Sun (E-SC-S) angle during the second Venus approach is 174 degrees. This angle occurs during a decreasing phase, with full solar conjunction occurring 9 days prior to the Venus flyby date. Previous studies \cite{morabito2002communicating, srivastava2016mars} have established that communication degradation typically begins at solar conjunction angles of approximately $3^\circ$--$5^\circ$ ($177^\circ$--$175^\circ$) for S-band, $2^\circ$ ($178^\circ$) for X-band, and ($179^\circ$) for Ka-band. The obtained value of 174 degrees therefore falls well within acceptable communication limits.

For spacecraft following 1:1 resonant orbits, the E-SC-S angle during the second Venus flyby consistently exceeds 90 degrees. This geometric constraint arises from the orbital dynamics: optimal $\Delta v$ trajectories typically yield E-SC-S angles between $0^\circ$ and $45^\circ$ during the initial Venus encounter. Following the gravity assist, the spacecraft completes one revolution ($2\pi$ radians) in approximately 224.7 days, while Earth traverses an orbital arc of $224.7 / 365.25 \times 2\pi \approx 3.86$ radians (221.5$^\circ$). This orbital phasing ensures that the E-SC-S angle at the second Venus encounter always exceeds 90 degrees. As shown in Table~\ref{tab:p5}, subsequent flybys occur at more favorable angles, significantly lower than the second encounter, thereby ensuring continuous radio communication with Earth throughout the mission.

\subsubsection{Determination of Trajectories for SC Return to Earth after Venus Gravity Assist and Asteroid Flybys}
\label{sec:sc23}

A potential variation of the E-V-A-V scheme is to return the SC to Earth after completing a series of Venus and asteroid flybys. In this scenario, the following flight scheme is considered:

\[
\text{Earth} \to \text{Venus} \to \text{1st Asteroid} \to \text{Venus} \to \cdots \to \text{Venus} \to \text{$p$-th Asteroid} \to \text{Venus} \to \text{Earth}.
\]

The trajectory search method and overall approach remain unchanged, but a new Venus-to-Earth transfer segment was incorporated into the optimization. This allows for the addition of the Venus-Earth segment of flight to the optimization problem established in Section \ref{sec:sc11}.

Let us first consider scenarios involving only one asteroid flyby. For the purpose of further analysis, let us use the results obtained in Section \ref{sec:sc22}. Subsequently, the analysis can be carried out in the following way: comparing the parameters obtained using the E-V-A-V flight scheme with those examined in this section.

In this context, it is important to highlight the overlap between schemes that allow for a flyby with or without a return to Earth. The study period remains the same, spanning from 2029 to 2031. A list of asteroids considered for flybys is provided in Table \ref{tab:evav}.

To compare both schemes, the following parameters are estimated in this section:

\begin{eqnarray*}
	\Delta^2 V = \Delta v^{EVAV}_{tot} - \Delta v^{EVAVE}_{tot} \text{,} \\
	\Delta^2 T = t^{EVAV}_{0} - t^{EVAVE}_{0} \text{,}
\end{eqnarray*}
where the parameters denoted by $EVAV$ and $EVAVE$ correspond to the E-V-A-V and E-V-A-V-E schemes, respectively.

The parameters $\Delta^2 V$  and  $\Delta^2 T$ represent the difference in total characteristic velocity and launch date, respectively, between the E-V-A-V-E (return) scheme and the baseline E-V-A-V (no return) scheme. Also, estimation is made in terms of changes in $\Delta v_{tot}$ instead of comparing each impulse given at specific points in both schemes. This is because the studied trajectories are impulse-free and one can assume that $\Delta v_{tot} \approx \Delta v_0$.

\begin{figure}[H]
	\centering
	\wmedia{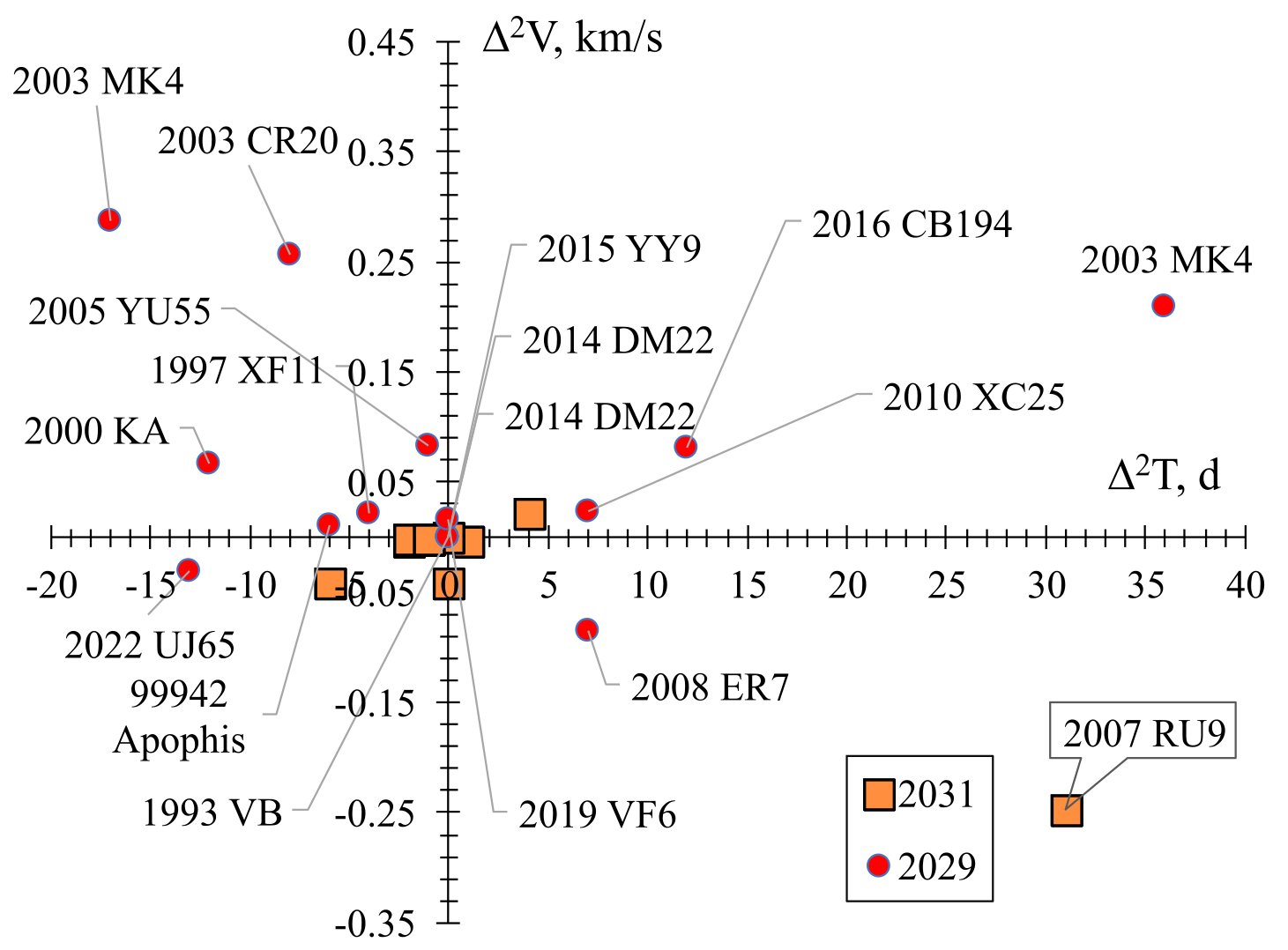} 
	\caption{The change in $\Delta^2 V$ and $\Delta^2 T$ values when implementing the E-V-A-V scheme.}
	\label{fig:return_earth}
\end{figure}

The results of the calculations are provided in Table \ref{tab:evave}. The estimated values of $\Delta^2 V$ and $\Delta ^2 T$ are shown in Fig. \ref{fig:return_earth}.

As a result of the study, 16 optimal return trajectories were found with launch dates in 2029 out of 35 previously obtained, and 11 out of 21 for launch in 2031. Each optimal trajectory corresponds to a unique asteroid from the selected set (see Table \ref{tab:evave}).

The results presented in Fig.~\ref{fig:return_earth} show that for several trajectories with a launch in 2029, the resulting changes in $\Delta v_{tot}$ are significantly smaller compared to those found in Section \ref{sec:sc221} (see Fig.~\ref{fig:deltaV0_comparison}). This reduction results from changes in the optimal duration and launch date for the Earth-Venus segment, while other trajectory characteristics remained similar to those in the previous section. Changes in the asteroid flyby velocity were also practically nonexistent, and the results fully replicate those shown in Fig.~\ref{fig:flyby_speed}. The difference in duration between Venus-to-Earth transfer segments is shown in Table A.\ref{tab:evave}, where this parameter varies from 74 to over 200 days.

For the launch in 2031, it can be seen that for 10 trajectories, the corresponding changes in $\Delta^2 V$ and $\Delta^2 T$ were negligible except for the 2007 RU9 asteroid. For the flyby of 2007 RU9 with a subsequent return to Earth after the Venus flyby, $\Delta v_0$ increased by 200 m/s while the trajectory was impulse free during the flybys of Venus and asteroids. This occurred because the optimal launch date shifted 35 days forward relative to the solution found in section \ref {sec:sc222}. A similar result was observed for a scheme including a flyby of the 2003 MK4 asteroid during the launch window in 2029. 

\subsubsection{Detailed Example of Constructing the SC Transfer Trajectory Including Venus and Asteroid Flybys with a Subsequent Return to Earth}
\label{sec:sc231}

Let us consider an example of constructing a return trajectory to Earth after a series of Venus and asteroids flybys for the case where two asteroid flybys are included (i.e., E-V-A-V-A-V). For this, the flybys of asteroids 1997 XF11 and 2013 BP73 from Table \ref{tab:num6} were considered. The choice of these asteroids was based on the fact that the flyby of the second asteroid occurred on a peculiar resonant orbit that has orbital parameters similar to those of Venus, but with a different inclination. Such orbits occur when $\gamma \approx 0^{\circ}$ or $180^{\circ}$, in the considered case $\gamma=179.53^{\circ}$. These types of orbits have been described in references \cite{golubev2019synthesizing, golubev2020adaptive, Konstantinov2024}. In this case, the SC enters Venus's SOI the second time before completing a full revolution around the Sun. In \cite{Konstantinov2024} such orbits are called $\pi$-resonances. Because of that, this scheme is worth constructing both in the patched-conic approach and in $n$-body simulations, which is shown below.

\begin{figure}[H]
	\centering
	\wmedia{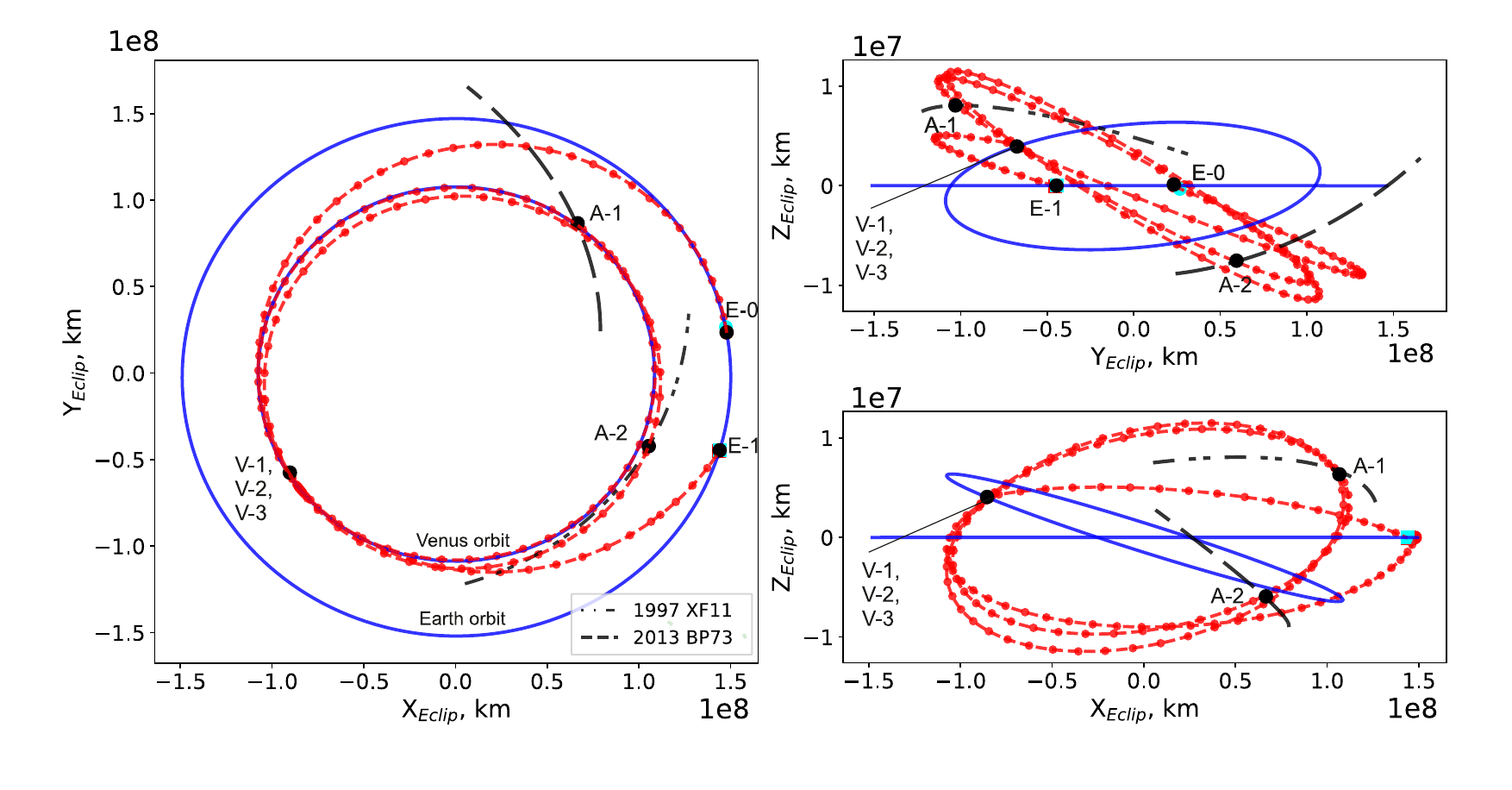} 
	\caption{Trajectory of SC flyby of Venus and two asteroids on a 1:1 resonant orbit}
	\label{fig:return_trajectory}
\end{figure}

The basic parameters of flight trajectory are given in  Table \ref{tab:return_earth1}. The transfer is shown in Fig.~\ref{fig:return_trajectory}.

\begin{table}[h]
	\centering
	\caption{Characteristics of trajectory E-V-A-V-A-V-E with a subsequent return to Earth}
	\label{tab:return_earth1}
	\scriptsize
	\begin{tabular}{@{} l l l c S[table-format=3.3] S[table-format=1.3] r S[table-format=2.2] S[table-format=2.3] @{}}
		\toprule
		\textbf{Segment} & \textbf{Celestial Body} & \textbf{Path} &
		{\makecell[c]{\textbf{Launch/} \\ \textbf{Maneuver Date}}} & {\makecell[c]{\textbf{Duration} \\ \textbf{(days)}}} & {\makecell[c]{\textbf{$\Delta v$} \\ \textbf{(km/s)}}} & {\makecell[c]{\textbf{Flyby Alt.} \\ \textbf{ $\tildH_{\pi}$ (km)}}} & {\makecell[c]{\textbf{Comm. Angle} \\ \textbf{(E-SC-S, deg)}}} & {\makecell[c]{\textbf{Flyby $v_{\infty}$} \\ \textbf{(km/s)}}} \\
		\midrule
		
		\textbf{Launch} & Earth & Earth (E-0) & \text{01.10.2029}	&{-} & \textbf{3.755} & 0.2 & {$\leq$ 1} & 3.459 \\
		1 & Venus & E-0 $\to$ V-1 & \text{17.03.2030}	&166.948 & 0.005 & 26513 & 42.69 & 5.070 \\
		2 & 1997 XF11 & V-1 $\to$ A-1 &\text{06.06.2030}	& 80.829 & 0.000 & {-} & 82.71 & 10.081 \\
		3 & Venus & A-1 $\to$ V-2 &\text{28.10.2030}	& 143.871 & 0.000 & 46098 & 174.79 & 5.069 \\
		4 & 2013 BP73 & V-2 $\to$ A-2 &\text{26.02.2031}	& 121.098 & 0.000 & {-} & 105.27 & 21.987 \\
		5 & Venus & A-2 $\to$ V-3 & \text{09.06.2031}	&103.602 & 0.001 & 6350 & 39.31 & 5.068 \\
		6 & Earth & V-3 $\to$ E-1 & \text{05.09.2031} & 88.038 & {-} & {-} & 42.69 & 7.799 \\
		
		\midrule
		
		\textbf{Total} & & & & \textbf{704.286} & \textbf{3.761} & \multicolumn{3}{l}{} \\
		& & & &  {\small(\textasciitilde 1.9 yrs)} & & \multicolumn{3}{l}{} \\
		
		\bottomrule
	\end{tabular}
\end{table}

Note that the segment up to maneuver V-3 in the constructed transfer trajectory completely replicates the results given in Table \ref{tab:p5}. In particular, all maneuvers, including V-3, are found to be almost impulse free. The value of the E-SC-S angle remains high. The trajectory enables a rapid return to Earth after the final maneuver near Venus. The overall transfer duration is about two years.

In Fig.~\ref{fig:return_trajectory}, it can be seen that the SC's flight during the asteroid flyby phase occurs in the vicinity of the Venus orbit. Note that the SC's return point to Earth orbit differs from its initial point after the Venus and asteroid flybys tour.

\subsection{Numerical Study of the Trajectory for a Flight to Venus with Asteroid Flybys and Subsequent Return to Earth}
\label{sec:sc232}	

This section validates the trajectory design methodology using a high-fidelity $n$-body dynamical model. A concrete example featuring complete state vectors and detailed flight characteristics to demonstrate the refinement of trajectories from the two-body approximation to full $n$-body dynamics is presented. Two distinct optimization approaches are investigated: direct $B$-plane targeting and multiple shooting. The selected test case incorporates multiple asteroid flyby arcs, motion along both regular and $\pi$-resonant 1:1 orbits with Venus, and concludes with a Venus-Earth return segment, providing a comprehensive validation of the proposed mission architecture.

\subsubsection{Dynamical Model of the Spacecraft Interplanetary Motion}
\label{sec:sc2321}

This section presents the numerical simulation of the SC trajectory using the $n$-body dynamical model, focusing on the Earth-Venus-asteroid flyby sequence with subsequent return to Earth.

The equations of motion for the SC, given in a vector form, are

\begin{equation}
	\mathbf{\ddot{r}} = -\frac{\mu}{r^{2}} \frac{\mathbf{r}}{r} - \sum\limits_{j=1}^{n} \frac{\mu_{j}}{r^{3}_{j-sc}} \left( \mathbf{r} + \frac{3 + 3q + q^{2}}{1 + (1 + q)^{3/2}} q \cdot \mathbf{r}_{j} \right)
	\label{eq:difeq}
\end{equation}

where $\mathbf{r}$ is the radius vector from the central attracting body to the SC; $\mathbf{r_j}$ is the heliocentric position vector of the $j$-th perturbing body; $q = \frac{\mathbf{r}}{r_{j}} \cdot \frac{1}{r_{j}} \left( \mathbf{r} - 2\mathbf{r}_{j} \right)$ is a function used to simplify calculations of perturbations on the right side of Eq. (\ref{eq:difeq}) when $\mathbf{r} - \mathbf{r_j}$ is small.

Note that this form of writing the equations using the function $q$ is adopted based on an analysis of the study \cite{Ivashkin2019}. When integrating Eq. (\ref{eq:difeq}), the gravitational influences of the Sun, Earth, Moon, Venus, and Jupiter were incorporated. Integration was performed for the complete problem using the Dormand-Prince 7(8)-th order method on heliocentric transfer segments and the Runge-Kutta 8(9)-th order method on planet-centric segments; both methods were implemented with a modification allowing for automatic step size control.

Note that when using Eq. (\ref{eq:difeq}), the transition from a planet-centric segment to a heliocentric one and vice versa should occur when the following inequality is satisfied: $\frac{\mu_{j}}{r^{2}_{j-sc}} / \frac{\mu}{r^{2}} > 0.001$. For Earth and Venus, this ratio is reached approximately 1.4-1.7 days into the flight. In this regard, a simplification was made in this work: the transition from a planet-centric segment to a heliocentric one and vice versa occurred 2 days after or before the periapsis approach for Earth and 1.5 days for Venus.

\subsubsection{Direct Approach Algorithm Based on $B$-plane Targeting}
\label{sec:sc2322}

The transition from the two-body problem to integrating the SC trajectory using Eq. (\ref{eq:difeq}) was performed using the direct approach algorithm based on $B$-plane targeting. This plane is always orthogonal to the incoming $\mathbf{v}_\infty$ at the planetary encounter. In this plane each approach trajectory of the SC can be described by parameters $B_R,B_T$. These parameters can be represented through $\mathbf{v}_\infty$  therefore they can be used for targeting purposes \cite{Chos2012, Cho2017}. In this plane The core of using approach is provided in \ref{alg:alg2}.

\scriptsize
\begin{algorithm}[H]
	\label{alg:alg2}
	\SetAlgoLined
	\KwIn{Number of segments $s$, minimum periapsis radius $r_{\pi,\text{min}}$}
	\KwOut{Optimized trajectory parameters $\mathbf{X}^*$, total correction impulse $\Delta v_{\text{cor}}$}
	
	\caption{ $B$-plane Targeting}
	
	\textbf{Step 1: Trajectory Segmentation}\;
	Divide trajectory into $s$ segments: Earth $\to$ Venus$_1$ $\to$ Asteroid$_1$ $\to$ $\cdots$ $\to$ Venus$_{s+1}$ $\to$ Earth\;
	
	\BlankLine
	\textbf{Step 2: Initial Conditions and B-Plane Parameters}\;
	\For{launch from Earth}{
		Compute $\mathbf{r}_E, \mathbf{v}_E = f_1(\mathbf{v}_{\infty,0}, r_{LEO}, i_0)$\;
	}
	\ForEach{gravity assist maneuver $j = 1,3,\dots,2s-1$}{
		Compute $\mathbf{B}_j = f_2(\mathbf{v}_{\infty,j}^{-}, \mathbf{v}_{\infty,j}^{+}, r_{\pi,j})$\;
	}
	
	\BlankLine
	\textbf{Step 3: Boundary Value Problem}\;
	\ForEach{impulse time $t_j$}{
		$\Delta\mathbf{v}_j = \mathbf{v}^{+}(t_j) - \mathbf{v}^{-}(t_j)$\;
		$\mathbf{r}^{-}(t_j) = \mathbf{r}^{+}(t_j)$\;
	}
	$\Delta v_{\text{cor}} = \sum \|\Delta\mathbf{v}_j\|$\;
	
	\BlankLine
	\textbf{Step 4: Parameter Optimization}\;
	Minimize $J(\mathbf{X}) = \Delta v_{\text{cor}}$ subject to $r_{\pi,j} \geq r_{\pi, \text{min}}$ using two-stage optimization method\ ;

	\Return{$\mathbf{X}^*, \Delta v_{\text{cor}}$}
\end{algorithm}
\normalsize

\subsubsection{Example of Determining the SC Flight Trajectory by Numerical Simulation}
\label{sec:sc233}

This section presents a concrete example of trajectory determination for the~E-V-A-V-A-V-E flight scheme using $n$-body dynamics. The initial conditions were obtained from the patched-conic approximation using the methods described in Sections~\ref{sec:sc11} and~\ref{sec:sc12}, with specific parameters derived from the analysis in Section~\ref{sec:sc22}. 

The SC state vector at LEO departure, calculated in the patched-conic approximation within the Earth's ecliptic inertial system (J2000), is given by:

\begin{equation}
	\{\textbf{r}^0_E, \textbf{v}^0_E, t^0_{JD, UTC}\} =
	\begin{bmatrix}
		-5433.285187\\
		3562.899618\\
		981.427222\\
		-5.098547\\
		-5.328694\\
		-8.881209\\
		2462411.308844\\
	\end{bmatrix}
	\label{eq:vect021}
\end{equation}

where $\{\textbf{r}^0_E, \textbf{v}^0_E\}$ represent the initial position and velocity vectors at periapsis, and $t^0_{JD,UTC}$ denotes the launch time from LEO. The superscript $^0$ indicates parameters calculated using the patched-conic approximation.

The initial time of flights on segment and periapsis radii s are given in Table  \ref{tab:return_earth1}, the $B$-plane components at each Venus flybys can be written as follows:
\begin{enumerate}
	\item 1st Venus flyby  ${B_T} = 36073.7$ km, ${B_R} = -24226.6$ km;
	\item 2nd Venus flyby  ${B_T} = 63359.7$ km, ${B_R} = -4809.2$ km;
	\item 3rd Venus flyby  ${B_T} = 4766.8$ km, ${B_R} = -21069.1$ km.
\end{enumerate}

The SC trajectory was computed from the given initial conditions, following the sequence of asteroid encounters and planetary flybys, terminating at the return to Earth. The SC trajectory was matched by targeting specific $B$-plane parameters, ensuring continuity of position and velocity across planet-centric and heliocentric segments as described in Algorithm \ref{alg:alg2}. During this process, impulses were applied at encounter points to maintain the feasibility of the trajectory within operational constraints. It should be noted that the boundary problems for asteroid targeting were relaxed in terms of encounter constraints. The distances from the SC to the asteroids at their closest encounter were obtained considering that the boundary problem had been solved with terminal conditions for close encounters.

The results of computation were verified using JPL's GMAT 2025 software\footnote{https://software.nasa.gov/software/GSC-19468-1 (Accessed 25.07.2025)} with the Dormand-Prince 7(8)-th order method and solving boundary value problems using the classical Newton method. The DE435 ephemeris\footnote{https://naif.jpl.nasa.gov/naif/ (Accessed 25.07.2025)} data were used to obtain the coordinates of the planets and the Moon.

The refined initial conditions of the SC at time of the transition from LEO to hyperbolic orbit, obtained after employing the trajectory optimization technique can be written as (given in the Earth ecliptic inertial system J2000):

\begin{equation*}
	\{\textbf{r}_E, \textbf{v}_E, t_{JD, UTC}\} =
	\begin{bmatrix}
		-5433.285186 \\
		3562.899617 \\
		981.427222 \\
		-5.098547 \\
		-5.328693 \\
		-8.881209 \\
		2462411.778099
	\end{bmatrix}
	\label{eq:vect1}
\end{equation*}

From comparing both (\ref{eq:vect021}) and (\ref {eq:vect}), it can be seen that the two vectors are almost identical, with only the launch date differing by 0.46 days.

\begin{figure}[H]
	\centering
	\wmedia{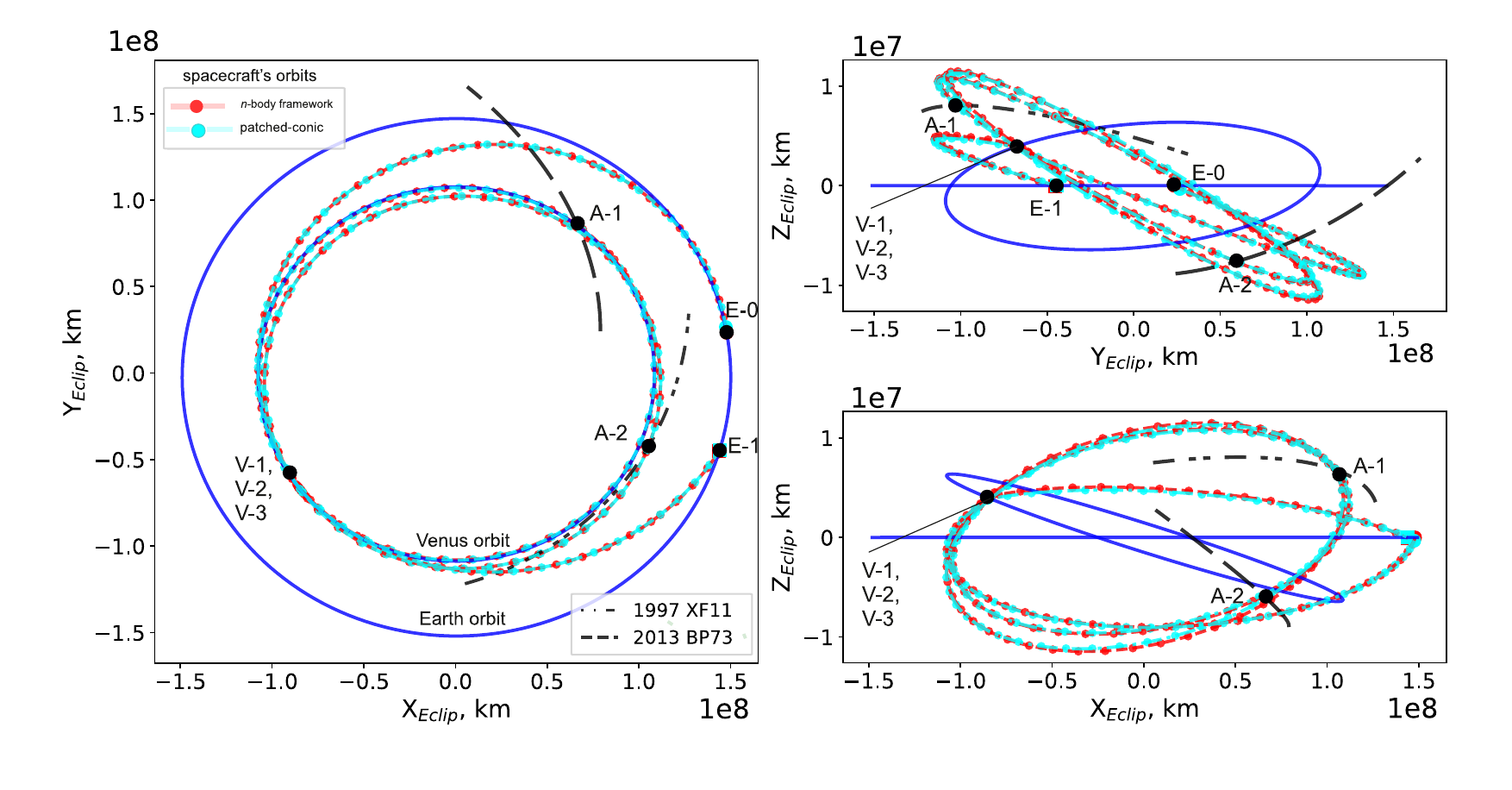}
	\caption{Complete spacecraft trajectory for Venus and dual-asteroid flyby sequence in the $n$-body model}
	\label{fig:trajectory_example2}
\end{figure}

\begin{figure}[H]
	\centering
	\wmedia{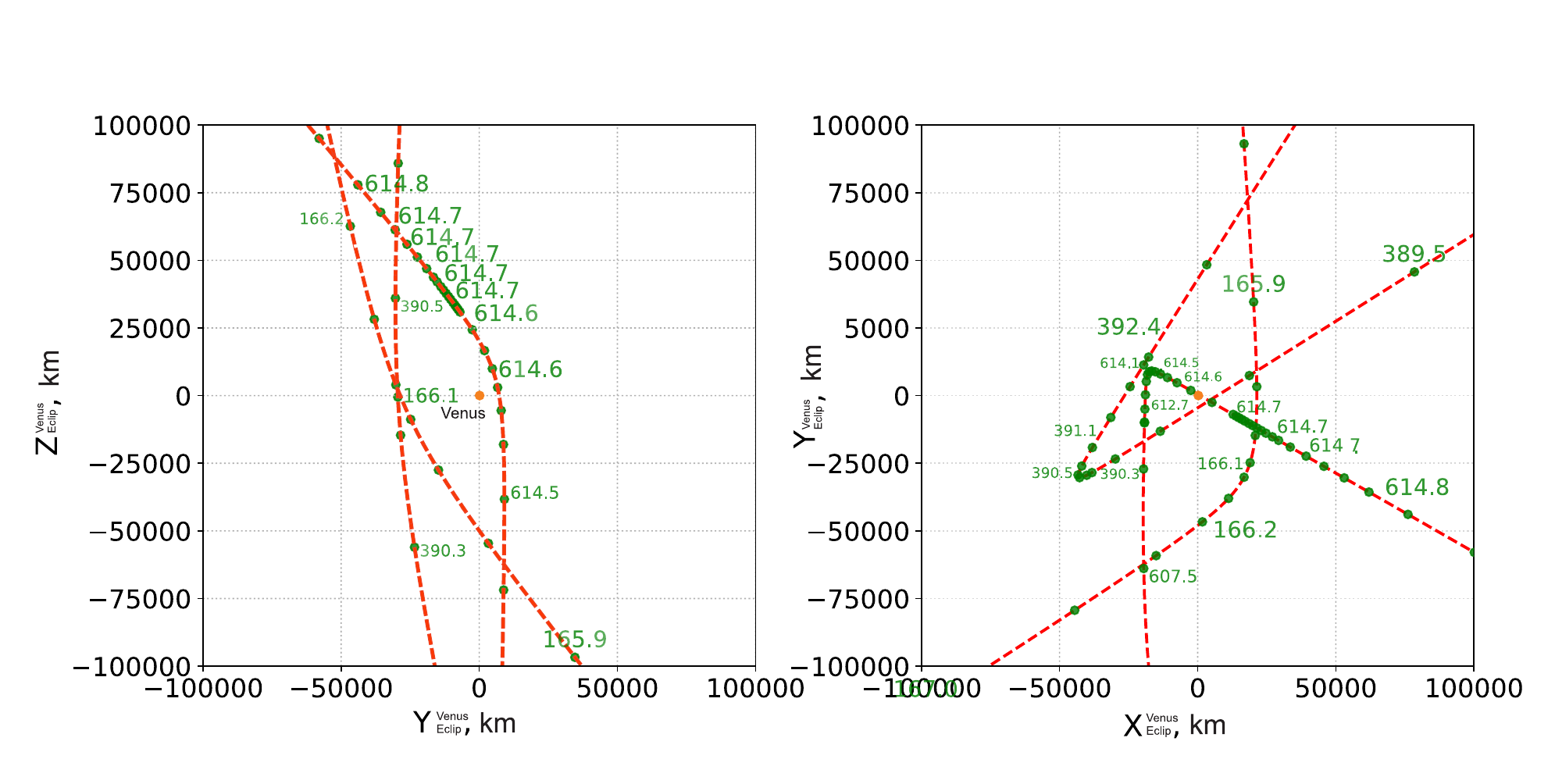}
	\caption{Hyperbolic trajectories during Venus flybys (green numbers indicate days since launch)}
	\label{fig:trajectory_example3}
\end{figure}

The optimized trajectory is illustrated in Fig.~\ref{fig:trajectory_example2}, which shows the complete E-V-A-V-A-V-E flight sequence, while Fig.~\ref{fig:trajectory_example3} provides a detailed view of the SC's motion within Venus's sphere of influence.

The key trajectory parameters obtained from the numerical simulation are summarized in Table~\ref{tab:trajectory_characteristics}.

\begin{table}[h]
	\centering
	\caption{Optimized trajectory characteristics for Earth-Venus-asteroid flyby sequence}
	\small
	\begin{tabular}{@{}lcccccc@{}}
		\toprule
		\textbf{Event} & \textbf{Date (UTC)} & \textbf{$\Delta t$ (days)} & \textbf{$B_T$ (km)} & \textbf{$B_R$ (km)} & \textbf{$r_{\pi}$ (km)} & \textbf{$\Delta v$ (km/s)} \\
		\midrule
		Earth Departure & 02.10.2029 & -- & -- & -- & 6571 & 3.755* \\
		Venus Flyby 1 & 17.03.2030 & 166.1 & 36130.0 & -23988.7 & 32453.0 & 0.027 \\
		1997 XF11 Flyby & 06.06.2030 & 80.9 & -- & -- & 11673.5 & 0.009 \\
		Venus Flyby 2 & 27.10.2030 & 143.4 & 60544.6 & -4023.0 & 48850.4 & 0.0 \\
		2013 BP73 Flyby & 26.02.2031 & 121.3 & -- & -- & 695035.1 & 0.053 \\
		Venus Flyby 3 & 08.06.2031 & 102.9 & 10933.3 & -19180.1 & 12419.0 & 0.0 \\
		\bottomrule
	\end{tabular}
	\label{tab:trajectory_characteristics}

\end{table}

Notice that $^*$ in Table \ref{tab:trajectory_characteristics} denotes $\Delta v_0$ required for transition from LEO to hyperbolic orbit.

Analysis of the simulation results reveals that while the periapsis radii remain closely aligned with predictions, the $B$-plane parameters exhibit deviations of 165 km, 2925 km, and 6442 km in the $B$-vector magnitude for the three Venus flybys, respectively. These discrepancies can be attributed to two primary factors:

\begin{enumerate}
	\item Slight temporal shifts in the Venus encounter dates and times within the $n$-body framework, which alter the $B$-plane orientation.
	
	\item The iterative refinement process of the trajectory, where initial $B$-plane predictions based on $\textbf{v}_{\infty}$ are subsequently corrected during optimization.
\end{enumerate}

Despite these deviations, the initial $B$-parameter estimates provide satisfactory first approximations for the $n$-body system.

Although the flyby distances vary slightly, the dates and times align precisely with the calculations, as well as the predicted periapsis radii. For example, in the case of the 1997 XF11 asteroid, the closest approach distance in the $n$-body simulation differs from the two-body prediction by 11673 kilometers. The closest approach for the 2013 BP73 asteroid differs by 695 thousand kilometers, which is a significant deviation from the originally intended close encounter. 

The observed deviations are particularly pronounced during the second Venus flyby, where the spacecraft enters a resonant orbit approaching a $\pi$-resonance configuration (as discussed in Section~\ref{sec:sc231}). In this case, the spacecraft's second encounter with Venus occurs at a periapsis distance of 523,510 km, significantly deviating from the two-body prediction and complicating the design of an impulse-free trajectory for the subsequent 2013 BP73 asteroid approach.

Figure~\ref{fig:trajectory_example3} illustrates the spacecraft's hyperbolic trajectories during all three Venus flybys, visually confirming the parameters presented in Table~\ref{tab:trajectory_characteristics}.

Let us analyze the impulses applied to correct SC trajectory at each flybys.
Table~\ref{tab:numres2} provides the components $[\Delta V_x, \Delta V_y, \Delta V_z]^T$ of the impulses applied during launch and at each celestial body flyby. These components represent corrections to the initial impulse values derived from the two-body dynamics analysis. To maintain consistency across all maneuvers, all impulse values are expressed in the heliocentric ecliptic coordinate system.

\begin{table}[h]
	\centering
	\caption{Maneuver $\Delta$V Dates and $XYZ$ Components}
	\small
	\begin{tabular}{@{}S[table-format=8.3]
			c
			c
			c
			c
			c
			c@{}}
		\toprule
		{Maneuver} &		
		{\textbf{Date (UTC)}} &
		{\textbf{Time (UTC)}} &
		{$\Delta V_x$} &
		{$\Delta V_y$} &
		{$\Delta V_Z$} \\
		&
		\textbf{(DD.MM.YYYY)} &
		\textbf{(HH:MM:SS)} &
		\textbf{(km/s)} &
		\textbf{(km/s)} &
		\textbf{(km/s)} \\
		\midrule
		\text{Earth Departure} & \text{03.10.2029} &06:40:27.838 & -0.0008864* & -0.0007187* & 0.0033121*  \\
		\text{1st Venus Flyby} & \text{17.03.2030} &   09:03:40.059 &-0.0207574&	0.0052086 &	0.0300977 \\
		\text{1997 XF11} & \text{06.06.2030} & 07:25:11.865 &  0.0085448 &	-0.0014008 &	-0.0035252  \\
		\text{2nd Venus Flyby} & \text{28.10.2030 }&   17:15:43.555& 0 & 0 & 0 \\
		\text{2013 BP73} & \text{26.02.2031 }&   00:20:15.191 & 0.0396911 &	-0.0332544 &	0.0134504\\
		\text{Departure to Earth} & \text{08.06.2031} &  22:24:20.483 & 0 & 0 & 0  \\
		\bottomrule

	\end{tabular}
	\label{tab:numres2}
\end{table}

Notice that $^*$ in Table \ref{tab:numres2} denotes the $\Delta v$ changes to the initial value, determined in the two-body approximation, $\Delta v_0$, required for the transition from a LEO to a hyperbolic orbit.

As shown in Table~\ref{tab:numres2}, the optimized trajectory achieves multiple asteroid encounters while maintaining $\Delta v$ expenditures within the mission budget. The absence of required impulses during the second Venus flyby and Earth departure segments results from the relaxed constraints on asteroid encounter distances, as discussed in the optimization methodology.

Therefore, the results presented in this section clearly depict that the use of the direct approach algorithm based on $B$-plane targeting became limited to the cases when $\gamma \to 0^\circ \text{ or } 180^{\circ}$. In case when solutions with such orbits appear it is necessary to use a more complex and reliable approach to refine the trajectory in an n-body model. Multiple shooting optimization techniques \cite{an2013direct, diedam2018global} were adopted and written using the GMAT script language \cite {hughes2017using}.

\subsubsection{Refining Trajectory Parameters Using a Multiple Shooting Approach}
\label{sec:sc234}

The core idea of the multiple-shooting approach is to determine the spacecraft states at specific nodes $\mathbf{Y}_i \in \mathbb{R}^6$ and epochs $\mathbf{T} \in \mathbb{R}^{g+1}$, with the condition that at these nodes the spacecraft's position is continuous while its velocity changes instantaneously through an impulse maneuver $\Delta\mathbf{V}_i \in \mathbb{R}^3$.

The spacecraft states are then propagated forward and backward in time. At each point connecting the backward- and forward-propagated states, the following continuity condition is enforced:

\begin{equation}
	\mathbf{D}_i(\mathbf{Y}_i^+, \mathbf{Y}_{i}^-) = \mathbf{Y}_{i}^+ - \mathbf{Y}_{i}^- = \mathbf{0}, \quad i=0,\dots,g-1
\end{equation}
where the $\pm$ superscripts denote states propagated forward ($+$) and backward ($-$) in time.

The constraints on the state vector at the initial point as well as at the terminal point can be written as:
\begin{equation}
		\mathbf{\Psi}_0(\mathbf{X}) = 0,  \\
		\mathbf{\Psi}_1(\mathbf{X}) = 0 
\end{equation}

The cost function in this case can be written as:

\begin{equation}
	\min_{\mathbf{X}} J(\mathbf{X}) = \Delta v_{cor} = \sum_{i=0}^{g} \|\Delta\mathbf{V}_i\|
\end{equation}
where
\begin{equation}
	\mathbf{X} = \left[ \mathbf{T}, \mathbf{Y}_0, \mathbf{Y}_1, \dots, \mathbf{Y}_g, \Delta\mathbf{V}_0, \Delta\mathbf{V}_1, \dots, \Delta\mathbf{V}_g \right]^T
\end{equation}

The total number of variables is: $N_x = (g+1) + 6(g+1) + 3(g+1) = 10(g+1)$.

The nonlinear constraints on the connected trajectory segments, as well as the initial and terminal points, are:

\begin{equation}
	\mathbf{C}(\mathbf{X}) = \begin{bmatrix}
		\mathbf{D}_0(\mathbf{X}) \\
		\mathbf{D}_1(\mathbf{X}) \\
		\vdots \\
		\mathbf{D}_{g-1}(\mathbf{X}) \\
		\mathbf{\Psi}_0(\mathbf{X})\\
		\mathbf{\Psi}_1(\mathbf{X}) \\
	\end{bmatrix} = \mathbf{0}
\end{equation}

The following algorithm was used to write a GMAT script for calculating trajectories using multiple shooting techniques.

\scriptsize
\begin{algorithm}[H]
	\caption{Multiple Shooting Optimization for N Segments}
	\KwIn{$g$ (number of segments), initial guess $\mathbf{X}_0$}
	\KwOut{Optimal trajectory $\mathbf{X}^*$}
	\Begin{
		$k \leftarrow 0$\;
		$\mathbf{X} \leftarrow \mathbf{X}_0$\;
		\While{$\|\nabla J(\mathbf{X}_k)\| > \epsilon$ and $k < k_{\text{max}}$}{
			Unpack decision variables: $\mathbf{T}, \mathbf{Y}, \Delta\mathbf{V} 
			\leftarrow \mathbf{X}_k$\;
			Apply maneuver at $t_0$: $\mathbf{Y}_0^+ \leftarrow \mathbf{Y}_0 + \Delta\mathbf{V}_0$\;
			Propagate segment forward ($+t$): $\mathbf{Y}_{1}^- \leftarrow \text{Integrate}(\mathbf{f}, \mathbf{Y}_0^+, t_0, t_{1})$\;
			\For{$i = 2$ \KwTo $g-1$}{
				Apply maneuver: $\mathbf{Y}_i^+ \leftarrow \mathbf{Y}_i + \Delta\mathbf{V}_i$;
				Propagate segment forward ($+t$): $\mathbf{Y}_{i+1}^- \leftarrow \text{Integrate}(\mathbf{f}, \mathbf{Y}_i^+, t_i, t_{i+1})$\;
				Propagate segment backward ($-t$): $\mathbf{Y}_{i-1}^+ \leftarrow \text{Integrate}(\mathbf{f}, \mathbf{Y}_i^+, t_i, t_{i+1})$\;
				Compute defect: $\mathbf{D}_i \leftarrow \mathbf{Y}_{i-1}^+ - \mathbf{Y}_{i-1}^-$\;
			}
			Compute constraints: $\mathbf{C}(\mathbf{X}_k) \leftarrow [\mathbf{D}_0, \dots, \mathbf{D}_{n-1}]$\;
			Compute objective: $J(\mathbf{X}_k) \leftarrow \sum_{i=0}^n \|\Delta\mathbf{V}_i\|$\;
			Compute gradient: $\nabla J(\mathbf{X}_k)$, $\nabla \mathbf{C}(\mathbf{X}_k)$\;
			Solve subproblem using sequential quadratic programming approach: $\Delta\mathbf{X} \leftarrow \arg\min \frac{1}{2}\Delta\mathbf{X}^T\mathbf{H}_k\Delta\mathbf{X} + \nabla J_k^T\Delta\mathbf{X}$\;
			Update: $\mathbf{X}_{k+1} \leftarrow \mathbf{X}_k + \alpha_k\Delta\mathbf{X}$\;
			$k \leftarrow k + 1$\;
		}
		$\mathbf{X}^* \leftarrow \mathbf{X}_k$\;
	}
\end{algorithm}

\normalsize

For numerical integration the previously established differential equations were used \ref{eq:difeq}.

\subsubsection{Calculated Trajectory Parameters under Multiple Shooting Approach}
\label{sec:sc235}

Let us consider the same problem as in \ref{sec:sc233}, i.e., the Earth-Venus-asteroid-Venus flight scheme. For this trajectory, the total number of variables is $N_x = 70$, with the number of nonlinear constraints equal to 34, including initial and terminal point constraints.

The refined initial conditions of the SC at time of the transition from LEO to hyperbolic orbit, obtained after employing the trajectory optimization technique from Section \ref{sec:sc234} can be written as (given in the Earth ecliptic inertial system J2000):

\begin{equation*}
	\{\textbf{r}_E, \textbf{v}_E, t_{JD, UTC}\} =
	\begin{bmatrix}
		-5434.309418 \\
	 3562.653721 \\
	981.289719 \\
	-5.127796 \\
	-5.338029 \\
	-8.843770  \\
		2462413.077214
	\end{bmatrix}
	\label{eq:vect2}
\end{equation*}

The refined parameters of the trajectory, calculated using the multiple shooting approach, are given in Table \ref{tab:numres3}. The required maneuvers are provided in Table \ref {tab:maneuver_data}, in the $VNB$ coordinate system, where $V$ is directed along the velocity vector relative to the centre body, $N$ in the normal direction and $B$ completes the right-handed set of vectors.

\begin{table}[h]
	\centering
	\caption{Characteristics of the obtained SC flight trajectory}
	\small
	\begin{tabular}{@{}c
			c
			c
			c
			c
			c
			c
			c@{}}
		\toprule
		Celestial Body &
		{Date (UTC)} &
		{${\Delta t}$} &
		{${B_T}$} &
		{${B_R}$} &
		{${r_{\pi}}$} &
		{$\Delta v$} \\
		- &
		{(DD.MM.YYYY)} &
		{(days)} &
		{(km)} &
		{(km)} &
		{(km)} &
		{(km/s)} \\
		\midrule
		Earth &  \text{03.10.2029}& 165.4 & \text{-} & \text{-} & 6571  & 3.755*  \\
		Venus & \text{17.03.2030}  & 80.6 & 36130.0 & -24078.2 & 32424.4&  0.003 \\
		1997 XF11  & \text{06.06.2030} &143.8  &  \text{-} & \text{-} & $\leq$ 1  & 0.001  \\
		Venus & \text{28.10.2030 }&120.9 & 	-5120.7 & 63453.7 & 52236.3  & 0.117  \\
		2013 BP73 & \text{26.02.2031} & 103.3 &    \text{-} & \text{-}& $\leq$ 1 & 0.010 \\
		Venus & \text{09.06.2031} & 87.8  & 4142.9 &  -19985.3 & 11422.8  & 0.037  \\
		\bottomrule
	\end{tabular}
	\label{tab:numres3}
\end{table}

Notice that $^*$ in Table \ref{tab:numres3} denotes $\Delta v_0$ required for transition from LEO to hyperbolic orbit.

\begin{table}[h]
	\centering
	\caption{Maneuver $\Delta$V Dates and $VNB$ Components}
	\small
	\begin{tabular}{@{}S[table-format=8.3]
			c
			c
			c
			c
			c
			c@{}}
		\toprule
		{Maneuver} &		
		{\textbf{Date (UTC)}} &
		{\textbf{Time (UTC)}} &
		{$\Delta V_V$} &
		{$\Delta V_N$} &
		{$\Delta V_B$} \\
		&
		\textbf{(DD.MM.YYYY)} &
		\textbf{(HH:MM:SS)} &
		\textbf{(km/s)} &
		\textbf{(km/s)} &
		\textbf{(km/s)} \\
		\midrule
		\text{Earth Departure} & \text{03.10.2029} & 13:51:11.324& 0.00001708*&	-0.00003267*	&0.00004652*	\\
		\text{1st Venus Flyby} & \text{17.03.2030} &  22:46:01.667 &-0.00048693&	-0.00063599	&0.00510856\\
		\text{1997 XF11} & \text{06.06.2030} &12:20:47.452 & 0.00660010	&-0.00486912&	-0.00051071\\
		\text{2nd Venus Flyby} & \text{28.10.2030 }&  06:37:18.494 & 0.05952555&	0.06808831&	0.07551561\\
		\text{2013 BP73} & \text{26.02.2031 }&  03:05:04.465 & 0.00512191&	-0.01271204&	-0.00516111\\
		\text{Departure to Earth} & \text{09.06.2031} &  11:07:40.146 &-0.00016309&	-0.00024640&	0.00432417\\
		\bottomrule
	\end{tabular}
	\label{tab:maneuver_data}
\end{table}

Notice that $^*$ in Table \ref{tab:maneuver_data} denotes the $\Delta v$ changes to the initial value, determined in the two-body approximation, $\Delta v_0$, required for the transition from a LEO to a hyperbolic orbit.

Tables \ref{tab:numres3} and \ref{tab:maneuver_data} show that the use of the multiple shooting method allows for a more precise restoration of the trajectory in the $n$-body model. The resulting $\Delta v_{cor}$ is 0.15 km/s, which is about twice as high as the result obtained in the previous section, which was 0.089 km/s. While in the previous method, large discrepancies were found at flyby distances from asteroids due to relaxed constraints, in this case, the impulses compensated for these discrepancies required an additional 100 m/s impulse. It should be noted that the comparison of data from the two-body approximation and the refined data in this case differs only slightly. The total difference between the components in the B-plane does not exceed 1000 km for all three approaches to Venus.

The Table \ref{tab:numres3} and \ref{tab:maneuver_data} analyses also show that the biggest impulse was applied at the 2013 BP73 asteroid flyby segment of the flight. This is similar to the discrepancies in asteroid flyby distances that were encountered in the previous section, as they also occur during the same flight segment.

\begin{figure}[H]
	\centering
	\wmedia{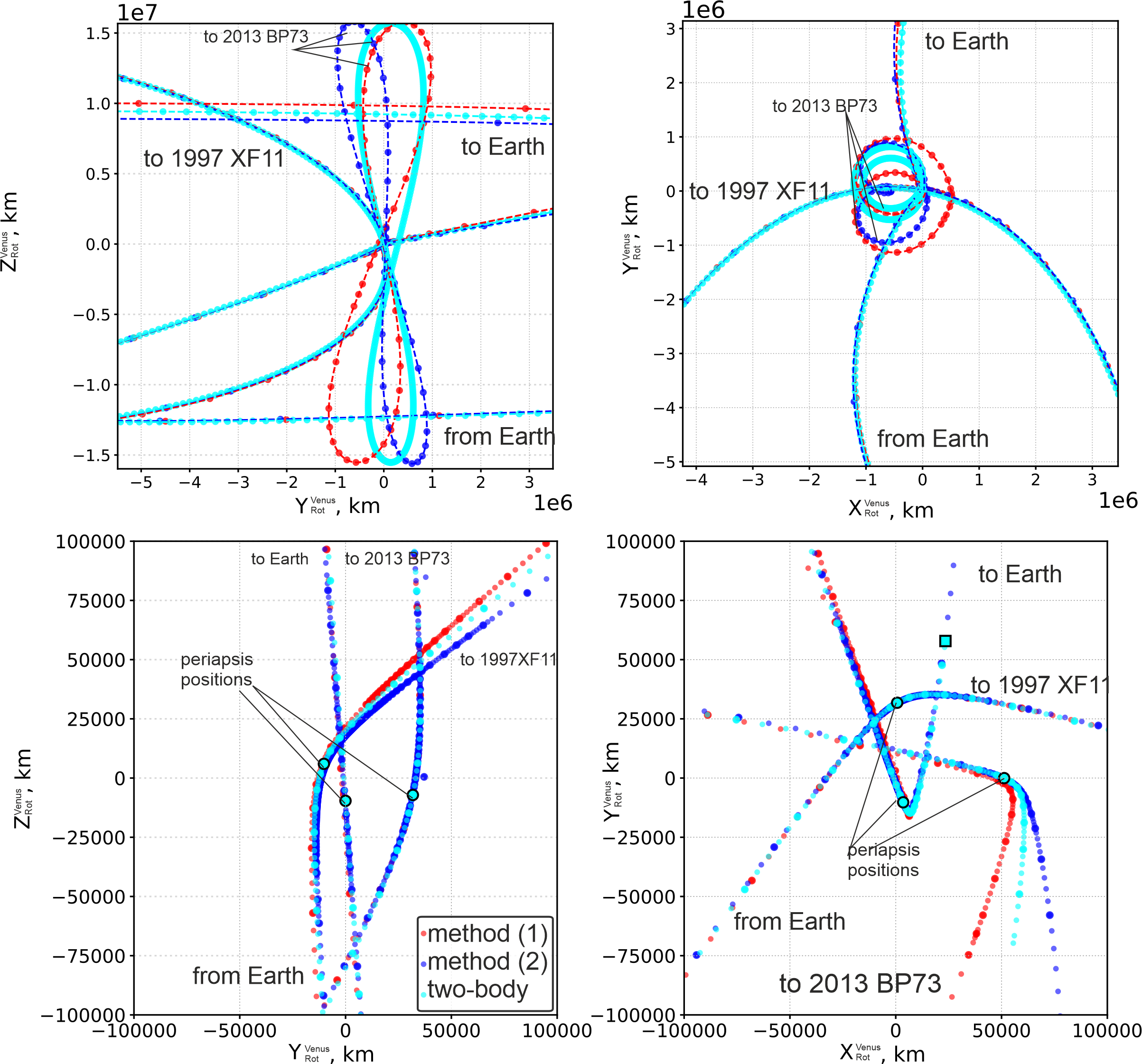}
	\caption{Hyperbolic trajectories during Venus flybys obtained using three discussed approaches in rotating coordinate system}
	\label{fig:trajectory_example4}
\end{figure}

Let us illustrate the three approaches used to calculate the trajectory discussed in Fig.~\ref{fig:trajectory_example4}. The figure shows the orbits of these asteroids in a rotating coordinate system, where the x-axis ($X_{rot}^{Venus}$) is aligned with the direction of the Sun-Venus, the z-axis ($Z_{rot}^{Venus})$ is directed along the orbital angular momentum, and the y-axis completes the three-dimensional basis ($Y_{rot}^Venus$). The "Venus" in upper script designates that the center of the coordinate system was placed at the center of Venus.

Analysis of Figure~\ref{fig:trajectory_example4} shows that the differences in trajectories calculated using the three discussed methods slightly deviate from each other starting with the first flyby and a significant deviation is indicated by impulses in the corresponding tables occurring during the flight to 2013 BP73 - entry into the $\pi$-resonant orbit.

\subsection{Combining multiple Venus gravity assists with flyby of planet's resonant asteroids and PHAs}
\label{sec:sc25}

To date, approximately 20 asteroids have been identified that are in resonance with Venus orbital motion \cite{pan2025attempt, carruba2025invisible}. These are rare objects, and their existence represents a unique phenomenon in celestial mechanics. Resonances near Venus are short-lived due to gravitational influences from giant planets and interactions with the Sun, which eventually disrupt these resonances, presenting a significant challenge to the stability of these asteroids' orbits. This study explores schemes involving flybys of potentially hazardous asteroids in conjunction with resonant asteroids.

\begin{figure}[H]
	\centering
	\wmedia{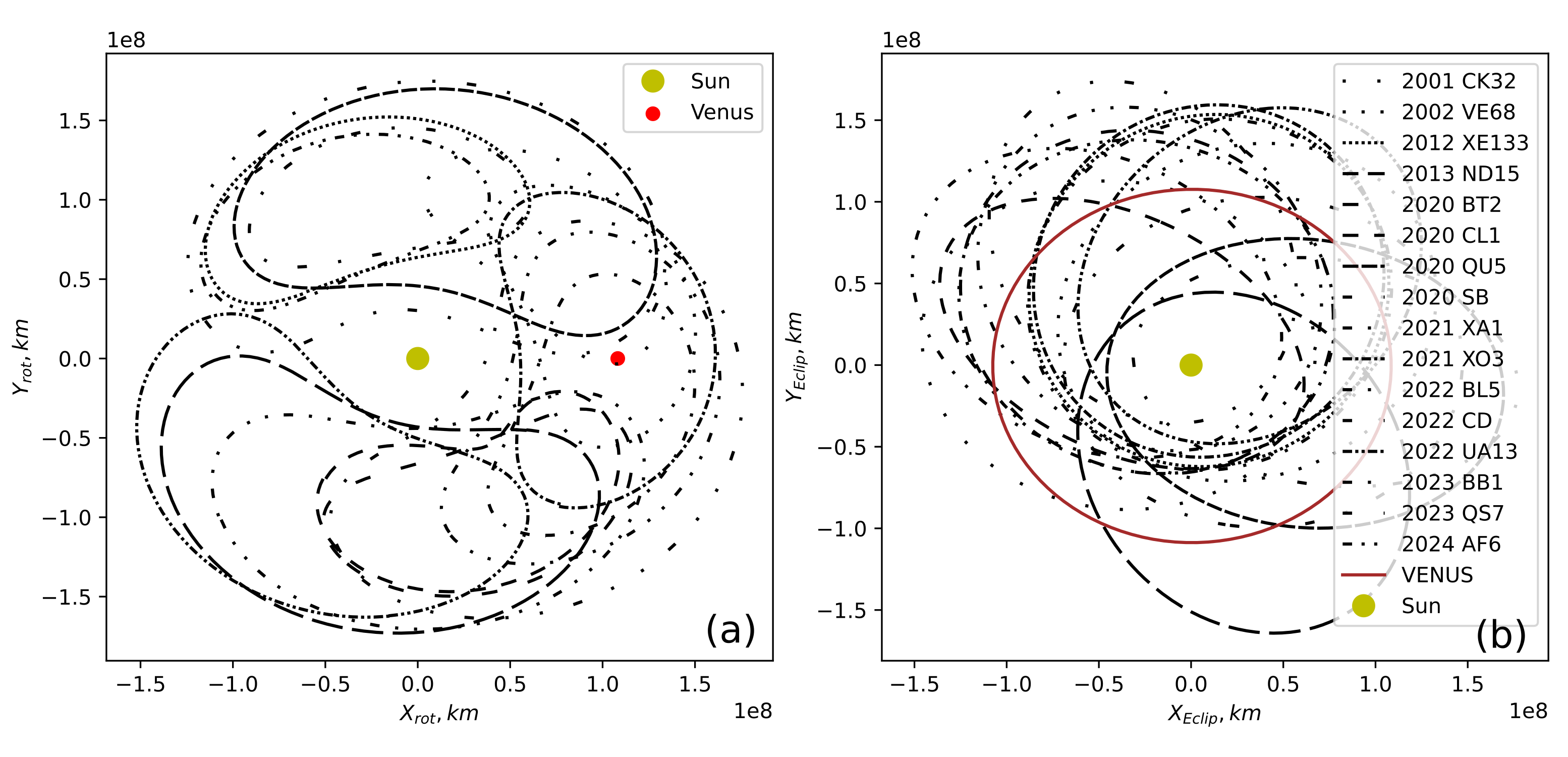} 
	\caption{Resonant asteroids orbiting Venus in rotating and inertial coordinate systems for the date period 2029-2030: (a) in rotating coordinate system; (b) in the Heliocentric Ecliptic J2000 coordinate system}
	\label{fig:resast2}
\end{figure}

Let us illustrate all known and suspected resonant asteroids in Fig.~\ref{fig:resast2}. The figure shows the orbits of these asteroids in a rotating coordinate system ($X_{rot}$, $Y_{rot}$). The remaining part of the image represents the orbits of the asteroids in the heliocentric, ecliptic J2000 frame ($X_{Eclip}, Y_{Eclip}$). 

\textbf{}

For the purpose of subsequent analysis, only those asteroids that make close encounters to Venus are considered. Therefore, horseshoe-type asteroids are not taken into account. Only 5 out of 20 asteroids satisfy the condition of a close encounter within 5 million km over a decade from 2029 until 2040. An example of constructing a flight trajectory using the 2001 CK32 asteroid as a target is provided. Note that the full set of PHAs was used for constructing the flight scheme.

\begin{figure}[H]
	\centering
	\wmedia{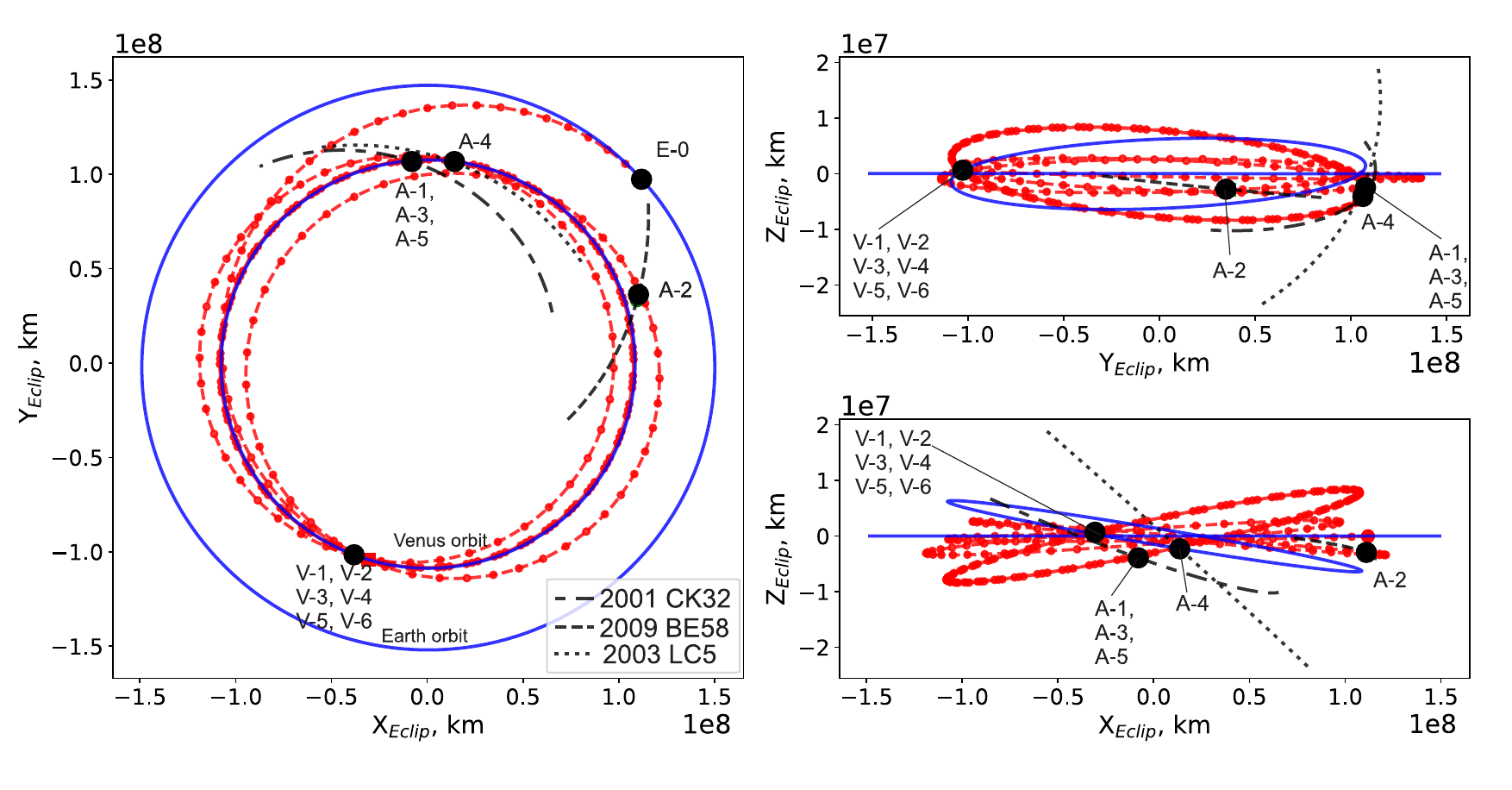} 
	\caption{Trajectory of SC flyby of Venus and two asteroids on a 1:1 resonant orbit}
	\label{fig:resasttrj}
\end{figure}

The trajectory found is depicted in Fig.~\ref{fig:resasttrj}. Some characteristics of flight trajectory are outlined in Table \ref{tab:resp5}.

\begin{table}[h]
	\centering
	\caption{Characteristics of transfer trajectories with flyby of co-orbital asteroids with Venus combined with PHAs}
	\label{tab:resp5}
	\scriptsize
	\begin{tabular}{@{} l l l c S[table-format=3.2] S[table-format=1.3] r S[table-format=2.2] S[table-format=2.3] @{}}
		\toprule
		\textbf{Segment} & \textbf{Celestial Body} & \textbf{Path} & {\makecell[c]{\textbf{Launch/} \\ \textbf{Maneuver Date}}} &{\makecell[c]{\textbf{Duration} \\ \textbf{(days)}}} & {\makecell[c]{\textbf{$\Delta v$} \\ \textbf{(km/s)}}} & {\makecell[c]{\textbf{Flyby Alt.} \\ \textbf{ $\tildH_{\pi}$ (km)}}} & {\makecell[c]{\textbf{Comm. Angle} \\ \textbf{(E-SC-S, deg)}}} & {\makecell[c]{\textbf{Flyby $v_{\infty}$} \\ \textbf{(km/s)}}} \\
		\midrule
		
		\textbf{Launch} & Earth & Earth (E-0) & \text{28.10.2029}  & {-} & \textbf{3.581} & 0.2 & {$\leq$ 1} & 2.823 \\
		1 & Venus & E-0 $\to$ V-1& \text{07.04.2030}  &160.48 & 0.062 & 6040 & 55.13 & 4.749 \\
		2 & 2001 CK32  & V-1 $\to$ A-1 &\text{11.08.2030} & 126.37 & 0.014 & {-} & 136.63 & 15.281 \\
		3 & Venus & A-1 $\to$ V-2 &\text{17.11.2030}& 98.32 & 0.001 & 1960 & 163.01 & 4.751 \\
		4 & 2009 BE58  & V-2 $\to$ A-2 &\text{22.02.2031} &96.72 & 0.000 & {-} & 134.29 & 23.274 \\
		5 & Venus & A-2 $\to$ V-3 &\text{30.06.2031} & 127.98 & 0.003 & 2080 & 26.05 & 4.758 \\
		6 & 2001 CK32  & V-3 $\to$ A-3 &\text{04.11.2031} & 126.51 & 0.001 & {-} & 54.13 & 15.332 \\
		7 & Venus & A-3 $\to$ V-4 &	\text{10.02.2032} & 98.18 & 0.050 & 14400 & 111.48 & 4.836 \\
		8 & (90367) 2003 LC5 & V-4 $\to$ A-4 &\text{26.05.2032} & 105.78 & 0.089 & {-} & 160.23 & 14.162 \\
		9 & Venus & A-4 $\to$ V-5 &\text{22.09.2032} &118.91 & 0.000 & 15000 & 106.96 & 4.767 \\
		10 & 2001 CK32  & V-5 $\to$ A-5 &\text{26.01.2033} & 126.55 & 0.001 & {-} & 31.36 & 15.356 \\
		11 & Venus & A-5 $\to$ V-6 &\text{04.05.2033} &98.14 & {-}& {-} & 28.06 & 4.760 \\
		
		\midrule
		
		\textbf{Total} & & & & \textbf{1285.14} & \textbf{3.802} & \multicolumn{3}{l}{} \\
		& & & &  {\small(\textasciitilde 3.5 yrs)} & & \multicolumn{3}{l}{} \\
		
		\bottomrule
	\end{tabular}
\end{table}

From Table \ref{tab:resp5}, it can be seen that the trajectory found is almost impulse-free. The average flyby velocity of the 2000 CK32 asteroid is 15.3 km/s, and the flyby velocity for the planet is 4.7 km/s. The maximum value of the E-SC-S angle is 163.01 deg, which possibly allows radar observation of the SC during the first gravity assist. The next flybys of Venus occur at slightly lower angle values, and at the last planet's flybys, the E-S-C angle takes a value of 28.03 deg.

\subsection{Determining Asteroid Flyby Trajectories on the Earth-Venus Transfer Segment}
\label{sec:sc24}
In the work \cite{simonov2025}, trajectories of flybys of asteroids whose orbits are in resonance with Venus were found for the Earth-Venus transfer segment. The study \cite{Zubko2023a} was unable to find trajectories for flybys of asteroids larger than 1 km in diameter in the same transfer segment. This work supplements these previous studies by investigating flybys of PHAs during the Earth-Venus transfer segment.

The search largely followed the methodology outlined in Section \ref {sec:sc11}. A significant simplification was made by considering only two transfer segments: from Earth to asteroid and from asteroid to Venus. Flybys of potentially hazardous objects from the main sample of 139 asteroids were considered. Launch periods between 2029 and 2050 were investigated. 

Some parameters of calculated transfer trajectories are presented in Table \ref {tab:eav}

\scriptsize
\begin{longtable}{@{}c l c c c c c c c c c c@{}}
	\caption{Parameters of the Earth-Asteroid-Venus transfer trajectories} \\
\toprule
Name & $t_0$ & $t_1$ & $t_2$ & $\Delta t_{E-V}$ & $\Delta v_0$ & $v_{\infty, 0}$ & $v_{\infty, 1}$ & $v_{\infty, 2}$ & $i_{eclip, 0}$\\
 - &
{(DD.MM.YYYY)} &-//-&-//-&	{(days)} &
{(km/s)} &
{(km/s)} &
{(km/s)} &
{(km/s)} &
{(deg)} \\
\midrule
\endfirsthead

\multicolumn{12}{c}{{\tablename\ \thetable{} -- continued}} \\
\toprule
Name & $t_0$ & $t_1$ & $t_2$ & $\Delta t_{E-V}$ & $\Delta v_0$ & $v_{\infty, 0}$ & $v_{\infty, 1}$ & $v_{\infty, 2}$ & $i_{eclip, 0}$\\
- & - &
{(DD.MM.YYYY)} &-//-&-//-&	{(days)} &
{(km/s)} &
{(km/s)} &
{(km/s)} &
{(km/s)} &
{(deg)} \\
\midrule
\endhead

\bottomrule
\endfoot

(367789) 2011 AG5& 25.05.2029 & 16.12.2029 & 10.08.2030 & 442 & 3.695 & 3.246 & 12.007 & 4.817 & 1.896 \\ 
(826936) 1998 HH49& 11.12.2030 & 13.07.2031 & 05.02.2032 & 421 & 3.625 & 2.994 & 9.474 & 2.733 & 2.842 \\ 
2018 BT6& 03.01.2031 & 27.05.2031 & 21.03.2032 & 443 & 3.905 & 3.927 & 30.581 & 4.363 & 5.414 \\  
(177049) 2003 EE16& 06.02.2031 & 18.06.2031 & 31.01.2032 & 359 & 3.947 & 4.05 & 20.003 & 6.868 & 1.183 \\ 
2010 JE88& 12.02.2031 & 07.08.2031 & 08.02.2032 & 361 & 3.884 & 3.747 & 31.978 & 7.776 & 0.489 \\  
(509456) 2007 LF& 03.04.2031 & 01.06.2031 & 17.10.2031 & 197 & 3.793 & 3.579 & 11.177 & 4.41 & 0.781 \\ 
(308635) 2005 YU55& 09.04.2031 & 13.09.2031 & 05.11.2031 & 210 & 4.008 & 4.226 & 15.818 & 5.702 & 1.142 \\ 
(143487) 2003 CR20& 25.11.2033 & 05.04.2034 & 20.03.2035 & 480 & 4.033 & 4.12 & 16.116 & 3.21 & 1.647 \\ 
2014 SP142& 24.03.2034 & 16.06.2034 & 12.06.2035 & 445 & 3.755 & 3.35 & 17.922 & 5.789 & 3.08 \\  
(141495) 2002 EZ11& 06.04.2034 & 01.11.2034 & 14.04.2035 & 373 & 4.071 & 4.136 & 22.492 & 8.439 & 2.883 \\ 
(101869) 1999 MM& 08.08.2034 & 28.11.2034 & 04.02.2035 & 180 & 3.796 & 3.59 & 21.571 & 6.086 & 3.246 \\ 
(177049) 2003 EE16& 22.09.2034 & 21.12.2034 & 09.03.2035 & 168 & 3.817 & 3.657 & 26.327 & 7.893 & 0.333 \\ 
2008 DJ& 21.09.2035 & 05.12.2035 & 31.10.2036 & 406 & 3.903 & 3.922 & 10.481 & 4.488 & 4.273 \\  
2021 SX3& 08.01.2036 & 31.07.2036 & 09.03.2037 & 426 & 4.065 & 4.382 & 23.181 & 6.978 & 4.125 \\  
2014 SP142& 29.08.2037 & 04.04.2038 & 04.11.2038 & 432 & 3.895 & 3.897 & 8.621 & 8.542 & 1.464 \\  
2003 MK4& 06.12.2040 & 04.01.2041 & 07.05.2041 & 152 & 3.598 & 2.886 & 13.907 & 3.07 & 1.255 \\  
2020 BC6& 09.04.2042 & 16.09.2042 & 11.04.2043 & 367 & 3.863 & 3.802 & 14.487 & 7.579 & 2.669 \\  
(85236) 1993 KH& 18.09.2045 & 20.11.2045 & 30.03.2046 & 193 & 3.895 & 3.898 & 14.208 & 5.539 & 0.871 \\ 
2013 ND24& 03.10.2045 & 14.01.2046 & 07.04.2046 & 186 & 3.796 & 3.593 & 29.243 & 5.668 & 0.045 \\  
(1864) Daedalus 1971 FA& 22.03.2047 & 13.08.2047 & 13.06.2048 & 449 & 4.033 & 4.252 & 20.892 & 8.868 & 0.993 \\
2008 MP1& 24.05.2047 & 31.08.2047 & 28.11.2047 & 188 & 3.893 & 3.733 & 9.81 & 6.456 & 3.019 \\  
2022 AY1& 22.04.2048 & 07.07.2048 & 11.08.2049 & 476 & 3.909 & 3.813 & 17.052 & 5.417 & 4.307 \\  
(89958) 2002 LY45& 20.06.2048 & 12.01.2049 & 28.08.2049 & 434 & 3.577 & 2.609 & 29.81 & 3.84 & 0.664 \\ 
2014 XL7& 30.06.2048 & 22.09.2048 & 20.09.2049 & 447 & 4.032 & 4.292 & 16.247 & 4.401 & 7.069 \\  
1997 XF11& 24.08.2048 & 16.03.2049 & 14.11.2049 & 447 & 3.833 & 3.462 & 18.768 & 6.438 & 3.165 \\ 

\bottomrule
\label{tab:eav}

\end{longtable}

\begin{center}
	\small Note: $t_i$, $i = 0,1,2$, corresponds to the characteristic dates of launch, asteroid flyby and Venus encounter; $\Delta t_{E-V}$ is the duration of the Earth-Venus transfer, d; $v_{\infty,i}$, $i=0,1$, correspond to asymptotic velocities at the times of launch and asteroid flybys. $i_{ecl, 0}$ is the inclination of the SC's orbit in the Earth-to-Venus section.
\end{center}

\normalsize

As a result of the study, the findings of which are presented in Table \ref{tab:eav}, it was established that even when using the limited sample of asteroids filtered by the MOID parameter value, considered in this work, it is possible to identify several impulse-free approach trajectories for the SC with an asteroid during an Earth-Venus transfer.

However, these opportunities are significantly limited compared to those considered in Section \ref{sec:sc22}. Only 8 out of the 25 identified trajectories for the launch period from 2029 to 2050 reached Venus in less than a single revolution. This is considerably fewer than the 56 trajectories found for the E-V-A-V scheme, which were identified for just the first two launch windows in the same period.

\section{Discussion}
\label{sec:disc}

The problem of multiple asteroid flybys has been previously studied. As stated in the introduction, many algorithms have been developed to address this issue. These algorithms include robust optimization techniques that are appropriate for the largest Global Trajectory Optimization Competition (GTOC). The need to create an asteroid tour with many nodes requires a large number of candidates to choose from. This explains the main focus of researchers on Main Belt asteroids and, from time to time, on NEAs \cite{ sukhanov1996trajectory, rivkin2020multi, sukhanov2024s, Bellome2024, Stikkle2024, olkin2024mission}.

Consequently, significant research interest exists for designing multi-asteroid encounter trajectories in the Main Belt region.  As a result, the space between Earth and Venus has remained underutilized and received less attention. The primary reasons for the limited exploration of asteroid flyby chains near Venus's orbit are the significantly smaller asteroid population compared to the Main Belt and the proximity to the Sun, which requires additional solutions to manage the thermal load on the SC during prolonged periods.

Growing interest in in-situ studies of Mercury and Venus, coupled with advances in small satellite technology presents opportunities to enhance the scientific return from inner Solar System missions.

To address this, the present work investigates schemes for designing SC trajectories that enable the simultaneous study of a planet and an asteroid through repeated flybys of both celestial bodies. The proposed approach uses Venus gravity assists to transition the spacecraft into resonant orbits, enabling subsequent asteroid flybys. 

Particular attention was paid to the 1:1 resonance scenario, which minimizes the time intervals between flybys. Potentially hazardous asteroids ($\rho_{MOID} \leq 1.05 \LD$) were chosen as objects of study by SC. A comparative analysis of two strategies was conducted within four flight schemes:
\begin{itemize}
	\item flight through the inner Solar System with a Venus gravity assist and movement along a resonant orbit;
	\item direct flight from Earth to an asteroid with a subsequent return to Earth.
\end{itemize}

The results obtained using the proposed approach show that the flight scheme involving Venus flybys and asteroid encounters:
\begin{itemize}
	\item did not significantly increase the total mission duration (from launch to the first encounter with the planet after the asteroid flyby);
	\item in some cases, allowed for a reduction in the required characteristic velocity for the asteroid flyby;
	\item expanded launch opportunities by providing shorter intervals between launch windows, while maintaining the necessary characteristic velocity for the flight to the asteroid;
	\item in several cases, the analysis revealed that the asymptotic velocity of an asteroid's flyby was lower than that for a direct flight.
\end{itemize}


The $n$-body simulation of a spacecraft trajectory, including flybys of 1997 XF11 and 2013 BP73 asteroids, demonstrated the feasibility of the developed method. This confirms the ability to construct trajectories that enable asteroid flybys while the spacecraft remains near Venus's orbit. The flyby of the second asteroid highlighted a significant challenge in applying the developed trajectory search technique. It was shown that asteroid flybys on $\pi$-resonant orbits complicate refinement in $n$-body simulation. Although most of the predicted parameters from the patched conic method differed only slightly, the closest encounter distance for the 2013 BP73 asteroid increased to 695 thousand kilometers. This was due to the second encounter with Venus on its way to the asteroid. Because of perturbations from that encounter, it was impossible to find a close approach trajectory using this method. The latter suggests that direct refinement of trajectories using $B$-plane targeting is possible if the initial solution has $0^\circ << \gamma << 180^\circ$. 

To handle the problem, disregarding the resonant orbit, the multiple shooting approach was used. The rationale for using it, despite its complexity, is that it uses patching between backward and forward time propagation, therefore being less sensitive to small perturbations caused by Venus and Earth in the heliocentric segment. The results of this technique showed that, in the case of strict restrictions on asteroid flyby distances, the total impulse for maneuvers increased to 0.15 km/s. The magnitude of the impulse needed for flight at $\gamma \approx 179^\circ$ is 0.11 km/s, while flight in all other segments requires less than 40 m/s in total. It can also be noted that the obtained results are close to the patched conic ones. It is known that a patched cone provides a lower limit on $\Delta v$, so the obtained result is about 5\% of $\Delta V_0^{lim}$, which means trajectories are feasible.

The importance of accounting for orbital uncertainties and their impact on flyby feasibility is acknowledged, with a comprehensive investigation of these aspects being planned for subsequent dedicated research. The development of techniques to validate trajectory solutions within high-fidelity $n$-body simulations that explicitly incorporate asteroid orbital uncertainties is intended for future work. Various methods, including Monte Carlo simulations and approaches based on min-max principles, will be explored to assess solution stability and resulting variations in $\Delta v_0$ requirements.

For the current study, a full uncertainty analysis was considered beyond the primary scope of establishing the fundamental feasibility of the proposed flight schemes. However, to partially address these concerns, an additional filter was incorporated into the asteroid selection process. Only asteroids with orbital element uncertainties below specific thresholds, as reported in the JPL and Minor Planet Center databases, were considered, ensuring that the candidate set was comprised of objects with well-determined orbits. Diagrams illustrating the distribution of these orbital errors for the selected asteroids have been included in Section~\ref{sec:sc21}, providing transparency regarding the quality of data utilized in the analysis.

The analysis presented was also extended from studying flybys of PHAs to co-orbital asteroids with Venus, a small population of objects that are in short-term resonance with the orbit of Venus. Trajectory construction, including a flyby of the 2001 CK32 asteroid, showed that such a scheme was feasible. This scheme shares similarities with the Lucy mission trajectory design \cite{olkin2024mission} but shifts focus from Jupiter Trojans to Venus co-orbitals and from Main Belt asteroids to PHAs. The provided example demonstrates that it is possible to combine flybys of co-orbital asteroids with Venus and PHAs with a study of Venus, thus possibly expanding the scientific value of the mission.

\section{Conclusions}
\label{sec:con}

This work presents a method for designing SC trajectories that enable the simultaneous study of a planet and an asteroid through repeated flybys of both celestial bodies. The proposed approach is based on the use of a Venus gravity assists to transition the SC into a resonant orbit, enabling subsequent asteroid flybys. To optimize calculations, a modified patched-conic method with a simplified procedure for searching for resonant orbits relative to Venus was used. For missions with multiple asteroids, geometric criteria for selecting trajectories based on orbital elements of the resonant orbit and asteroid were developed.

An advantage of the proposed flight schemes is their applicability to limited object sets. For instance, trajectories involving flybys of 1 to 3 objects were found within a set of 139 asteroids. Furthermore, trajectories involving flybys of more than two asteroids are easily found within an expanded set of selected asteroids. An analysis of the trajectory calculated for flybys of 1997 XF11, 2013 BP73, 2012 TO139, 2001 HY7, 2022 BJ asteroids in a launch window of 2029 revealed that the trajectories allow for impulse-free flybys, with the magnitude of the impulses required for planetary flybys being less than 100 m/s.

The possibility of returning the SC after a series of encounters with a planet and asteroids was investigated. The trajectories of the E-V-A-V scheme were taken and the analysis revealed that for about half of them it was possible to construct a return segment from Venus to Earth, although adding this segment  to optimization changes the optimal launch date and increases $\Delta v_0$ while maintaining impulse-free asteroid flybys. The trajectory was calculated including the flybys of 1997 XF11, 2013 BP73 and return to Earth in both two-body and $n$-body dynamics frameworks.

 It was revealed that flybys of asteroids on the trajectories closely aligned with 1:1 resonant orbits having $\gamma \approx 0^{\circ} \text{ or } 180^{\circ}$ lead to a significant deviation in asteroid flyby distance which was shown in 2013 BP73 case. While such deviations occurred, the impulses as well as $B$-plane parameters in $n$-body dynamics remained closely aligned with the patched-conic solution in terms of the time of flight, $\Delta v$, for maneuver predictions and periapsis radii at Venus flybys. The refined $B$-plane parameters, though different from initial approximations, provide good starting points for  $n$-body framework optimization.

To address the issue of finding a solution to the problem of encountering with asteroids while flying along $\pi$-resonant orbits, a multiple-shooting approach was investigated. The results of this method showed that under strict restrictions on the distance of asteroid flybys, the total impulse for maneuvers in the studied flight scheme, including flybys of 1997 XF11 and 2013 BP73, and return to Earth, was increased to 0.15 km/s.

The feasibility of a joint flyby of co-orbital asteroids with Venus and PHAs was also investigated to increase the scientific return. An example of the construction of a multiple flyby scheme for the 2001 CK32 asteroid shows that such a scheme requires a launch from Earth to Venus at a characteristic velocity about 3.581 km/s, with subsequent asteroid approaches done in low $\Delta v$. The total $\Delta v$ spent excluding the launch is no more than 100 m/s. 

The investigation into the transfer between Earth and Venus revealed that, in certain instances, impulse-free asteroid flybys are feasible. However, such scenarios are uncommon even for a predefined group of asteroids. For example, 25 feasible trajectories were identified from the set of 139 asteroids over a 20-year study period, compared to 56 options identified in two years of launches studied using the proposed schemes.

\begin{appendices}
\section{Characteristics of the Earth-Asteroid-Earth and Earth-Venus-Asteroid-Venus flight trajectories}
\label{sec:app}

Tables \ref{tab:evav} and \ref{tab:eae} contain a set of parameters describing the SC's flight trajectories to asteroids using schemes E-V-A-V (Table \ref{tab:evav}) and E-A-E (Table \ref{tab:eae}). Table \ref{tab:evave} shows data on optimal flight trajectories found using the E-V-A-V-E scheme.

\renewcommand{\thetable}{\thesection.\arabic{table}}
\setcounter{table}{0} %
Tables \ref{tab:evav}, \ref{tab:eae} contain some parameters characterizing the SC's flight trajectories to asteroids and subsequent return to the planet.

\scriptsize
\begin{longtable}{@{}l c c c c c c c c c c c@{}}
	\caption{Some parameters of the Earth-Venus-Asteroid-Venus flight trajectories} \\
	\toprule
	Name & $t_0$ & $t_1$ & $t_2$ & $t_3$  &$P_{sc}$  &$\Delta v_0$  &$\tilde{H}$  &$v_{\infty, 0}$ & $v_{\infty, 1}$  &$v_{\infty, 2}$ & $i_{eclip, 1}$ \\
	- & 
	&   & & & {( day)} & 
	{(km/s)} &
	{(km)} & 
	{(km/s)} &
	{(km/s)} &
	{(km/s)} &
	{(deg)} \\
	\midrule
	\endfirsthead
	
	\multicolumn{12}{c}{{\tablename\ \thetable{} -- continued}} \\
	\toprule
	Name & $t_0$ & $t_1$ & $t_2$ & $t_3$  &$P_{sc}$  &$\Delta v_0$  &$\tilde{H}$  &$v_{\infty, 0}$ & $v_{\infty, 1}$  &$v_{\infty, 2}$ & $i_{eclip, 1}$ \\
	- & 
	& & & & {( day)} & 
	{(km/s)} &
	{(km)} & 
	{(km/s)} &
	{(km/s)} &
	{(km/s)} &
	{(deg)} \\
	\midrule
	\endhead
	
	\bottomrule
	\endfoot
	(35396) 1997 XF11 & 29.11.2029 & 17.05.2030 & 12.08.2030 & 09.08.2031 & 449.4 & 3.881 & 3.959 & 3.859 & 6.82 & 5.026 & 3.723 \\ 
	(37638) 1993 VB & 02.10.2029 & 12.03.2030 & 11.09.2030 & 05.06.2031 & 449.4 & 4.029 & 0.501 & 4.161 & 6.11 & 16.328 & 3.907 \\ 
	(85236) 1993 KH & 29.10.2029 & 17.02.2030 & 28.10.2030 & 23.12.2031 & 337.06 & 3.886 & 18.528 & 3.874 & 4.072 & 16.057 & 3.838 \\ 
	(99942) Apophis 2004 MN4 & 30.09.2029 & 10.03.2030 & 09.02.2031 & 25.08.2032 & 449.43 & 4.059 & 4.571 & 4.357 & 6.185 & 10.252 & 2.545 \\
	(143487) 2003 CR20 & 01.10.2029 & 04.04.2030 & 08.05.2031 & 07.02.2032 & 337.06 & 3.759 & 2.164 & 3.471 & 5.195 & 14.236 & 1.232 \\ 
	(164207) Cardea 2004 GU9 & 27.10.2029 & 10.02.2030 & 22.10.2030 & 16.12.2031 & 337.06 & 3.821 & 2.344 & 3.673 & 4.691 & 12.63 & 4.052 \\
	(308635) 2005 YU55 & 17.12.2029 & 23.05.2030 & 29.12.2031 & 07.11.2032 & 449.4 & 4.039 & 4.51 & 4.311 & 7.242 & 2.315 & 3.704 \\ 
	(484506) 2008 ER7 & 06.09.2029 & 11.03.2030 & 24.04.2031 & 14.01.2032 & 337.05 & 3.909 & 42.626 & 3.941 & 4.603 & 26.404 & 2.524 \\ 
	(614433) 2009 KK & 20.10.2029 & 28.04.2030 & 30.12.2031 & 13.10.2032 & 449.4 & 3.947 & 3.008 & 4.052 & 6.527 & 9.313 & 1.668 \\ 
	2003 MK4 & 10.10.2029 & 21.03.2030 & 28.11.2030 & 24.01.2032 & 336.93 & 3.732 & 22.317 & 3.306 & 5.155 & 16.616 & 2.857 \\  
	2011 JA & 11.11.2029 & 14.04.2030 & 19.07.2031 & 18.02.2032 & 337.05 & 3.641 & 26.545 & 3.057 & 5.005 & 22.293 & 3.915 \\  
	2014 DM22 & 29.10.2029 & 28.04.2030 & 29.01.2031 & 21.07.2031 & 449.4 & 3.802 & 0.501 & 3.609 & 6.114 & 17.137 & 4.708 \\  
	2016 FG60 & 08.11.2029 & 31.03.2030 & 22.12.2030 & 03.02.2032 & 337.06 & 3.748 & 47.55 & 3.431 & 5.568 & 27.328 & 2.308 \\  
	2022 UJ65 & 30.10.2029 & 04.02.2030 & 12.11.2030 & 11.12.2031 & 337.05 & 3.837 & 26.791 & 3.722 & 5.409 & 10.958 & 3.578 \\  
	(35396) 1997 XF11 & 29.11.2029 & 17.05.2030 & 12.08.2030 & 09.08.2031 & 449.4 & 3.881 & 3.955 & 3.859 & 6.813 & 5.027 & 3.723 \\ 
	(37638) 1993 VB & 03.10.2029 & 12.03.2030 & 11.09.2030 & 05.06.2031 & 449.4 & 4.029 & 0.501 & 4.155 & 6.122 & 16.331 & 3.902 \\ 
	(85236) 1993 KH & 30.10.2029 & 17.02.2030 & 27.10.2030 & 23.12.2031 & 337.06 & 3.883 & 18.999 & 3.865 & 4.055 & 16.058 & 3.808 \\ 
	(101869) 1999 MM & 27.10.2029 & 05.04.2030 & 15.11.2030 & 08.02.2032 & 337.06 & 3.578 & 1.224 & 2.811 & 4.84 & 21.854 & 1.289 \\ 
	(141495) 2002 EZ11 & 23.10.2029 & 18.01.2030 & 19.02.2031 & 23.11.2031 & 337.05 & 4.038 & 17.133 & 4.311 & 8.844 & 41.59 & 2.66 \\ 
	(143487)  2003 CR20 & 28.10.2029 & 11.04.2030 & 06.05.2031 & 14.02.2032 & 337.05 & 3.601 & 5.37 & 2.902 & 5.013 & 14.442 & 1.255 \\ 
	(144898) 2004 VD17 & 19.10.2029 & 15.02.2030 & 15.10.2030 & 21.12.2031 & 337.05 & 3.924 & 27.028 & 3.986 & 4.598 & 23.293 & 3.524 \\ 
	(152685) 1998 MZ & 27.10.2029 & 11.02.2030 & 27.03.2031 & 18.12.2031 & 337.54 & 3.857 & 0.501 & 3.697 & 4.559 & 10.698 & 3.416 \\ 
	(308635) 2005 YU55 & 06.12.2029 & 23.05.2030 & 29.12.2031 & 07.11.2032 & 449.4 & 3.945 & 4.19 & 4.045 & 7.233 & 2.317 & 3.705 \\ 
	(468468) 2004 KH17 & 25.10.2029 & 23.01.2030 & 12.04.2030 & 18.04.2031 & 449.4 & 3.91 & 3.408 & 3.944 & 7.583 & 20.453 & 4.214 \\ 
	(471240) 2011 BT15 & 27.10.2029 & 06.04.2030 & 27.05.2031 & 09.02.2032 & 337.05 & 3.579 & 42.03 & 2.816 & 4.851 & 14.416 & 1.527 \\ 
	(484506) 2008 ER7 & 13.10.2029 & 12.02.2030 & 22.04.2031 & 18.12.2031 & 337.05 & 3.942 & 5.127 & 4.039 & 4.973 & 26.216 & 4.38 \\ 
	(614433) 2009 KK & 19.10.2029 & 27.04.2030 & 30.12.2031 & 12.10.2032 & 449.4 & 3.941 & 2.423 & 4.034 & 6.466 & 9.374 & 1.554 \\ 
	2016 CB194 & 14.11.2029 & 02.05.2030 & 06.08.2031 & 06.03.2032 & 337.05 & 3.754 & 5.781 & 3.454 & 6.022 & 13.248 & 4.124 \\  
	(741081) 2005 LW3 & 28.10.2029 & 14.02.2030 & 04.04.2031 & 20.12.2031 & 337.06 & 3.854 & 17.614 & 3.774 & 4.281 & 19.705 & 4.374 \\ 
	2000 KA & 19.10.2029 & 12.02.2030 & 18.10.2030 & 18.12.2031 & 337.05 & 3.861 & 26.911 & 3.798 & 4.7 & 7.995 & 3.408 \\  
	2002 XO14 & 22.10.2029 & 01.04.2030 & 16.11.2030 & 08.01.2032 & 334.7 & 3.634 & 20.474 & 2.835 & 4.835 & 10.066 & 3.4 \\  
	2010 XC25 & 12.12.2029 & 30.05.2030 & 02.09.2031 & 03.04.2032 & 337.22 & 4.008 & 21.787 & 4.184 & 7.648 & 16.918 & 3.353 \\  
	2011 JA & 04.11.2029 & 15.04.2030 & 19.07.2031 & 18.02.2032 & 337.05 & 3.622 & 26.409 & 2.984 & 5.097 & 22.268 & 3.801 \\  
	2012 TO139 & 08.11.2029 & 29.03.2030 & 10.10.2031 & 13.09.2032 & 449.41 & 3.915 & 3.136 & 3.957 & 6.33 & 19.681 & 1.073 \\  
	2013 ED28 & 20.10.2029 & 18.01.2030 & 28.04.2030 & 31.08.2030 & 224.7 & 4.003 & 16.59 & 4.212 & 8.554 & 29.02 & 4.822 \\  
	2014 DM22 & 29.10.2029 & 28.04.2030 & 29.01.2031 & 21.07.2031 & 449.4 & 3.801 & 0.506 & 3.608 & 6.115 & 17.137 & 4.708 \\  
	2014 SP142 & 16.12.2029 & 01.06.2030 & 18.08.2031 & 06.04.2032 & 337.05 & 4.012 & 10.788 & 4.235 & 7.812 & 15.349 & 5.702 \\  
	2014 XL7 & 23.11.2029 & 05.05.2030 & 24.11.2031 & 19.10.2032 & 449.4 & 3.796 & 0.502 & 3.555 & 6.026 & 6.839 & 3.802 \\  
	2015 YY9 & 13.11.2029 & 28.04.2030 & 03.02.2031 & 22.07.2031 & 224.7 & 3.717 & 8.061 & 3.328 & 5.744 & 11.469 & 8.094 \\  
	2016 FG60 & 21.10.2029 & 25.03.2030 & 22.12.2030 & 28.01.2032 & 337.06 & 3.663 & 43.372 & 3.133 & 5.211 & 27.348 & 2.397 \\  
	2016 XA2 & 20.11.2029 & 06.05.2030 & 16.08.2031 & 10.03.2032 & 337.05 & 3.779 & 27.532 & 3.536 & 6.151 & 24.381 & 2.237 \\  
	2019 VF6 & 21.10.2029 & 05.02.2030 & 25.03.2031 & 11.12.2031 & 337.06 & 3.797 & 5.048 & 3.596 & 5.324 & 7.508 & 5.814 \\  
	2022 UJ65 & 21.10.2029 & 05.02.2030 & 12.11.2030 & 11.12.2031 & 337.05 & 3.797 & 22.529 & 3.596 & 5.318 & 10.97 & 3.575 \\  
	(85640) 1998 OX4 & 12.07.2031 & 30.12.2031 & 11.05.2032 & 24.03.2033 & 449.4 & 3.936 & 3.993 & 4.019 & 8.18 & 18.865 & 4.032 \\ 
	(137108) 1999 AN10 & 21.05.2031 & 17.10.2031 & 28.10.2032 & 03.04.2034 & 449.41 & 4.002 & 1.656 & 4.21 & 7.105 & 34.755 & 0.829 \\ 
	(141495) 2002 EZ11 & 29.05.2031 & 21.10.2031 & 05.03.2032 & 12.01.2033 & 449.4 & 3.823 & 4.667 & 3.678 & 6.253 & 25.315 & 0.696 \\ 
	(163243) 2002 FB3 & 23.05.2031 & 18.10.2031 & 04.01.2033 & 22.08.2033 & 337.06 & 4.062 & 24.614 & 4.375 & 7.341 & 16.038 & 5.357 \\ 
	(216985) 2000 QK130 & 24.05.2031 & 30.11.2031 & 01.11.2032 & 17.05.2034 & 449.41 & 3.913 & 1.267 & 3.811 & 6.448 & 15.562 & 3.418 \\ 
	(332446) 2008 AF4 & 10.06.2031 & 07.12.2031 & 16.07.2033 & 23.05.2034 & 449.4 & 3.813 & 1.165 & 3.648 & 6.525 & 16.574 & 5.587 \\ 
	(371660) 2007 CN26 & 07.06.2031 & 03.12.2031 & 23.11.2032 & 20.05.2034 & 449.41 & 3.782 & 0.507 & 3.546 & 6.23 & 17.814 & 5.176 \\ 
	(405212) 2003 QC10 & 31.07.2031 & 14.01.2032 & 27.06.2032 & 08.04.2033 & 449.4 & 4.014 & 3.989 & 4.241 & 9.234 & 19.456 & 2.562 \\ 
	(438908) 2009 XO & 17.05.2031 & 06.11.2031 & 06.01.2033 & 10.09.2033 & 337.06 & 3.582 & 8.186 & 2.827 & 4.341 & 7.745 & 1.432 \\ 
	(523662) 2012 MU2 & 24.05.2031 & 13.11.2031 & 21.01.2033 & 17.09.2033 & 337.06 & 3.631 & 5.711 & 3.018 & 4.726 & 15.712 & 2.132 \\ 
	(530520) 2011 LT17 & 28.07.2031 & 11.01.2032 & 13.05.2033 & 14.05.2034 & 443.84 & 4.041 & 3.954 & 4.2 & 9.033 & 22.651 & 3.391 \\ 
	(614433) 2009 KK & 02.06.2031 & 02.12.2031 & 07.05.2033 & 19.05.2034 & 449.41 & 3.807 & 1.858 & 3.626 & 6.284 & 24.285 & 4.955 \\ 
	2006 AR3 & 22.05.2031 & 05.09.2031 & 20.02.2033 & 20.02.2034 & 449.4 & 3.814 & 1.046 & 3.65 & 6.333 & 6.661 & 4.845 \\  
	2007 JY2 & 04.06.2031 & 23.10.2031 & 05.06.2033 & 09.04.2034 & 449.41 & 3.91 & 3.309 & 3.943 & 6.506 & 25.155 & 0.11 \\  
	2007 RU9 & 04.06.2031 & 22.09.2031 & 16.11.2031 & 03.05.2032 & 224.7 & 3.808 & 0.501 & 3.63 & 3.938 & 11.226 & 5.091 \\  
	2009 XT6 & 24.05.2031 & 12.11.2031 & 23.05.2032 & 03.02.2033 & 224.7 & 3.624 & 16.37 & 2.993 & 4.656 & 20.903 & 2.986 \\  
	2012 TY52 & 28.04.2031 & 03.09.2031 & 24.07.2032 & 17.02.2034 & 449.4 & 3.972 & 0.501 & 4.124 & 6.381 & 23.269 & 5.053 \\  
	2012 UR158 & 01.08.2031 & 15.01.2032 & 24.02.2033 & 19.11.2033 & 337.05 & 4.019 & 17.554 & 4.255 & 9.295 & 21.038 & 2.443 \\  
	2015 YY9 & 27.07.2031 & 11.01.2032 & 22.06.2033 & 28.06.2034 & 449.4 & 3.996 & 1.694 & 4.192 & 8.989 & 25.532 & 5.449 \\  
	2016 FG60 & 28.05.2031 & 04.12.2031 & 09.04.2032 & 26.02.2033 & 449.4 & 3.908 & 1.893 & 3.925 & 6.702 & 22.899 & 3.913 \\  
	2021 MK1 & 04.08.2031 & 18.01.2032 & 15.02.2033 & 23.11.2033 & 337.05 & 4.04 & 10.058 & 4.313 & 9.513 & 11.488 & 3.809 \\  
	2022 CQ3 & 15.05.2031 & 14.11.2031 & 04.07.2032 & 19.09.2033 & 337.06 & 3.687 & 3.322 & 3.222 & 5.113 & 13.583 & 2.934 \\  
	2022 CY & 20.04.2031 & 15.11.2031 & 19.08.2032 & 06.02.2033 & 224.62 & 4.078 & 0.501 & 4.374 & 6.251 & 18.958 & 7.502 \\  
	(809875) 2020 BX12 & 26.04.2031 & 16.11.2031 & 29.08.2032 & 07.02.2033 & 449.4 & 3.966 & 1.304 & 4.109 & 6.044 & 22.483 & 4.641 \\ 
	2007 RU9 & 06.05.2031 & 15.10.2031 & 16.11.2031 & 26.05.2032 & 224.69 & 3.614 & 24.624 & 2.952 & 5.03 & 11.842 & 4.943 \\  
	(85640) 1998 OX4 & 21.07.2031 & 30.12.2031 & 11.05.2032 & 24.03.2033 & 449.39 & 3.98 & 4.39 & 4.147 & 8.188 & 18.867 & 4.026 \\ 
	(216985) 2000 QK130 & 19.05.2031 & 28.11.2031 & 01.11.2032 & 15.05.2034 & 449.41 & 3.893 & 0.624 & 3.894 & 6.361 & 15.557 & 3.197 \\ 
	(438908) 2009 XO & 22.05.2031 & 07.11.2031 & 06.01.2033 & 11.09.2033 & 337.05 & 3.585 & 8.891 & 2.838 & 4.301 & 7.718 & 1.491 \\ 
	(523662) 2012 MU2 & 29.05.2031 & 13.11.2031 & 21.01.2033 & 17.09.2033 & 337.06 & 3.636 & 5.832 & 3.036 & 4.692 & 15.721 & 2.143 \\ 
	2006 AR3 & 25.05.2031 & 05.09.2031 & 20.02.2033 & 20.02.2034 & 449.4 & 3.816 & 0.825 & 3.658 & 6.321 & 6.653 & 4.864 \\  
	2016 FG60 & 30.05.2031 & 05.12.2031 & 09.04.2032 & 27.02.2033 & 449.4 & 3.918 & 2.256 & 3.919 & 6.767 & 22.928 & 3.945 \\  
	2022 SF11 & 29.07.2031 & 13.01.2032 & 11.09.2032 & 06.04.2033 & 449.4 & 4.006 & 3.805 & 4.218 & 9.126 & 5.065 & 3.109 \\  
	
	\bottomrule
	\label{tab:evav}
	
\end{longtable}

\small $t_i, i = 0,1,2,3$ correspond to the characteristic dates of launch, transit of Venus and asteroid, return to Venus; $P_{sc}$ is the period of the SC's orbit; $v_{\infty, i}, i = $0,1,2 correspond to the asymptotic velocities at the time of launch, transit of Venus and the asteroid; $i_{eclip, 1}$ is the inclination of the SC's orbit to the ecliptic after a Venus gravity assist. All dates are given in the UTC format (DD.MM.YYYY).

\normalsize

\scriptsize
\begin{longtable}{l c c c c c c c c}
	\caption{Some parameters of the Earth-Asteroid-Earth flight trajectories} \\
	\toprule
	Name & $t_0$ & $t_1$ & $t_2$  &$P_{sc}$  & $\Delta v_0$   & $v_{\infty, 0}$ & $v_{\infty, 1}$ & $i_{eclip, 0}$ \\
	- & 
	&  & & {( day)} & 
	{(km/s)} &
	{(km/s)} &
	{(km/s)} &
	{(deg)} \\
	\midrule
	\endfirsthead
	
	\multicolumn{9}{c}{{\tablename\ \thetable{} -- continued}} \\
	\toprule
	Name & $t_0$ & $t_1$ & $t_2$  &$P_{sc}$  & $\Delta v_0$   & $v_{\infty, 0}$ & $v_{\infty, 1}$ & $i_{eclip, 0}$ \\
	- & 
	&   & & {( day)} & 
	{(km/s)} &
	{(km/s)} &
	{(km/s)} &
	{(deg)} \\
	\midrule
	\endhead
	
	\bottomrule
	\endfoot
	(35396) 1997 XF11 & 11.12.2029 & 31.10.2031 & 11.12.2032 & 547.89 & 3.722 & 3.347 & 18.243 & 1.111 \\ 
	(37638) 1993 VB & 31.08.2029 & 12.12.2030 & 01.09.2031 & 365.24 & 3.547 & 2.68 & 7.782 & 4.587 \\ 
	(90416) 2003 YK118 & 26.11.2029 & 27.12.2030 & 26.11.2032 & 547.88 & 3.908 & 3.937 & 17.856 & 3.249 \\ 
	(99942) Apophis 2004 MN4 & 14.01.2029 & 13.04.2029 & 12.01.2030 & 365.28 & 3.225 & 0.013 & 7.36 & 0.005 \\
	(101869) 1999 MM & 17.12.2029 & 27.07.2030 & 17.12.2031 & 365.25 & 3.342 & 1.611 & 16.701 & 0.358 \\ 
	(141495) 2002 EZ11 & 15.10.2029 & 01.03.2030 & 15.10.2030 & 365.24 & 3.468 & 2.325 & 27.218 & 2.95 \\ 
	(152685) 1998 MZ & 13.07.2029 & 06.12.2030 & 13.07.2031 & 365.15 & 3.485 & 2.324 & 15.632 & 0.171 \\ 
	(164207) Cardea 2004 GU9 & 03.06.2029 & 22.10.2029 & 04.06.2030 & 365.25 & 3.32 & 1.453 & 8.184 & 0.36 \\
	(177049) 2003 EE16 & 29.09.2029 & 29.10.2031 & 29.09.2032 & 547.88 & 3.734 & 3.388 & 11.286 & 0.897 \\ 
	(269690) 1996 RG3 & 20.07.2029 & 29.10.2030 & 20.07.2031 & 365.25 & 3.345 & 1.63 & 14.48 & 2.951 \\ 
	(279744) 1998 KM3 & 12.09.2029 & 13.10.2030 & 12.09.2032 & 547.88 & 3.845 & 3.747 & 11.285 & 2.789 \\ 
	(367789) 2011 AG5 & 14.10.2029 & 09.07.2031 & 14.10.2032 & 547.85 & 3.848 & 3.753 & 16.663 & 1.614 \\ 
	(373135) 2011 SD173 & 13.10.2029 & 30.04.2030 & 13.10.2030 & 365.26 & 3.434 & 2.157 & 10.675 & 0.12 \\ 
	(380636) 2004 XN14 & 30.08.2029 & 25.01.2031 & 30.08.2031 & 365.45 & 3.629 & 2.875 & 6.725 & 3.176 \\ 
	(415745) 2000 GV147 & 20.10.2029 & 20.05.2030 & 20.10.2030 & 365.25 & 3.275 & 1.052 & 11.791 & 1.362 \\ 
	(434734) 2006 FX & 07.05.2029 & 06.05.2031 & 07.05.2032 & 547.87 & 3.748 & 3.434 & 17.324 & 1.274 \\ 
	(444193) 2005 SE71 & 01.01.2029 & 01.06.2029 & 01.01.2030 & 365.25 & 4.065 & 4.381 & 14.743 & 3.827 \\ 
	(471240) 2011 BT15 & 09.02.2029 & 03.02.2031 & 10.02.2032 & 547.88 & 3.709 & 3.299 & 9.922 & 0.373 \\ 
	(484506) 2008 ER7 & 15.02.2029 & 05.07.2029 & 16.02.2030 & 365.27 & 3.909 & 3.941 & 22.819 & 1.759 \\ 
	(523654) 2011 SR5 & 08.03.2029 & 22.08.2029 & 08.03.2030 & 365.26 & 3.949 & 4.059 & 29.64 & 0.167 \\ 
	(549948) 2011 WL2 & 08.06.2029 & 03.11.2030 & 09.06.2031 & 365.02 & 3.302 & 0.883 & 12.129 & 0.837 \\ 
	2016 WJ1 & 10.07.2029 & 03.12.2030 & 10.07.2031 & 365.14 & 3.417 & 1.962 & 16.607 & 0.48 \\  
	2021 NQ5 & 02.08.2029 & 07.02.2030 & 02.08.2030 & 365.25 & 3.577 & 2.807 & 14.851 & 0.144 \\  
	(741081) 2005 LW3 & 10.02.2029 & 30.06.2029 & 10.02.2030 & 365.27 & 3.594 & 2.872 & 12.24 & 3.016 \\ 
	(826936) 1998 HH49 & 31.01.2029 & 18.05.2029 & 31.01.2030 & 365.25 & 3.853 & 3.773 & 14.635 & 6.364 \\ 
	2003 MK4 & 08.11.2029 & 18.06.2030 & 08.11.2030 & 365.26 & 3.723 & 3.349 & 12.326 & 1.919 \\  
	2008 DJ & 01.01.2029 & 29.01.2030 & 30.12.2031 & 547.81 & 3.781 & 3.437 & 10.416 & 1.686 \\  
	2010 KR10 & 12.05.2029 & 26.10.2029 & 12.05.2030 & 365.27 & 3.513 & 2.533 & 14.652 & 1.77 \\  
	2013 ED28 & 19.07.2029 & 23.08.2030 & 19.07.2032 & 547.88 & 3.809 & 3.632 & 22.671 & 2.357 \\  
	2013 ND24 & 15.02.2029 & 21.01.2031 & 16.02.2032 & 547.89 & 3.743 & 3.415 & 25.223 & 1.629 \\  
	2014 SP142 & 16.01.2029 & 15.07.2029 & 16.01.2030 & 365.26 & 3.344 & 1.627 & 11.608 & 0.146 \\  
	2014 XL7 & 01.01.2029 & 29.05.2029 & 01.01.2030 & 365.25 & 3.289 & 1.192 & 15.749 & 0.548 \\  
	2018 BT6 & 28.10.2029 & 26.08.2031 & 28.10.2032 & 547.88 & 3.728 & 3.367 & 18.932 & 0.931 \\  
	2018 LK & 07.08.2029 & 07.12.2030 & 07.08.2031 & 365.24 & 3.44 & 2.178 & 11.688 & 1.189 \\  
	2019 BC1 & 18.09.2029 & 12.02.2031 & 18.09.2031 & 365.47 & 3.737 & 3.261 & 17.767 & 4.872 \\  
	2019 LZ1 & 01.01.2029 & 31.05.2029 & 01.01.2030 & 365.25 & 3.972 & 4.124 & 35.6 & 0.679 \\  
	2020 HJ4 & 06.06.2029 & 04.11.2029 & 06.06.2030 & 365.26 & 3.373 & 1.811 & 14.221 & 1.681 \\  
	2021 TA8 & 26.12.2029 & 12.05.2031 & 27.12.2031 & 365.27 & 3.307 & 1.344 & 14.975 & 2.06 \\  
	2021 XL6 & 27.10.2029 & 23.11.2030 & 25.03.2031 & 366.21 & 3.227 & 0.183 & 14.016 & 0.004 \\  
	2022 BX1 & 01.01.2029 & 31.01.2030 & 01.01.2032 & 547.94 & 3.983 & 4.144 & 15.37 & 2.678 \\  
	2022 CQ3 & 14.08.2029 & 07.02.2030 & 14.08.2030 & 365.25 & 3.298 & 1.273 & 16.788 & 0.208 \\  
	2022 GY2 & 04.12.2029 & 04.12.2031 & 04.12.2032 & 547.88 & 3.814 & 3.648 & 20.734 & 2.687 \\  
	(35396) 1997 XF11 & 18.03.2031 & 15.05.2032 & 18.03.2034 & 547.9 & 3.791 & 3.576 & 19.082 & 2.417 \\ 
	(89958) 2002 LY45 & 21.09.2031 & 24.03.2032 & 20.09.2032 & 365.26 & 3.238 & 0.541 & 31.949 & 0.165 \\ 
	(90416) 2003 YK118 & 12.03.2031 & 25.03.2033 & 12.03.2034 & 547.9 & 3.872 & 3.827 & 9.08 & 3.162 \\ 
	(141495) 2002 EZ11 & 09.11.2031 & 09.04.2032 & 08.11.2032 & 365.26 & 4.026 & 4.274 & 32.122 & 0.086 \\ 
	(144898) 2004 VD17 & 25.09.2031 & 01.05.2032 & 24.09.2032 & 365.26 & 3.226 & 0.184 & 18.367 & 0.157 \\ 
	(177049) 2003 EE16 & 04.03.2031 & 24.12.2032 & 03.03.2034 & 547.9 & 3.718 & 3.332 & 23.094 & 0.459 \\ 
	(231937) 2001 FO32 & 03.10.2031 & 17.04.2032 & 02.10.2032 & 365.25 & 3.721 & 3.344 & 31.246 & 0.21 \\ 
	(267337) 2001 VK5 & 12.05.2031 & 01.11.2031 & 12.05.2032 & 365.22 & 3.402 & 1.969 & 17.392 & 0.041 \\ 
	(279744) 1998 KM3 & 05.01.2031 & 09.12.2032 & 05.01.2034 & 547.89 & 3.839 & 3.726 & 21.062 & 3.034 \\ 
	(297300) 1998 SC15 & 29.09.2031 & 20.03.2032 & 29.09.2032 & 365.25 & 3.492 & 2.439 & 12.307 & 0.126 \\ 
	(308635) 2005 YU55 & 26.04.2031 & 18.10.2031 & 25.04.2032 & 365.25 & 3.541 & 2.66 & 11.019 & 0.403 \\ 
	(371660) 2007 CN26 & 16.11.2031 & 05.10.2033 & 15.11.2034 & 547.87 & 3.776 & 3.526 & 12.603 & 2.261 \\ 
	(380636) 2004 XN14 & 17.07.2031 & 02.01.2032 & 16.07.2032 & 365.27 & 3.489 & 2.426 & 11.801 & 0.03 \\ 
	(405212) 2003 QC10 & 21.02.2031 & 19.12.2032 & 21.02.2034 & 547.89 & 4.047 & 4.332 & 14.567 & 5.064 \\ 
	(436724) 2011 UW158 & 26.12.2031 & 04.11.2033 & 26.12.2034 & 547.88 & 4.036 & 4.302 & 15.608 & 4.855 \\ 
	(468468) 2004 KH17 & 01.01.2031 & 21.06.2031 & 01.01.2032 & 365.26 & 3.719 & 3.333 & 11.139 & 4.848 \\ 
	(487577) 2014 YQ15 & 06.06.2031 & 19.12.2031 & 05.06.2032 & 365.25 & 3.247 & 0.694 & 9.794 & 0.02 \\ 
	(509456) 2007 LF & 13.01.2031 & 06.06.2031 & 14.01.2032 & 365.26 & 3.296 & 1.254 & 9.448 & 1.683 \\ 
	2019 LZ1 & 18.05.2031 & 17.12.2031 & 17.05.2032 & 365.26 & 3.49 & 2.431 & 17.408 & 0.755 \\  
	2016 WJ1 & 20.02.2031 & 28.05.2032 & 20.02.2034 & 548.36 & 3.764 & 3.442 & 8.531 & 0.889 \\  
	(741081) 2005 LW3 & 09.03.2031 & 01.12.2032 & 09.03.2034 & 546.87 & 3.898 & 3.797 & 19.585 & 1.592 \\ 
	2000 TU28 & 19.10.2031 & 28.04.2032 & 19.10.2032 & 365.26 & 3.409 & 2.02 & 10.612 & 0.016 \\  
	2003 MK4 & 19.02.2031 & 06.01.2032 & 22.08.2032 & 366.86 & 3.279 & 0.243 & 13.036 & 0.402 \\  
	2003 RS1 & 28.02.2031 & 02.09.2031 & 29.02.2032 & 365.25 & 3.323 & 1.478 & 8.872 & 0.457 \\  
	2005 WY55 & 26.09.2031 & 20.06.2033 & 26.09.2034 & 547.88 & 3.801 & 3.607 & 25.803 & 2.375 \\  
	2007 RU9 & 06.02.2031 & 14.09.2031 & 06.02.2032 & 365.25 & 3.658 & 3.119 & 18.847 & 2.99 \\  
	2009 XT6 & 22.03.2031 & 27.06.2032 & 22.03.2034 & 547.9 & 3.784 & 3.552 & 29.988 & 1.055 \\  
	2011 DV & 21.04.2031 & 29.10.2032 & 15.05.2034 & 552.81 & 3.899 & 3.632 & 6.975 & 0.006 \\  
	2012 TO139 & 14.02.2031 & 23.09.2031 & 14.02.2032 & 365.25 & 3.227 & 0.244 & 32.763 & 0.057 \\  
	2012 UR158 & 10.07.2031 & 23.11.2032 & 09.07.2033 & 365.26 & 3.647 & 3.079 & 27.35 & 3.153 \\  
	2013 JL22 & 19.09.2031 & 07.11.2032 & 19.09.2034 & 547.89 & 3.752 & 3.448 & 19.658 & 1.522 \\  
	2014 EG45 & 17.01.2031 & 17.02.2032 & 16.01.2034 & 547.88 & 3.934 & 4.016 & 11.693 & 4.13 \\  
	2014 SP142 & 10.08.2031 & 23.12.2032 & 10.08.2033 & 365.25 & 4.02 & 4.258 & 17.007 & 0.729 \\  
	2014 XL7 & 26.02.2031 & 23.10.2031 & 26.02.2032 & 365.26 & 3.699 & 3.265 & 14.455 & 4.259 \\  
	2018 LK & 26.05.2031 & 12.01.2032 & 26.05.2033 & 365.26 & 3.765 & 3.49 & 14.666 & 0.803 \\  
	2018 LB1 & 26.05.2031 & 05.09.2032 & 26.05.2033 & 365.25 & 3.246 & 0.681 & 8.656 & 0.63 \\  
	2019 LZ1 & 10.09.2031 & 20.10.2032 & 10.09.2034 & 547.89 & 3.789 & 3.57 & 34.359 & 2.233 \\  
	
	\bottomrule
	\label{tab:eae}
	
\end{longtable}

\small Note: $t_i, i = 0,1,2$ correspond to the characteristic dates of launch, passage of the asteroid and return to Earth; $P_{sc}$ is the period of the SC's orbit; $v_{\infty, i}, i = 0,1$ correspond to the asymptotic velocities at the moment of launch and passage of the asteroid; $i_{eclip, 0}$ is the inclination of the SC's orbit to the ecliptic. All dates are given in the UTC format (DD.MM.YYYY).	

\scriptsize
\begin{longtable}{@{}c c c c c c c c c c c c c@{}}	
	\caption{Parameters of the Earth-Venus-Asteroid-Venus-Earth flight paths} \\
	\toprule
	Name & $t_0$ & $t_1$ & $t_2$ & $t_3$ & $t_4$ & $\Delta v_0$ & $\tilde{H}$ & $v_{\infty, 0}$ & $v_{\infty, 1}$ & $v_{\infty, 2}$ \\
	- &  & &  & &  & (km/s) & (km) & (km/s) & (km/s) & (km/s) \\
	\midrule
	\endfirsthead
	
	\multicolumn{12}{c}{{\tablename\ \thetable{} -- continued}} \\
	\toprule
	Name & $t_0$ & $t_1$ & $t_2$ & $t_3$ & $t_4$ & $\Delta v_0$ & $\tilde{H}$ & $v_{\infty, 0}$ & $v_{\infty, 1}$ & $v_{\infty, 2}$ \\
	- &  & &  & &  & (km/s) & (km) & (km/s) & (km/s) & (km/s) \\
	\midrule
	\endhead
	
	\bottomrule
	
	(35396) 1997 XF11 & 25.11.2029 & 08.05.2030 & 14.08.2030 & 31.07.2031 & 27.04.2032 & 3.902 & 0.503 & 3.644 & 6.062 & 5.274 \\ 
	(143487) 2003 CR20 & 20.10.2029 & 21.03.2030 & 12.05.2031 & 24.01.2032 & 31.01.2033 & 3.857 & 2.511 & 3.781 & 6.075 & 14.004 \\ 
	2003 MK4 & 15.11.2029 & 01.04.2030 & 30.11.2030 & 04.02.2032 & 18.02.2033 & 3.942 & 33.723 & 4.033 & 6.272 & 15.626 \\ 
	2014 DM22 & 29.10.2029 & 28.04.2030 & 29.01.2031 & 21.07.2031 & 26.11.2031 & 3.801 & 0.529 & 3.608 & 6.116 & 17.137 \\ 
	2022 UJ65 & 17.10.2029 & 05.02.2030 & 12.11.2030 & 11.12.2031 & 20.11.2032 & 3.806 & 20.203 & 3.626 & 5.394 & 10.959 \\ 
	(99942) Apophis 2004 MN4 & 06.10.2029 & 13.03.2030 & 10.02.2031 & 06.06.2031 & 27.01.2032 & 4.017 & 5.131 & 4.25 & 6.264 & 10.209 \\
	(37638) 1993 VB & 03.10.2029 & 13.03.2030 & 11.09.2030 & 05.06.2031 & 26.08.2031 & 4.029 & 0.501 & 4.154 & 6.126 & 16.332 \\ 
	(143487) 2003 CR20 & 20.10.2029 & 21.03.2030 & 12.05.2031 & 24.01.2032 & 29.01.2033 & 3.857 & 2.507 & 3.784 & 6.077 & 14.004 \\ 
	(484506) 2008 ER7 & 20.10.2029 & 21.03.2030 & 24.04.2031 & 24.01.2032 & 30.01.2033 & 3.857 & 6.406 & 3.785 & 6.071 & 26.96 \\ 
	2003 MK4 & 23.09.2029 & 08.03.2030 & 28.11.2030 & 12.01.2032 & 10.01.2033 & 4.02 & 7.038 & 4.254 & 5.795 & 17.461 \\ 
	2014 DM22 & 29.10.2029 & 28.04.2030 & 29.01.2031 & 21.07.2031 & 10.04.2032 & 3.801 & 0.512 & 3.608 & 6.115 & 17.137 \\ 
	(37638) 1993 VB & 03.10.2029 & 13.03.2030 & 11.09.2030 & 05.06.2031 & 26.08.2031 & 4.029 & 0.501 & 4.153 & 6.127 & 16.333 \\ 
	(308635) 2005 YU55 & 05.12.2029 & 25.05.2030 & 25.12.2031 & 11.11.2032 & 31.05.2034 & 4.028 & 4.132 & 4.083 & 7.354 & 2.134 \\ 
	(484506) 2008 ER7 & 20.10.2029 & 21.03.2030 & 24.04.2031 & 24.01.2032 & 30.01.2033 & 3.857 & 6.406 & 3.784 & 6.071 & 26.96 \\ 
	2016 CB194 & 26.11.2029 & 12.05.2030 & 07.08.2031 & 16.03.2032 & 22.04.2033 & 3.836 & 9.319 & 3.721 & 6.501 & 12.961 \\ 
	2000 KA & 07.10.2029 & 07.02.2030 & 17.10.2030 & 13.12.2031 & 24.11.2032 & 3.928 & 17.461 & 3.999 & 5.399 & 7.865 \\ 
	2010 XC25 & 19.12.2029 & 30.05.2030 & 02.09.2031 & 03.04.2032 & 02.05.2033 & 4.031 & 21.789 & 4.267 & 7.604 & 16.876 \\ 
	2014 DM22 & 29.10.2029 & 28.04.2030 & 29.01.2031 & 21.07.2031 & 26.11.2031 & 3.801 & 0.508 & 3.608 & 6.115 & 17.137 \\ 
	2015 YY9 & 13.11.2029 & 28.04.2030 & 03.02.2031 & 22.07.2031 & 06.12.2031 & 3.717 & 7.981 & 3.328 & 5.751 & 11.481 \\ 
	2019 VF6 & 21.10.2029 & 04.02.2030 & 25.03.2031 & 10.12.2031 & 19.11.2032 & 3.813 & 5.004 & 3.595 & 5.391 & 7.463 \\ 
	(141495) 2002 EZ11 & 30.05.2031 & 21.10.2031 & 06.03.2032 & 12.01.2033 & 03.04.2033 & 3.824 & 4.8 & 3.681 & 6.232 & 25.326 \\ 
	(163243) 2002 FB3 & 25.05.2031 & 19.10.2031 & 04.01.2033 & 23.08.2033 & 27.07.2034 & 4.065 & 23.372 & 4.379 & 7.343 & 16.077 \\ 
	(216985) 2000 QK130 & 20.05.2031 & 29.11.2031 & 01.11.2032 & 20.02.2033 & 05.05.2033 & 3.892 & 0.782 & 3.892 & 6.381 & 15.558 \\ 
	(371660) 2007 CN26 & 08.06.2031 & 03.12.2031 & 23.11.2032 & 25.02.2033 & 16.06.2033 & 3.783 & 0.659 & 3.546 & 6.242 & 17.812 \\ 
	(530520) 2011 LT17 & 27.07.2031 & 12.01.2032 & 13.05.2033 & 13.05.2034 & 23.10.2034 & 4.045 & 3.939 & 4.203 & 9.043 & 22.672 \\ 
	2007 RU9 & 04.05.2031 & 10.10.2031 & 22.12.2031 & 01.01.2033 & 24.08.2033 & 4.056 & 0.921 & 4.36 & 7.127 & 6.532 \\ 
	2009 XT6 & 30.05.2031 & 11.11.2031 & 24.05.2032 & 02.02.2033 & 09.09.2033 & 3.666 & 14.635 & 3.008 & 4.47 & 21.145 \\ 
	2012 TY52 & 28.04.2031 & 03.09.2031 & 24.07.2032 & 24.11.2032 & 03.05.2033 & 4.014 & 0.506 & 4.135 & 6.382 & 23.268 \\ 
	2022 CY & 22.04.2031 & 16.11.2031 & 19.08.2032 & 07.02.2033 & 05.08.2033 & 4.08 & 0.501 & 4.339 & 6.263 & 18.97 \\ 
	2016 FG60 & 28.05.2031 & 04.12.2031 & 09.04.2032 & 26.02.2033 & 06.06.2033 & 3.907 & 1.78 & 3.927 & 6.684 & 22.89 \\ 
	(809875) 2020 BX12 & 27.04.2031 & 16.11.2031 & 29.08.2032 & 07.02.2033 & 22.10.2033 & 3.967 & 1.342 & 4.107 & 6.049 & 22.478 \\ 
	
	\bottomrule
	\label{tab:evave}
	
\end{longtable}

\small Note: $t_i, i = 0,1,2,3$ correspond to the characteristic dates of launch, passage of Venus and asteroid, and Venus, and return to Earth; $v_{\infty, i}, i = $0,1,2 correspond to the asymptotic velocities at the time of launch, transit of Venus and the asteroid; $i_{eclip, 1}$ is the inclination of the SC's orbit to the ecliptic after a Venus gravity assist. All dates are given in the UTC format (DD.MM.YYYY).

\normalsize
\end{appendices}

\section{Data availability}
Full catalogs of optimal trajectories are available in the appendix.

\section*{Acknowledgments}

This research was funded by the Russian Science Foundation, Grant No. 25-79-00042, \url{https://rscf.ru/en/project/25-79-00042/}

\section*{Declaration of competing interest}

The authors have no competing interests to declare that are relevant to
the content of this article.

\bibliography{references}

\end{document}